\documentclass[aps,english,prb,floatfix,superscriptaddress,twocolumn,amsmath,amssymb,longbibliography,nofootinbib]{revtex4-2}
\usepackage[T1]{fontenc}
\usepackage{color}
\usepackage{amsmath}
\usepackage{amssymb}
\usepackage{graphicx}
\usepackage{dsfont}

\makeatletter

\providecommand{\tabularnewline}{\\}

\PassOptionsToPackage{caption=false}{subfig} 
\usepackage{hyperref}
\hypersetup{
breaklinks=true,
colorlinks=true,
citecolor=blue,
linkcolor=blue,
filecolor=blue,
urlcolor=blue
}
\IfFileExists{lmodern.sty}{\usepackage{lmodern}}{}

\makeatother

\usepackage{babel}
\begin{document}
\title{Signatures of infinite randomness in transport properties of strongly disordered spin chains}
\author{L. F. C. Faria}
\affiliation{Gleb Wataghin Institute of Physics, The University of Campinas (Unicamp), 13083-859 Campinas, SP,
Brazil}
\author{Victor L. Quito \thanks{vquito@ifsc.usp.br}}
\affiliation{Department of Physics and National High Magnetic Field Laboratory,
Florida State University, Tallahassee, Florida 32306, USA}
\affiliation{S\~{a}o Carlos Institute of Physics, University of S\~{a}o Paulo, IFSC – USP, 13566-590, S\~{a}o Carlos, SP, Brazil.}

\author{Jo\~{a}o C. Getelina}
\affiliation{S\~{a}o Carlos Institute of Physics, University of S\~{a}o Paulo, IFSC – USP, 13566-590, S\~{a}o Carlos, SP, Brazil.}

\affiliation{
 Department of Physics, Missouri University of Science and Technology, Rolla, MO 65409, USA}
\author{Jos\'{e} A. Hoyos}
\affiliation{S\~{a}o Carlos Institute of Physics, University of S\~{a}o Paulo, IFSC – USP, 13566-590, S\~{a}o Carlos, SP, Brazil.}

\affiliation{Max Planck Institute for the Physics of Complex Systems, N\"othnitzer Str. 38, 01187 Dresden, Germany}
\author{E. Miranda}
\affiliation{Gleb Wataghin Institute of Physics, The University of Campinas (Unicamp), 13083-859 Campinas, SP,
Brazil}
\date{\today}
\begin{abstract}

We study the spin transport properties of some disordered spin chains with a special focus on the distribution of the frequency-dependent spin conductivity. In the cases of interest here, the systems are governed by an effectively infinite disorder at low energies. A hallmark of this behavior is the wide discrepancy between the average and the typical values of some physical quantities, which are described by extremely broad distributions.  We show that such is also the case of the spin conductivity, whose average value is metallic but whose typical one, the physically relevant quantity, is insulating. This solves the apparent contradiction between the prediction of a spin metallic phase of the spin-1/2 disordered XX chain and its known localized behavior (after a mapping to free fermions). Our results are based on analytical and numerical implementations of a strong-disorder renormalization group as well as exact diagonalization studies. We present our analyses in very general terms, valid for systems of any spin $S$ value, but the cases of $S=1/2$ and 1 are studied in greater detail. 

\end{abstract}
\maketitle

\section{Introduction \label{sec:Introduction}}

In one-dimensional spin systems, the interplay of quantum fluctuations and disorder leads to distinctive behavior~\cite{igloi-review,Igloi2018,Vojtareview2006}. Broadly speaking, in a wide class of systems and regimes, which include the spin-1/2 Heisenberg and XXZ models, the effective disorder grows without limit at increasingly lower energies and longer length scales. 
This unique feature allows for unprecedented detail and precision in their theoretical description, which is characterized by many universal features, for a broad range of initial bare coupling constant distributions. For example, thermodynamic properties can be obtained exactly at low temperatures $T$: the specific heat and the magnetic susceptibility behave as $c_V(T)\sim \left|\ln T\right|^{-\left(1+1/\psi\right)}$ and  $T \chi(T) \sim \left|\ln T\right|^{-1/\psi}$, where $\psi$ is a universal exponent. The most striking signature of this Infinite Randomness Fixed Point (IRFP)~\cite{fisher94-xxz}, however, is the extremely broad distribution of some physical quantities. For example, the spin correlation function in the ground state, $C(x)=\left\langle S(0)S(x)\right \rangle$, shows a power-law decay in its arithmetic average value $C_{\text{av}} \sim x
^{-2}$, whereas its geometric average (typical) value is a stretched exponential $C_{\text{typ}} \sim e^{-\sqrt{x/\xi^\prime}}$ ~\cite{fisher94-xxz,Getelina2020,XavierHoyosMiranda_PRB_2018}. All of these results stem from a rather simple approximate physical picture of the ground state and the low-energy excitations. According to this picture, the ground state consists of a collection of singlets formed by spins at random positions and arbitrarily large distances~\cite{fisher94-xxz}, the so-called ``random singlet phase''. The low-energy excitations amount to breaking the most weakly bound random singlets. All of these results were obtained through a Strong Disorder Renormalization Group (SDRG) method, which leads to exact results when the flow of the effective disorder is governed by an IRFP
~\cite{igloi-review,Igloi2018}. 

Dynamical properties of several random spin $S$ chains (with $S=1/2$ and 1) have also been studied with the SDRG ~\cite{DamleMotrunichHusePRL,MotrunichPRB2001}. In particular, the dynamical spin correlation function and the frequency-dependent spin conductivity of these chains have been analyzed in the low frequency and long wavelength limits. A surprising result of this analysis was the observation of a spin-metallic phase (i.e., with a diverging spin conductivity in the thermodynamic limit) for all the random singlet phases of these systems. As is well known, single-particle states are always localized in one dimension in the presence of uncorrelated disorder \cite{MottTwose_1961,Borland_1963}, but interaction effects may well lead to a metallic behavior. An intriguing feature, however, is the fact that one model found to be a spin metal is the spin-1/2 XX chain~\cite{DamleMotrunichHusePRL,MotrunichPRB2001}. This system can be mapped into a free-fermion system with a Wigner-Jordan transformation, which would seem to contradict the localized nature of single-particle states in one dimension. The equivalent free-fermion model has off-diagonal disorder only, which translates into particle-hole (chiral) symmetry. In this case, the localization properties of the state at $E=0$, $\psi
^{\mathrm{ph}}_{E=0}(x)$, reflect the additional symmetry and are anomalous. More specifically, its envelope is a stretched exponential $\psi
^{\mathrm{ph}}_{E=0}(x)\sim e^{-\sqrt{x/\xi^\prime}}$ in contrast to the exponentially localized behavior found in the generic situation $\psi(x)\sim e^{-x/\xi}$ \cite{FleishmanLicciardello_JPC_1977,Abrahamsetal_PRB_1994}. This anomaly has led to a great deal of confusion in the literature, with claims of a putative extended nature of the state~\cite{EggarterRiedinger_PRB_1978,TheodorouCohen_PRB_1976}. These have been dismissed, however, since they were largely based on a definition of the localization length $\xi$ that assumes incorrectly an exponentially localized state $\psi^{\mathrm{ph}}_{E=0}(x)\sim e^{-x/\xi}$, only to conclude that $\xi\to\infty$. 

This calls into question the diverging low-frequency spin conductivity found with the SDRG. A clue is the nature of the distribution of DC conductance $g$ in one dimension, which is known to be extremely large not only in the usual case of localization \cite{Anderson1980,Abrikosov1981,markos_2006} but also for the particle-hole symmetric free-fermion model of interest here \cite{SoukoulisEconomou_PRB_1981,MudryFurusaki_PRB_2000}. In the latter case and for a system of length $L$, it has a broad distribution with the average conductance behaving as $g_\text{av}\sim L^{-1/2}$, the standard deviation as $\Delta g \sim L^{-1/4}$, while the typical (geometric average) decays as $g_\text{typ} \sim e^{-\gamma\sqrt{L}}$. Therefore, the average DC conductivity diverges with the system size as $\sigma_\text{av}\sim  \sqrt{L}$, its standard deviation goes like $\Delta \sigma \sim L^{3/4}$, and its typical value is strongly suppressed $\sigma_\text{typ}\sim L e^{-\gamma\sqrt{L}}$, reflecting the anomalous localization of the single-particle state. In other words, the distribution is strongly non-self-averaging with both $\Delta g/g_\text{av} =\Delta \sigma/\sigma_\text{av} \to \infty$ and  $g_\text{av}/g_\text{typ} =\sigma_\text{av}/\sigma_\text{typ} \to \infty$ as $L\to\infty$. Note that this DC behavior can be captured within the SDRG~\cite{Mard_PRB_2014,Mard_PRB_2014}.
The Drude weight of the fermionic model (which translates into the spin stiffness of the XX chain) confirms these conclusions~\cite{LaflorencieRieger_EJPB_2004}. Since the results for the frequency-dependent conductivity were obtained for its average value, it is important to revisit this question from the perspective of the full distribution.

In this work, we show results for the zero-temperature frequency-dependent spin conductivity of spin-$S$ random chains in their random singlet phases. We pay particular attention to their distributions in general and to their average, variance, and typical values in particular. The distribution is shown to become increasingly broader as the frequency decreases. As a result, although the average spin conductivity diverges logarithmically as $\omega \to 0$ in the thermodynamic limit, its geometric average, which we take to be representative of its typical value, vanishes with decreasing frequency as expected from the DC behavior. This applies both to the random singlet phase of the spin-1/2 XXZ model and the two random singlet phases of the SU(2)-symmetric spin-1 chain \cite{Quito_PhysRevLett.115.167201}. In addition, our analytical results suggest that the same behavior should be expected for a much broader class of systems, i.e., any rotationally invariant spin-$S$ chain governed by an IRFP.~\citep{quitoprb2016}

Our analysis is based, for the most part, on an SDRG treatment, but we also show exact diagonalization results for the spin-1/2 XX chain, where the free-fermion mapping allows for an exact numerical treatment~\cite{Lieb1961}. We develop the necessary SDRG tools in a form that is valid for any spin-$S$ Hamiltonian with a conserved total magnetization in a given direction (taken to be the $z$-direction), although our main focus is on the particular cases of $S=1/2$ and 1. We thus generalize to any $S$ the tools previously formulated for $S=1/2$ and 1~\cite{DamleMotrunichHusePRL,MotrunichPRB2001}. Our implementation of the SDRG combines analytical calculations and numerical simulations. 

This paper is organized as follows. In Sec.~\ref{sec:Model}, we
introduce the class of spin models we are going to study, devoting special
attention to conserved quantities, in particular the $z$ component of the angular momentum. We review the SDRG technique
and the thermodynamic properties of the spin chains studied here in
Sec.~\ref{sec:Method}. The derivation of the current renormalization
is addressed in detail in Sec.~\ref{sec:Current-renormalization}.
We present analytical results for the average and the standard deviation
of the conductivity distribution in Sec.~\ref{sec:Conductivity-distribution}.
The numerical SDRG flow is presented in Sec.~\ref{sec:Numerical_results},
where the distributions of conductivity are probed for both anisotropic
spin-1/2 chains and bilinear-biquadratic spin-1 chains. Finally, in
Sec.~\ref{sec:conclusion} we summarize our findings and propose
future directions.

\section{Model~\label{sec:Model}}

We consider the most general random one-dimensional SU(2)-symmetric
spin-$S$ model with nearest-neighbor interactions,

\begin{equation}
H^{\text{SU(2)}}=\sum_{j}\sum_{k=1}^{2S}J_{j}^{\left(k\right)}\left(\mathbf{S}_{j}\cdot\mathbf{S}_{j+1}\right)^{k},\label{eq:generic-SU2-hamilt}
\end{equation}
where $J_{j}^{\left(k\right)}$ are independent random variables, and $\mathbf{S}_{j}$ are the usual spin operators.
The sum over $k$ runs from 1 to $2S$~\cite{quitoprb2016}, resulting
in $2S$ possible linearly independent terms. In the particular case of spin-1/2, we allow for anisotropy in the $z$ direction, yielding the so-called XXZ model, whose Hamiltonian is given by
\begin{equation}
H^{\text{XXZ}}=\sum_{j}J_{j}^{\perp}\left[\left(S_{j}^{x}S_{j+1}^{x}+S_{j}^{y}S_{j+1}^{y}\right)+\Delta_{j}S_{j}^{z}S_{j+1}^{z}\right],\label{eq:spin-1/2-H}    
\end{equation}
with the random variables $J_{j}^{\perp}>0$ and $-1/2 \le  \Delta_{j} \le 1$ obtained from independent distributions. We use units such that $\hbar=1$.

To analyze the transport properties, the first step is to identify the constants of the motion. Given the symmetry of the Hamiltonian in Eq.~\eqref{eq:generic-SU2-hamilt} under global rotations, the three components of the total angular momentum are conserved. As for Eq.~\eqref{eq:spin-1/2-H}, the SO(2) symmetry group of rotations around the $z$ axis guarantees that only the $z$ component of the total angular momentum is conserved. Moreover, since these Hamiltonians are independent of time, the energy is also conserved. In this work, we focus on spin transport only, leaving heat transport for future work.

In order to unify the approach for both kinds of systems, we
focus on the $z$ component of spin, whose density is $n_{j}=\frac{1}{a}S_{j}^{z}$, where $a$ is the lattice spacing. A consequence of the above symmetries is the local conservation of the spin current (density), described by the continuity equation

\begin{equation}
\frac{\partial n_{j}}{\partial t}+\partial_{x}\tau_{j}=0.
\end{equation}
Here, $\tau_{j}$ represents the current flowing from site $j$ to $j+1$ and $\partial_{x}\tau_{j}$ is a short-hand notation for the lattice derivative $\partial_{x}\tau_{j}=\frac{\tau_{j+1}-\tau_{j}}{a}$. Using the Heisenberg equation of motion, $\frac{\partial n_{j}}{\partial t}$ is related to the commutator of $n_{j}$ with the Hamiltonian, yielding

\begin{equation}
\partial_{x}\tau_{j}=i\left[n_{j},H\right].\label{eq:diff-eq-current}
\end{equation}

By writing the Hamiltonians in Eqs.~\eqref{eq:generic-SU2-hamilt}
and \eqref{eq:spin-1/2-H} generically as $H=\sum_{j}H_{j,j+1}$,
Eq.~\eqref{eq:diff-eq-current} can be solved for the current operator on a given bond,
\begin{align}
\tau_{j} & =i\left[S_{j}^{z},H_{j,j+1}\right],\nonumber \\
 & =-i\left[S_{j+1}^{z},H_{j,j+1}\right].\label{eq:tau-def}
\end{align}
The two equivalent ways of writing \eqref{eq:tau-def} can be easily verified by subtracting one from the other and noticing that $\left[S_{j}^{z}+S_{j+1}^{z},H_{j,j+1}\right]=0$,
given the SO(2) symmetry of all the models considered here. We will take advantage of these two equivalent forms to simplify the calculations in the next sections. The total current is found by summing over all sites, $\tau=\sum_{j}\tau_{j}$.

The commutators in Eq.~\eqref{eq:tau-def} can be explicitly calculated
for both Eqs.~\eqref{eq:generic-SU2-hamilt}~and~\eqref{eq:spin-1/2-H}.
For the spin 1/2 case~\cite{MotrunichPRB2001,DamleMotrunichHusePRL}, 
we find
\begin{equation}
\tau_{j}^{\text{XXZ}}=i\frac{J_{j}^{\perp}}{2}\left(S_{j}^{+}S_{j+1}^{-}-S_{j}^{-}S_{j+1}^{+}\right).\label{eq:spin-1/2-current}
\end{equation}
For the generic SU(2)-symmetric class of spin-$S$ Hamiltonians in Eq.~\eqref{eq:generic-SU2-hamilt}, it is convenient
to rewrite the dot product operators $\left(\mathbf{S}_{j}\cdot\mathbf{S}_{j+1}\right)^{k}$
in terms of irreducible spherical tensors, whose simple commutation
relations with $S_{j}^{z}$ make the calculations dramatically easier~\cite{quitoprb2016}.
We do not take this route here, as the explicit form of $\tau_{j}$
is unnecessary for the future sections. However, we devote particular
attention to the $S=1$ case, which is analyzed in greater detail
later on. In this case, going back to Eq.~\eqref{eq:generic-SU2-hamilt},
we call $J_{j}=J_{j}^{\left(1\right)}$, $D_{j}=J_{j}^{\left(2\right)}$
and write the Hamiltonian as
\begin{equation}
H^{S=1}=\sum_{j}\left[J_{j}\mathbf{S}_{j}\cdot\mathbf{S}_{j+1}+D_{j}\left(\mathbf{S}_{j}\cdot\mathbf{S}_{j+1}\right)^{2}\right].\label{eq:spin-1-H}
\end{equation}
This case is simple enough for the commutator in Eq.~\eqref{eq:tau-def}
to be computed without introducing the spherical tensors. A straightforward
calculation then yields
\begin{align}
\tau_{j}^{S=1} & =i\frac{J_{j}}{2}\left(S_{j}^{+}S_{j+1}^{-}-S_{j}^{-}S_{j+1}^{+}\right)\nonumber \\
 & +i\frac{D_{j}}{2}\left[\left(S_{j}^{+}S_{j+1}^{-}-S_{j}^{-}S_{j+1}^{+}\right)\mathbf{S}_{j}\cdot\mathbf{S}_{j+1}-\mathrm{h}.\mathrm{c}.\right].\label{eq:spin-1-current}
\end{align}

The approach we take to analyze the transport is to implement an SDRG
procedure, which allows us to numerically calculate the conductivity distribution while computing its average and standard deviation
analytically. This distribution ultimately determines the insulating
or metallic character of the disordered chain. Before addressing the
transport properties, we review in the next section the SDRG procedure
for these systems, as well as their SDRG flow.

\section{The SDRG method~\label{sec:Method}}

\begin{figure}
\includegraphics[width=1\columnwidth]{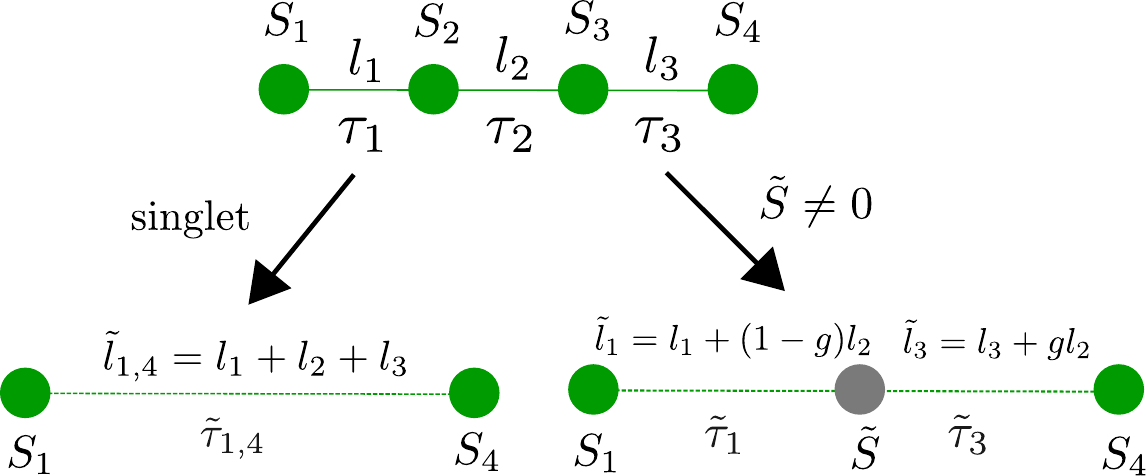}\caption{
Schematic real-space renormalization of the spin chain and its effect on the current operator for arbitrary spins $S_{i}$. We assume that the largest local gap comes from sites 2 and 3. There are two types of processes, depending on the total angular momentum of the ground state of the most strongly coupled pair. If the ground state is a singlet, the two sites are removed from the chain, leading to an effective current operator between sites $1$ and $4$ ($\tilde{\tau}_{1,4}$), which are separated by $\tilde{l}_{1,4}$. If the ground state is not a singlet, an effective spin $\tilde{S}$ is added to mimic the ground state multiplet. The current operators $\tau_{1,3}$ are renormalized and the distance between spins $2$ and $3$ is distributed among $\tilde{l}_1$ and $\tilde{l}_3$. In the cases studied in detail in this work, $\tilde{S}=S$, implying $g=1/2$ and the distance $l_2$ getting distributed equally. \label{fig:current-renorm}}
\end{figure}

In order to analyze the transport properties, we implement the SDRG
approach to the models defined in Eqs.~\eqref{eq:generic-SU2-hamilt}
and \eqref{eq:spin-1/2-H}. For future reference, we will briefly review how the method works, especially as applied to the systems of interest here.

The basic SDRG step consists of replacing
the most strongly coupled pairs of sites by their lowest-lying multiplet while
renormalizing the neighboring couplings. For each bond $\left(j,j+1\right)$,
there is an associated local gap between its ground state and the first excited state; see Fig.~\ref{fig:current-renorm}. Let us assume, for
concreteness, that the largest local gap of the chain, called $\Omega$, is formed
by the pair of sites 2 and 3, and that this bond is much stronger than the neighboring ones.
Throughout this section, we split the Hamiltonian into two parts, i.e., $H=H_0+V$, with

\begin{equation}
H_{0}=H_{2,3}, \quad V=H_{1,2}+H_{3,4},\label{eq:H_broken}
\end{equation}
indicating that the bonds connected to sites 2 and 3 are small perturbations
to the strongly coupled pair. The bonds not connected to sites
2 and 3 do not change to leading order in perturbation theory, and are thus ignored in the decimation step.

\begin{figure}
\includegraphics[width=1\columnwidth]{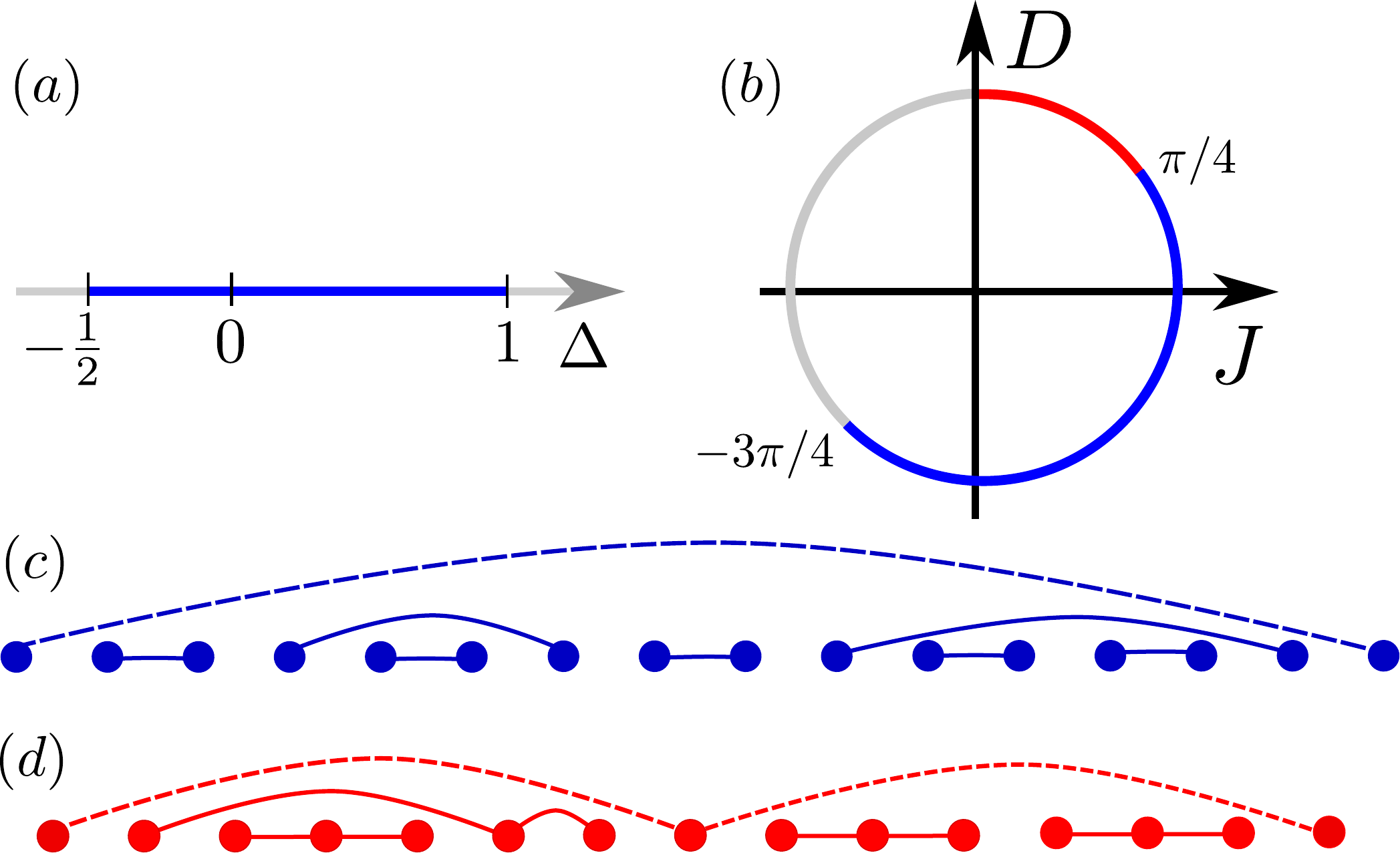}\caption{
Phase diagram of (a) random spin-1/2 (XXZ) and (b) random spin-1 chains. The blue and red regions correspond to mesonic (mRSP) and baryonic (bRSP) random singlet phases, respectively. The gray regions are outside the scope of this work. (c) For the mRSP, the ground state consists of singlets formed by pairs of spins located at random positions. The singlets can be formed by spins arbitrarily far apart. Typically, long singlets are weakly bound and are the first ones to be broken at temperatures comparable to their effective coupling (dashed line). (d) For spin-1 chains, there is also a bRSP in which the ground state consists of singlets formed by spin trios (or with higher multiples of 3).  \label{fig:GS-excitations}}
\end{figure}

In the spin-1/2 XXZ case [Eq.~\eqref{eq:spin-1/2-H}], $\Omega=J^{\perp}_{2}(1+\Delta_{2})/2$.
The pair of spins 2 and 3 is then frozen
in its singlet ground state. A new coupling between spins 1 and 4 is found
in second-order perturbation theory, originating in virtual excitations
of the frozen singlet~\cite{fisher94-xxz}, and is given by
\begin{align}
\tilde{J}^{\perp}_{14}&=\frac{J^{\perp}_{1}J^{\perp}_{3}}{J^{\perp}_{2}\left(1+\Delta_{2}\right)}
=\frac{J^{\perp}_{1}J^{\perp}_{3}}{2\Omega}.
\label{eq:J14_spin_half}
\end{align}
On the other hand, the anisotropy renormalizes as
\begin{align}
\tilde{\Delta}_{14}&=\Delta_{1}\Delta_{3}\frac{\left(1+\Delta_{2}\right)}{2}.
\label{eq:Delta-spin-half}
\end{align}
For the spin-1 case [Eq.~\eqref{eq:spin-1-H}],  the decimation rules are richer, as the local ground state can be either a singlet, a triplet, or a quintuplet. These possibilities are represented in Fig.~\hyperref[fig:GS-excitations]{\ref{fig:GS-excitations}(b)}. For this analysis, it proves convenient to define polar variables $(r_j,\theta_j)$  in the $(J_j,D_j)$ plane: $r_j=\sqrt{J_j^2+D_j^2}$ and $\tan\theta_{j}=\frac{D_{j}}{J_{j}}$. The different local ground states of spins 2 and 3 are then determined by the value of the angle $\theta_2$. For $-3\pi/4<\theta_{2}<\arctan\frac{1}{3}$,
the two spins are locked in a singlet
ground state. The neighboring couplings are renormalized in second-order
perturbation theory according to~\cite{PhysRevLett.80.4562,Quito_PhysRevLett.115.167201}
\begin{align}
\tilde{J}_{14} & =\frac{4\left(J_{1}-D_{1}/2\right)\left(J_{3}-D_{3}/2\right)}{3\left(J_{2}-3D_{2}\right)}-\frac{D_{1}D_{3}}{9\left(J_{2}-D_{2}\right)},\label{eq:J14-order-2}\\
\tilde{D}_{14} & =-\frac{2D_{1}D_{3}}{9\left(J_{2}-D_{2}\right)}.\label{eq:D14-order-2}
\end{align}
If $\arctan\frac{1}{3}<\theta_{2}<\pi/2$, then the local ground state is instead a triplet, the pair of spins is replaced by an effective spin-$1$ degree of freedom, and the adjacent couplings are renormalized
according to

\begin{align}
\tilde{J}_{j} & =\frac{J_{j}}{2}+\frac{D_{j}}{4},\label{eq:J1-order1-triplet}\\
\tilde{D}_{j} & =-\frac{D_{j}}{2}.\label{eq:D1-order1-triplet}
\end{align}
In this work, we do not consider flows in the region of the phase diagram that require decimations where $\frac{\pi}{2}<\theta_{2}<\frac{5\pi}{4}$, when the pair of spins 2 and 3 is replaced by an effective spin-2 degree of freedom. In this case, the SDRG generates spins of all sizes $S>1$~\cite{quitoprb2016}, and the technique used below to compute the transport properties needs to be generalized.

We now review the flow of both the spin-1/2 XXZ and the spin-1 models. From the structure of the SDRG transformations, it is obvious that the ground state consists of a collection of singlets formed by spins at arbitrary distances, a so-called random singlet phase (RSP)~\cite{fisher94-xxz,linetal98}. The ground-state structures of the models studied here are shown in  Fig.~\hyperref[fig:GS-excitations]{\ref{fig:GS-excitations}(c)} and \hyperref[fig:GS-excitations]{(d)}.

We consider the spin-1/2 XXZ model with random $J_\perp$ and fixed $\Delta$. In this case, the anisotropy is an irrelevant perturbation around the $\Delta=0$ point, whose basin of attraction extends from -1/2 to 1, as depicted in Fig.~\hyperref[fig:GS-excitations]{\ref{fig:GS-excitations}(a)}~\cite{fisher94-xxz}. Asymptotically, the energy scale $\Omega$ is determined, therefore, only by the values of $J_j^{\perp}$. The Heisenberg point $\Delta_{i}=1, \forall i$, is special in that it exhibits an enlarged SU(2)-symmetry, which is preserved by the SDRG flow. As a consequence, $\Delta_{i}$ remains equal to one throughout the flow, as can be checked from Eq.~\eqref{eq:Delta-spin-half}. In either case, the ground state is a collection of random singlet pairs, as shown schematically in Fig.~\hyperref[fig:GS-excitations]{\ref{fig:GS-excitations}(c)}.

The SDRG flow of the spin-1 chain, Eq.~\eqref{eq:Ham_spin1_polar}, is more easily analyzed by working in polar variables. In this case,
the Hamiltonian is rewritten as 
\begin{equation}
H^{S=1}=\sum_{j}r_{j}\left[\cos\theta_{j}\left(\mathbf{S}_{j}\cdot\mathbf{S}_{j+1}\right)+\sin\theta_{j}\left(\mathbf{S}_{j}\cdot\mathbf{S}_{j+1}\right)^{2}\right]\label{eq:Ham_spin1_polar}
\end{equation}
In this paper, we focus only on initial bare distributions with uniform angles $\theta_{j}=\theta_{0}, \forall j$, whereas $r_{j}$ is assumed to be random with a sufficiently broad distribution. The ground state and thermodynamic properties are determined by the initial angle $\theta_{0}$. When $-3\pi/4<\theta_{0}<\pi/4$ [see the blue sector of Fig.~\hyperref[fig:GS-excitations]{\ref{fig:GS-excitations}(b)}], the ground state consists of a collection of random singlets formed by pairs of spins, a structure identical to the spin-1/2 case [Fig.~\hyperref[fig:GS-excitations]{\ref{fig:GS-excitations}(c)}]. The low-energy excitations consist of free spins-1 found by thermally breaking such singlets. The ground state and low-energy excitations are SU(3)-symmetric, an emergent symmetry not present in the bare Hamiltonian~\cite{Quito_PhysRevLett.115.167201}. This is a consequence of a generic feature of SO(N) random spin chains, which exhibit enlarged SU(N) symmetric random-singlet ground states~\cite{QuitoLopes2019PRB,QuitoLopes2020EJPB}. Because the singlet pairs can be viewed as bound states of SU(3)-symmetric quarks and antiquarks, in analogy with hadronic physics, we call this phase \emph{mesonic}. On the other hand, for $\pi/4<\theta_{0}<\pi/2$ [the red sector in Fig.~\hyperref[fig:GS-excitations]{\ref{fig:GS-excitations}(b)}], the ground state consists of a collection of singlets formed by integer multiples of three spins, as shown schematically in Fig.~\hyperref[fig:GS-excitations]{\ref{fig:GS-excitations}(d)}. Again, the ground and low-lying excited states formed by breaking the singlets are SU(3)-symmetric. This phase is referred to as \emph{baryonic}, as the singlet trios can be seen as bound states of three SU(3) quarks, here realized as spins-1~\cite{Quito_PhysRevLett.115.167201}.
Even though the initial angle $\theta_{j}$ is assumed to be the same for all bonds, as the SDRG proceeds, the angles also become disordered. At low energies, however, the distribution of $\theta_{j}$ flows back to a delta function, and the angles tend to a unique fixed point value. These are the
angular fixed points~\cite{Quito_PhysRevLett.115.167201,quitoprb2016}. Also importantly, at low energies, the flows of $\theta$ and $r$
decouple from each other.

The SDRG flow of the  SU(2)-symmetric spin $S>1$ systems of Eq.~\eqref{eq:generic-SU2-hamilt}
have been studied in Ref.~\cite{quitoprb2016}.  As mentioned before, generic SU(2)-symmetric spin-S chains have $2S$ independent couplings per bond. Similarly to the $S=1$ case, the SDRG rules are simplified when recast in terms of radial and angular variables. A collection of angular fixed-points are found~\cite{quitoprb2016}, a generalization of the ones found for the spin-1 case. For the generic SDRG rules, we refer the reader to Ref.~\cite{quitoprb2016}. When the spin sizes do not grow under the SDRG flow,  different types of random singlet phases are found for $S>1$ which are all fairly similar to the mesonic spin-1 phase described above.

For both the XXZ spin-1/2 and the spin-1 case [in fact, for all the SU(2)-invariant models], if the initial disorder is strong enough, the effective disorder strength, measured by the ratio of the standard deviation to the average of the renormalized distribution, grows without bounds under the SDRG transformations ~\cite{fisher94-xxz,igloi-review}. The SDRG method becomes, therefore, asymptotically exact, and the physics is governed by an infinite randomness fixed point (IRFP). The RSPs are distinguished by the tunneling exponent $\psi$, which governs the activated dynamical scaling of the energy with the length of excitations \cite{fisher94-xxz}: $\ln(\Omega)\sim - L^\psi$. In the models of our interest, $\psi=1/k$, where $k$ is the minimal number of spins forming singlets in that phase,  thus, $k=2$ for the XXZ and mesonic phase and $k=3$ for the baryonic phase (see Fig.~\ref{fig:GS-excitations}). Physical observables, such as specific heat and magnetic susceptibility, are determined by $\psi$. For instance, the magnetic susceptibility reads $\chi\sim\frac{1}{T \log T^{1/\psi}}$.

For transport properties, besides the coupling distribution, it is also necessary to follow how the bond lengths flow. Given the structure of the SDRG flow, it is generically clear that if a pair of sites $j$ and $j+1$ is removed (singlet decimations), the length $l_{j}$ is renormalized as 

\begin{equation}
\tilde{l}_{j}=l_{j-1}+l_{j}+l_{j+1}.\label{eq:length-order-2}
\end{equation}

For first-order decimations, in which the pair $j$ and $j+1$ is replaced by a new effective spin, the new lengths are

\begin{align}
\tilde{l}_{j-1} & =l_{j-1}+\frac{1}{2}l_{j},\label{eq:length-order-1a}\\
\tilde{l}_{j+1} & =l_{j+1}+\frac{1}{2}l_{j},\label{eq:length-order-1b}
\end{align}
i.e, we assume the length of the decimated bond $l_{j}$ is equally split
between its neighbors. For a schematic representation of the SDRG step, see Fig.~\ref{fig:current-renorm}. Generically, the characterization of
the SDRG flow involves the joint distribution of couplings and lengths.
These distributions remain correlated to each other, even asymptotically~\cite{fisher94-xxz}.
This feature has important consequences
for transport properties and we will come back to it in Secs.~\ref{sec:Current-renormalization}~and~\ref{sec:Conductivity-distribution}.
After revising the SDRG scheme, we are now ready to address the transport
properties of the spin chains.

\section{Current renormalization~\label{sec:Current-renormalization}}

In the spirit of the SDRG, we need to work out the renormalization
of the current operator when a given pair of sites is decimated, following
the decimation steps described in Sec.~\ref{sec:Method}. This was first done for $S=1/2$ and $1$
in Refs.~\cite{DamleMotrunichHusePRL,MotrunichPRB2001}.
We repeat this procedure here with three goals in mind. First, we want to
generalize it to higher-spin Hamiltonians. Second, we will use a slightly
different technique which is more flexible and also generalizable
to other transport properties. Finally, this approach will allow us to reach an important conclusion: the current \emph{retains its form under renormalization} [see, e.g., Eqs.~\eqref{eq:spin-1/2-current} and \eqref{eq:spin-1-current}] with the difference that the coupling constants are the same ones as those that appear in the renormalized Hamiltonian [Eqs.~\eqref{eq:J14_spin_half}, \eqref{eq:Delta-spin-half}, \eqref{eq:J14-order-2} and \eqref{eq:D14-order-2}].

We start by focusing on the $T=0$ Kubo formula
for the frequency-dependent spin conductivity in its Lehmann representation~\cite{MotrunichPRB2001,DamleMotrunichHusePRL}
\begin{equation}
\sigma(\omega)=\frac{1}{\omega L}\sum_{m}|\langle m|\sum_{j}\tau_{j}|0\rangle |^{2}\delta\left(\omega-E_{m}\right),\label{eq:Kubo-formula}
\end{equation}
where $\langle m|\sum_{j}\tau_{j}|0\rangle $
is the matrix element of the total current operator between the many-body
ground state $\left|0\right\rangle $ and excited states $\left|m\right\rangle $ with energy
$E_{m}$ (measured with respect
to the ground state), and $L$ is the system size.

The SDRG decimation contains, as any renormalization group procedure,
two elements: (i) a truncation of the Hilbert space (elimination of
some high-energy states), and (ii) a change of the truncated Hamiltonian
(a ``renormalization'') that incorporates into it some information
from the eliminated states. This renormalization is performed in order
to retain, up to some desired accuracy, the form of the low-energy
spectrum in the truncated Hilbert space. Analogously, when focusing
on some physical operator such as the current, it should also suffer
a renormalization so that, after truncation, the operator retains
the same action in the truncated Hilbert space as it had before (again,
up to the desired accuracy). Thus, we should determine how the current
operator is renormalized at each SDRG decimation step. In order to
do this, we again assume that the decimated, strongest local gap is
between sites 2 and 3. The current operator across the four sites
considered here can be written as

\begin{equation}
\tau_{1,4}=l_{1}\tau_{1}+l_{2}\tau_{2}+l_{3}\tau_{3},\label{eq:tau_14_before}
\end{equation}
where $l_{j}$ is the length of the $j$-th bond. These lengths are
equal to 1 in the original chain but, as will be shown later, renormalized current operators add algebraically just like the bond lengths under the SDRG flow. Therefore, we need to use the form in Eq.~\eqref{eq:tau_14_before} if we want to follow the current renormalization. The other
current operators $\tau_{j}$, with $j\ne1,2,3$ are not affected
by the decimation step.

The renormalized current has contributions from each term of
Eq.~\eqref{eq:tau_14_before}, leading to an effective current between
sites 1 and 4 given by
\begin{equation}
\tilde{\tau}_{1,4}\equiv l_{1}\tilde{\tau}_{1}+l_{2}\tilde{\tau}_{2}+l_{3}\tilde{\tau}_{3},\label{eq:tau_14_after}
\end{equation}
where we define $\tilde{\tau}_{j}$ as the contribution of $\tau_{j}$
to the renormalized $\tilde{\tau}_{1,4}$ . It is important to note
that each of these three contributions must be written in terms of
the degrees of freedom in the truncated Hilbert space. Therefore, in this case,
none of them can involve the spins 2 and 3. The details of
the calculations of these contributions are shown in Appendix~\ref{sec:current-operator-derivation}.
Here, we only summarize the results.

Let us consider the case in which the ground state of the strongest coupled spin pair is a singlet [see Fig.~\ref{fig:current-renorm}]. 
$\tilde{H}_{1,4}$ represents the effective
Hamiltonian connecting sites 1 and 4 after the first SDRG step (remember that sites 2 and 3 were removed from the chain). Note that $\tilde{H}_{1,4}$ is written in terms of the renormalized coupling constants {[}e.g., as in Eqs.~\eqref{eq:J14-order-2} and \eqref{eq:D14-order-2}{]}. In this case, the three contributions $\tilde{\tau}_{j}\ \left(j=1,2,3\right)$ are all equal (see Appendix~\ref{sec:current-operator-derivation})
and can be conveniently written as a commutator involving only objects that live in the truncated Hilbert space {[}cf. Eq.~\eqref{eq:tau-def}{]}

\begin{align}
\tilde{\tau}_{1}=\tilde{\tau}_{2}=\tilde{\tau}_{3} & =i\left[S_{1}^{z},\tilde{H}_{1,4}\right],\label{eq:tau-renorm-result}\\
 & =-i\left[S_{4}^{z},\tilde{H}_{1,4}\right].
\end{align}
The renormalized current operator has retained its form but it now involves the effective couplings present in the renormalized Hamiltonian
$\tilde{H}_{1,4}$. 
From Eqs.~\eqref{eq:tau-renorm-result}  and \eqref{eq:tau_14_after} we obtain, in general,

\begin{equation}
\tilde{\tau}_{1,4}=\left(l_{1}+l_{2}+l_{3}\right)\tilde{\tau}_{1}\equiv\tilde{l}_{1}\tilde{\tau}_{1}.
\end{equation}
Note how the lengths that multiply the current operator are renormalized in the same way as the bond lengths themselves, as given in Eq.~\eqref{eq:length-order-2}.
This result is quite general and valid for \emph{any} spin-$S$ Hamiltonian
(with a conserved total $S^{z}$). For the XXZ spin-1/2 model, the
decimation procedure has been previously derived in Refs.~\cite{MotrunichPRB2001,DamleMotrunichHusePRL},
and

\begin{equation}
\tilde{\tau}_{1}^{\text{XXZ}}=i\frac{\tilde{J}_{1,4}^{\perp}}{2}\left(S_{1}^{+}S_{4}^{-}-S_{1}^{-}S_{4}^{+}\right), \label{eq:tau_ren_XXZ}
\end{equation}
with $\tilde{J}^{\perp}_{14}$ given in Eq.~\eqref{eq:J14_spin_half}. As
to the spin-1 chain,

\begin{align}
\tilde{\tau}_{1}^{S=1} & =i\frac{\tilde{J}_{1,4}}{2}\left(S_{1}^{+}S_{4}^{-}-S_{1}^{-}S_{4}^{+}\right)+\nonumber \\
 & +i\frac{\tilde{D}_{1,4}}{2}\left[\left(S_{1}^{+}S_{4}^{-}-S_{1}^{-}S_{4}^{+}\right)\mathbf{S}_{1}\cdot\mathbf{S}_{4}-\mathrm{h}.\mathrm{c}.\right].\label{eq:spin-1-current-1}
\end{align}
Notice that, as anticipated, Eqs.~\eqref{eq:tau_ren_XXZ} and~\eqref{eq:spin-1-current-1} retain the functional forms of Eqs.~\eqref{eq:spin-1/2-current} and \eqref{eq:spin-1-current}. 

When the ground state of the strongest coupled pair is not a singlet but rather a spin $\tilde{S}$ multiplet [see Fig.~\ref{fig:current-renorm}], the SDRG step involves, as mentioned before, removing sites 2 and 3 and replacing them with an effective site with spin $\tilde{S}$. Here, $\tilde{S}$ represents the ground-state multiplet of the spin pair. The renormalized Hamiltonian after the
SDRG step can be written as $\tilde{H}_{1,4}=\tilde{H}_{1,\left(2,3\right)}+\tilde{H}_{\left(2,3\right),4}$,
where we use $\left(2,3\right)$ to denote the effective spin-$\tilde{S}$.
The calculation in Appendix~\ref{sec:current-operator-derivation}
shows that

\begin{equation}
\tilde{\tau}_{1,4}=\left(l_{1}+(1-g)l_{2}\right)\tilde{\tau}_{1}+\left(l_{3}+gl_{2}\right)\tilde{\tau}_{3},\label{eq:tau-renorm-result2}
\end{equation}
with 
\begin{align}
\tilde{\tau}_{1} & =i\left[{S}_{1}^{z},\tilde{H}_{1,\left(2,3\right)}\right],\label{eq:tau-renorm-result3}\\
\tilde{\tau}_{3} & =-i\left[S_{4}^{z},\tilde{H}_{\left(2,3\right),4}\right],\label{eq:tau-renorm-result4}
\end{align}
where 
\begin{equation}
g=\frac{\tilde{S}\left(\tilde{S}+1\right)+S_{2}\left(S_{2}+1\right)-S_{3}\left(S_{3}+1\right)}{2\tilde{S}\left(\tilde{S}+1\right)}. \label{eq:Lande2}
\end{equation}
For the cases considered in this work, $S_2=S_3$ implying that $g=1/2$. 
Once again, $\tilde{\tau}_{1}$ and $\tilde{\tau}_{3}$ retain their original form and 
are conveniently written as commutators involving only objects in
the truncated Hilbert space. Moreover, this result is also valid for
any spin-$S$ Hamiltonian (with a conserved total $S^{z}$). Particularizing
to $S=1$,
\begin{align}
\tilde{\tau}_{1}^{S=1} & =i\frac{\tilde{J}_{1}}{2}\left(S_{1}^{+}\tilde{S}^{-}-S_{1}^{-}\tilde{S}^{+}\right)\nonumber \\
 & +i\frac{\tilde{D}_{1}}{2}\left[\left(S_{1}^{+}\tilde{S}^{-}-S_{1}^{-}\tilde{S}^{+}\right)\mathbf{S}_{1}\cdot\tilde{\mathbf{S}}-\mathrm{h}.\mathrm{c}.\right],\label{eq:spin-1-current-2}
\end{align}
where $\tilde{\mathbf{S}}$ is the spin operator in the
ground multiplet $\tilde{S}$ and
$\tilde{J}_{1}$ and $\tilde{D}_{1}$ are given in Eqs.~\eqref{eq:J1-order1-triplet}
and \eqref{eq:D1-order1-triplet} ($\tilde{\tau}_{3}$ follows analogously).

The decimation process is depicted in Fig.~\ref{fig:current-renorm}.
By iterating the above procedure, it becomes clear that, as the energy
scale is lowered from $\Omega_{0}$ to $\Omega$,

\begin{equation}
\sum_{j}\tau_{j}\rightarrow\tilde{\sum}_{j}\tilde{l}_{j}\tilde{\tau}_{j},
\end{equation}
with the sum on the r.h.s restricted to the non-decimated sites and
the length scales $\tilde{l}_{j}$ and operators $\tilde{\tau}_{j}$
calculated at the RG scale $\Omega$ after sequentially applying Eqs.~\eqref{eq:length-order-2}, and~\eqref{eq:tau-renorm-result},
for second-order decimations, or Eqs.~\eqref{eq:length-order-1a}, \eqref{eq:length-order-1b}, \eqref{eq:tau-renorm-result3}, and \eqref{eq:tau-renorm-result4}, for first-order decimations.

We now address how the Kubo formula is used in conjunction with the
SDRG procedure. In order to find $\sigma\left(\omega\right)$, one
runs the SDRG until the cutoff energy scale $\Omega=\omega$, where
the Kubo formula becomes
\begin{equation}
\sigma(\omega)=\frac{1}{\omega L}\sum_{m}\left|\left\langle m\left|\tilde{\sum}_{j}\tilde{l}_{j}\tilde{\tau}_{j}\right|0\right\rangle \right|^{2}\delta\left(\omega-\tilde{E}_{m}\right),\label{eq:Kubo-formula-renorm}
\end{equation}
where $\tilde{E}_{m}$ are the excitation energies in the renormalized spectrum. The decimations now occur at bonds with $\tilde{E}_{m}\sim\omega$. When applied to these bonds, the current operators $\tilde{\tau}_{j}$  connect the ground multiplets to some of these excited state multiplets (see Appendix~\ref{sec:Matrix-elements-current}), thus contributing to the conductivity $\sigma\left(\omega\right)$ at scale $\omega$. The only task left is to compute the matrix element of the current operators in Eq.~\eqref{eq:Kubo-formula-renorm} between the ground and excited multiplets of the decimated bonds. As this is relevant for the conductivity distribution, we leave this calculation for the next section.

\section{Features of the conductivity distribution~\label{sec:Conductivity-distribution}}

In this section, we obtain analytical expressions for the average and standard deviation of the spin conductivity distribution. This distribution will be confronted against a numerical implementation of the SDRG method and a numerical exact diagonalization in the next section. Our main result is that, while both the average and standard deviation decrease throughout the SDRG flow, the average decreases\emph{ faster}, implying that the distribution broadens indefinitely. In other words, the infinite randomness fixed point is reflected in the conductivity distribution as well, which becomes arbitrarily broad. This is not surprising, as the DC conductivity (or conductance) distribution of one-dimensional disordered non-interacting quantum particles is known to be extremely broad~\cite{Anderson1980,Abrikosov1981}. As a consequence, at low energies, the average \emph{does not} give a reliable indicator of the phase of the system. Rather, one should look at the typical or the geometric average of the conductivity as the correct diagnostic tool~\cite{Anderson1980}. In addition, since, as we will show, the conductivity distribution can be obtained directly from the universal fixed-point distributions of couplings, it is also universal.

The first step of the calculation is to compute the matrix elements
in the Kubo formula, Eq.~\eqref{eq:Kubo-formula-renorm}. For that, we have to highlight some important ingredients. First, generically,
after the initial transient, the system flows towards a fixed point completely
described by one set of random couplings. For the XXZ case, these
variables are $J_{j}^{\perp}$ as the $\Delta_{j}$ flow to
zero at the RSP considered here, with the exception of the SU(2)-symmetric point
$\Delta_{j}=1$, which is described by a single
set of couplings from the beginning. As for the SU(2)-symmetric spin-$S$
chains, the angular fixed points (in the $S=1$ case, there is only one angle $\theta=\theta_{\text{FP}}$) fix
the \emph{ratio} of distinct couplings and the distribution of couplings
can again be described by a single parameter~\cite{quitoprb2016}.
Let us call this generic coupling $K_{j}$. 

Second, we point out that the $S_z$ spin current is one component of a vector in spin space (the other components being the $S_x$ and $S_y$ spin currents). In fact, it is the zeroth component of a rank-1 irreducible spherical tensor. Thus, from the Wigner-Eckart
theorem~\cite{cohen1977quantum,Edmondsbook}, its only non-vanishing matrix elements are between states whose total spins differ by 1. In other words, 
if the ground state in the matrix elements of Eq.~\eqref{eq:Kubo-formula} has spin $\tilde{S}$, then only excited states with spin $\tilde{S}\pm1$ will contribute. For a singlet ground state, only the triplet is accessed.

We can now proceed to the matrix element calculation. The local two-site gaps $\Delta E_{j,\tilde{S},\pm}$ separating the
ground state multiplet of spin $\tilde{S}$ from the excited states of
$\tilde{S}\pm1$ of the pair $(j,j+1)$ are obviously proportional to the coupling
$K_{j}$. If the proportionality constant is $\alpha_{\tilde{S},\pm}$,
then $\Delta E_{j,\tilde{S},\pm}=\alpha_{\tilde{S},\pm}K_{j}$. 
As shown in Appendix~\ref{sec:Matrix-elements-current}, the matrix elements of $\tau_{j}$ are proportional to the local gaps $\Delta E_{j,\tilde{S},\pm}$ and can be parametrized as $\beta_{\tilde{S},\pm,M}\omega$ [because of the delta function in Eq.~\eqref{eq:Kubo-formula-renorm}] with
\begin{equation}
\beta_{\tilde{S},\pm,M}=\left\langle \tilde{S},M\left|S_{2}^{z}\right|\tilde{S}\pm1,M\right\rangle.\label{eq:betadef}
\end{equation}
In Appendix~\ref{sec:Matrix-elements-current}, we list the values of $\beta_{S,\pm,M}$. Note that while the gaps are independent of $M$, the $z$ component of the angular momentum of the states involved, the matrix elements
do depend on this value.
We list here the $\beta_{S,\pm,M}$ relevant for the cases we focus on.
For the XXZ case, we only need $\beta_{0,+,0}=\frac{1}{2}$~\cite{MotrunichPRB2001,DamleMotrunichHusePRL}
(since $\tilde{S}=0$, only $\tilde{S}+1=1$ is accessed, and $M=0$).
For the spin-1 chain, the relevant values are $\beta_{1,+,0}=\frac{1}{\sqrt{3}}$ and $\beta_{1,+,\pm1}=\frac{1}{2}$, when the different $M$ states of spin-1 ground multiplet are connected to components of the $S=2$ quintuplet. Finally, $\beta_{1,-,0}=\beta_{0,+,0}=\sqrt{\frac{2}{3}}$ when the spin-1 ground multiplet is excited to the singlet or the ground singlet is excited to
the triplet (by necessity $M=0$). Plugging the matrix elements back into
the Kubo formula,
\begin{equation}
\sigma\left(\omega\right)=\frac{\omega}{L}\sum_{j=1}^{N}l_{j}^{2}\sum_{\tilde{S},M,k=\pm}\beta_{\tilde{S},k,M}^{2}\delta\left(\omega-\alpha_{\tilde{S},k}K_{j}\right).\label{eq:sigma2}
\end{equation}

Now, since the couplings $K_{j}$ are random variables, the conductivity
will vary from one disorder realization to the other. Let us call
$Q\left(\sigma\right)$ the conductivity probability distribution (we omit the $\omega$-dependence in order to simplify the notation). The distribution $Q\left(\sigma\right)$ is found by integrating over all possible values of $l_{j}$ and $K_{j}$ with weights given by the joint probability distribution of couplings and bond lengths $P\left(K_{j},l_{j}|\omega\right)$ (with cutoff $\Omega=\omega$)~\footnote{Strictly speaking, $P\left(J_{1},l_{1},\ldots J_{L},l_{L}\right)\ne\prod_{i}P\left(J_{i},l_{i}\right)$. This is because the lengths are correlated as they sum to $L$. We are neglecting these correlations as they lead only to subleading corrections~\cite{Mard_PRB_2014}.}~\cite{fisher94-xxz},

\begin{align}
Q\left(\sigma\right) & =\int\left[\prod_{j=1}^{N\left(\omega\right)}dl_{j}dK_{j}P\left(K_{j},l_{j}|\omega\right)\right]\times\nonumber \\
\times & \delta\left[\sigma-\frac{\omega}{L}\sum_{j=1}^{N\left(\omega\right)}l_{j}^{2}\sum_{\tilde{S},M,k=\pm}\beta_{\tilde{S},k,M}^{2}\delta\left(\omega-\alpha_{\tilde{S},k}K_{j}\right)\right],\label{eq:distr_conductivity}
\end{align}
where we assumed a chain of length $L$ and $N\left(\omega\right)$
is the number of non-decimated spins at the scale $\omega$, which
asymptotically goes like 

\begin{equation}
    N\left(\omega\right)\sim L/\Gamma_{\omega}^{\frac{1}{\psi}}, \label{eq:N_omega}
\end{equation} 
where $\Gamma_{\omega}=\ln\left(\Omega_{0}/\omega\right)$ and $\Omega_{0}$ is the initial cutoff. We also used the fact that correlations between
random variables on different bonds are absent asymptotically.

The expression for the conductivity distribution $Q\left(\sigma\right)$, Eq.~\eqref{eq:distr_conductivity}, includes two delta functions, one to relate a particular value of $\sigma$ to the random couplings (we call it ``\textit{outside}'' delta), while the other one connects the couplings at particular sites to the frequency $\omega$ (``\textit{inside}'' delta function). The ``\textit{outside}'' delta function presents no challenge, as its integral representation
\begin{equation}
\delta\left(x\right)=\int_{-\infty}^{+\infty}\frac{d\lambda}{2\pi}e^{i\lambda x}
\end{equation}
comes in handy. The expression for $Q\left(\sigma\right)$ becomes
\begin{align}
Q\left(\sigma\right) & =\int_{-\infty}^{+\infty}\frac{d\lambda}{2\pi}e^{i\lambda\sigma}\prod_{j=1}^{N\left(\omega\right)}dl_{j}dK_{j}P\left(K_{j},l_{j}|\omega\right)\nonumber \\
 & \times\exp\left[-\frac{i\lambda\omega}{L}l_{j}^{2}\sum_{\tilde{S},M,k=\pm}\beta_{\tilde{S},k,M}^{2}\delta\left(\omega-\alpha_{\tilde{S},k}K_{j}\right)\right],\\
 & =\int_{-\infty}^{+\infty}\frac{d\lambda}{2\pi}e^{i\lambda\sigma}\left[\mathcal{Q}(\lambda)\right]^{N\left(\omega\right)},\label{eq:fourier-transform-cond}
\end{align}
where 
\begin{align}
\mathcal{Q}(\lambda) & =\int dldKP\left(K,l|\omega\right)\times\nonumber \\
 & \times\exp\left[-\frac{i\lambda\omega}{L}l^{2}\sum_{\tilde{S},M,k=\pm}\beta_{\tilde{S},k,M}^{2}\delta\left(\omega-\alpha_{\tilde{S},k}K\right)\right].\label{eq:R_lambda}
\end{align}

We now focus on the mean and variance of $\sigma$. In order to obtain the asymptotic scaling behavior and in the spirit of the SDRG, factors of order one like $\beta_{\tilde{S},k,M}^{2}$ and $\alpha_{\tilde{S},k}$ will be ignored in the following. In the next section, we implement the SDRG method numerically and keep all the numerical prefactors. Within this approximation,

\begin{equation}
\mathcal{Q}(\lambda)=\int dldKP\left(K,l|\omega\right)\exp\left[-\frac{i\lambda\omega}{L}l^{2}\delta\left(\omega-K\right)\right].\label{eq:R_lambda-1}
\end{equation}
The average conductivity and its variance can be directly related
to the derivatives of $\mathcal{Q}$ calculated at $\lambda=0$,

\begin{align}
\left\langle \sigma\right\rangle  & =\int d\sigma Q\left(\sigma\right)\sigma=iN\left(\omega\right)\left.\frac{d\mathcal{Q}(\lambda)}{d\lambda}\right|_{\lambda=0},\label{eq:average_sigma}\\
\text{Var}\,\sigma & =N\left(\omega\right)\left[\left(\left.\frac{d\mathcal{Q}(\lambda)}{d\lambda}\right|_{\lambda=0}\right)^{2}-\left.\frac{d^{2}\mathcal{Q}(\lambda)}{d\lambda^{2}}\right|_{\lambda=0}\right].\label{eq:variance_sigma}
\end{align}
The derivatives can be explicitly expressed as
\begin{align}
\left.\frac{d\mathcal{Q}(\lambda)}{d\lambda}\right|_{\lambda=0} & =-\frac{i\omega}{L}\int dl\,l^{2}\int dKP\left(K,l|\omega\right)\delta\left(\omega-K\right),\label{eq:first_derivative}\\
\left.\frac{d^{2}\mathcal{Q}(\lambda)}{d\lambda^{2}}\right|_{\lambda=0} & =-\frac{\omega^{2}}{L^{2}}\int dl\,l^{4}\int dKP\left(K,l|\omega\right)\delta^{2}\left(\omega-K\right).\label{eq:second_derivative}
\end{align}

\begin{figure}
\includegraphics[width=1\columnwidth]{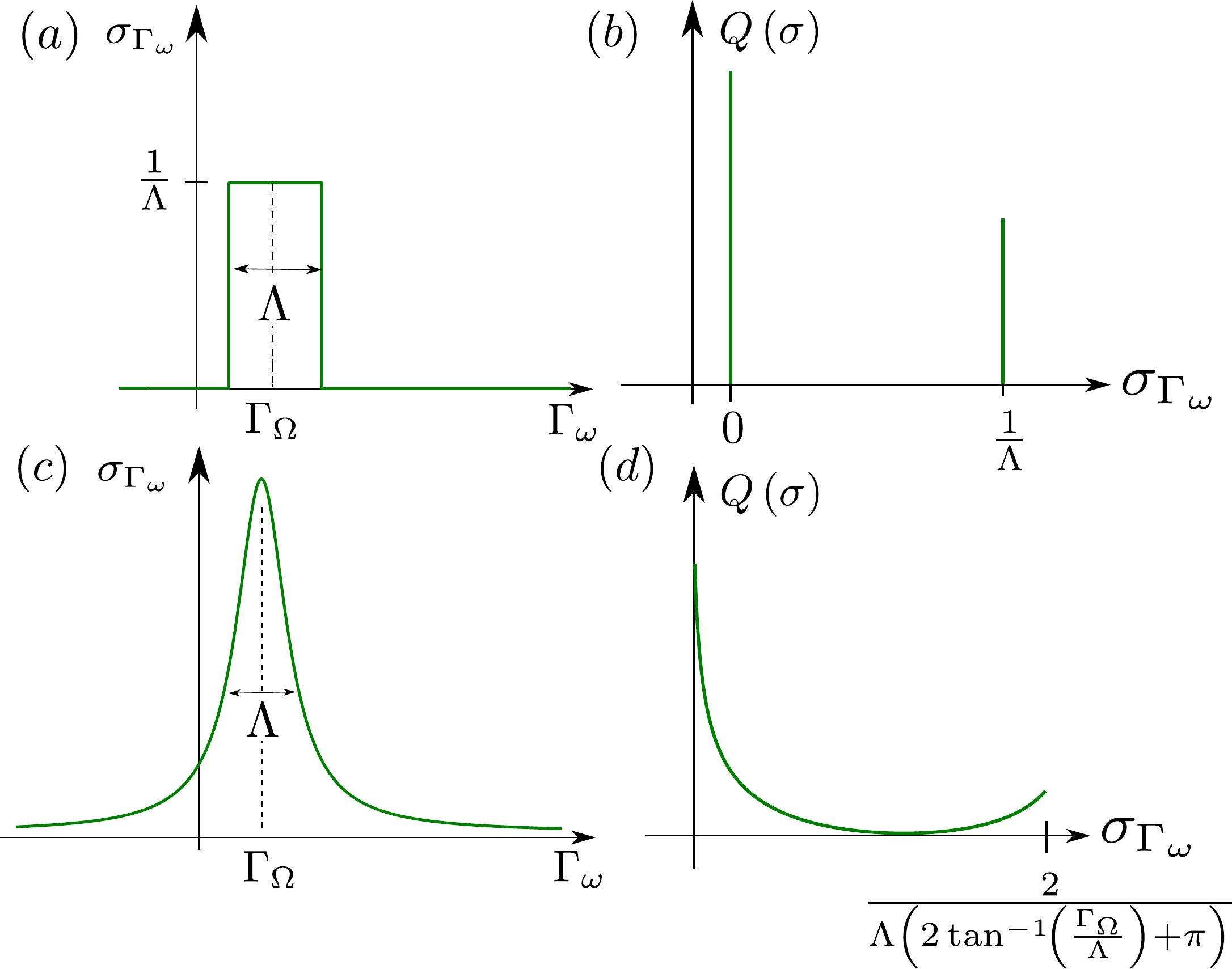}
\caption{Two choices for broadening the inside delta function in Eq.~\eqref{eq:distr_conductivity}: it is replaced by (a) a box or by (c) a Lorentzian, both with width $\Lambda$ and centered in $\Gamma_{\Omega}$. In (b) and (d), the corresponding distribution $Q(\sigma)$ generated by this particular decimation. For the box choice, only $\sigma=0,1/\Lambda$ are present in the distribution. The peak in $0$ does not allow for the calculation of typical values. For the Lorentzian choice, this is not an issue. 
\label{fig:delta_schemes}}
\end{figure}

The squared delta function in Eq.~\eqref{eq:second_derivative} seems troubling, but here we note an important feature of the conductivity distribution.
In general, for finite-sized isolated localized systems, the Kubo formula in
Eq.~\eqref{eq:Kubo-formula} leads to ill-controlled distributions since it consists of a \emph{non-dense} set of isolated delta functions.  Therefore, to obtain physically acceptable results, some sort of averaging or broadening must be introduced \cite{Stein1979,Fisher1981,Thouless_IOP_1981}. Suppose, for simplicity, that the sum in Eq.~\eqref{eq:sigma2} contains only one term, i.e., a single delta function. If we use a box-type representation with a broadening $\Lambda$, as in Fig.~\hyperref[fig:delta_schemes]{\ref{fig:delta_schemes}(a)}, the distribution $Q\left(\sigma\right)$ will consist of two two delta functions, as shown in Fig.~\hyperref[fig:delta_schemes]{\ref{fig:delta_schemes}(b)}. On the other hand, a Lorentzian representation of the delta function, as in Fig.~\hyperref[fig:delta_schemes]{\ref{fig:delta_schemes}(c)}, will smooth out the features of $Q\left(\sigma\right)$, see Fig.~\hyperref[fig:delta_schemes]{\ref{fig:delta_schemes}(d)}. In general, $Q(\sigma)$ encodes the information of the image of the function $\sigma_{\Gamma_{\omega}}$.

In real systems, inelastic processes \emph{extrinsic} to our model (coming, e.g., from phonons or external leads) will broaden these delta functions and regularize the distribution (see, e.g., Ref.~\cite{DelMaestro_PRL_2010}). We will consider several possibilities for the broadened delta function in our numerical calculations, as explained in the next section. 

For the analytical discussion that follows, we will simply impose one of the delta functions, which will, unavoidably, lead to powers of $\delta(0)$ for the higher moments of the conductivity distribution.  For $\delta^2$, for instance, we make the replacement
\begin{equation}
    \delta^{2}\left(\omega-K\right)\to C\delta\left(0\right)\delta\left({\omega}-K\right), \label{eq:delta2}
\end{equation}
bearing in mind that an object like $\delta\left(0\right)$ should be understood as a distribution with a finite $\omega$-dependent width determined by intrinsically inelastic scattering processes. For the following derivations, the precise analytical form of $\delta(0)$ is not needed. For a Lorentzian, for instance, it is proportional to the inverse of the decay rate of the state.
In Appendix~\ref{sec:delta_n}, by using the form of the broadened $\delta^2$, we show that, indeed, this replacement is justified, leading to the proper scaling behavior. The constant $C$ in Eq.~\eqref{eq:delta2}, in fact, depends on the choice of the regularized $\delta$ function. For a Lorentzian distribution, we show in Appendix~\ref{sec:delta_n} that $C=1/2$ (for a box, in contrast, $C=1$.)

The average and variance are then found by combining Eqs.~\eqref{eq:first_derivative} and \eqref{eq:second_derivative} with \eqref{eq:average_sigma} and \eqref{eq:variance_sigma}, using also the form of $N(\omega)$ from Eq.~\eqref{eq:N_omega}: 

\begin{align}
\left\langle \sigma\right\rangle  & =\frac{\omega}{4\Gamma_{\omega}^{1/\psi}}\left.\left\langle l^{2}\right\rangle \right|_{\omega},\label{eq:sigma-avg}\\
\text{Var}\,\sigma & =\frac{\omega^{2}}{16L\Gamma_{\omega}^{1/\psi}}\left[{C}\delta\left(0\right)\left.\left\langle l^{4}\right\rangle \right|_{\omega}-\left(\left.\left\langle l^{2}\right\rangle \right|_{\omega}\right)^{2}\right].\label{eq:sigma-variance}
\end{align}
with
\begin{equation}
\left.\left\langle l^{n}\right\rangle \right|_{\omega}=\int dl\,l^{n}\,P\left(\omega,l|\omega\right),
\end{equation}
being the average of $l^{n}$, with $l$ the length of the singlets decimated at the energy scale $\omega$.



It is convenient to work with log-variables $\zeta=\ln\left(\frac{\omega}{J}\right)$ and $\Gamma_{\omega}=\ln\left(\frac{\Omega_{0}}{\omega}\right)$~\cite{fisher94-xxz}.
Then, the fixed-point distribution 
can be written as 
\begin{equation}
P\left(J,l|\omega\right)=\frac{e^{\zeta}}{\omega\Gamma_{\omega}^{1+\frac{1}{\psi}}}R\left(\zeta/\Gamma_{\omega},l/\Gamma_{\omega}^{\frac{1}{\psi}}\right),\label{eq:scaling-form-distribution}
\end{equation}
in terms of the scaling function
$R\left(\zeta/\Gamma_{\omega},l/\Gamma_{\omega}^{\frac{1}{\psi}}\right)$, where 
we made use of the scaling of lengths $l$ as $\sim \Gamma_{\omega}^{\frac{1}{\psi}}$~\cite{fisher94-xxz,Damle2002,HoyosMiranda_PRB_2004}. The pre-factor $e^{\zeta}/\omega\Gamma_{\omega}^{1+\frac{1}{\psi}}$ is associated with the Jacobian after the change of variables.
For $J=\omega$, which implies $\zeta=0$, Eq.~\eqref{eq:scaling-form-distribution} becomes
\begin{equation}
P\left(\omega,l|\omega\right)=\frac{1}{\omega\Gamma_{\omega}^{1+\frac{1}{\psi}}}R\left(0,l/\Gamma_{\omega}^{\frac{1}{\psi}}\right).
\end{equation}
The distribution $P\left(\omega,l|\omega\right)$ has been calculated explicitly for the spin-1/2 XX model in Ref.~\cite{Hoyos2007PRB}. Its explicit form is not needed, however, as the value of $\left.\left\langle l^{n}\right\rangle \right|_{\omega}$
can be easily calculated
\begin{align}
\int dl\,l^{n}\,P\left(\omega,l|\omega\right)= & \int dl\,l^{n}\,\frac{1}{\omega\Gamma_{\omega}^{1+\frac{1}{\psi}}}R\left(0,l/\Gamma_{\omega}^{\frac{1}{\psi}}\right),\nonumber \\
 =& \frac{\Gamma_{\omega}^{\frac{n}{\psi}-1}}{\omega}\int dy\,y^{n}\,R\left(0,y\right)\equiv\frac{\Gamma_{\omega}^{\frac{n}{\psi}-1}}{\omega}\left\langle y^{n}\right\rangle ,
\end{align}
where $\left\langle y^{n}\right\rangle $ is \emph{independent of
$\omega$}. For our purposes, we need
\begin{align}
\left.\left\langle l^{2}\right\rangle \right|_{\omega} & =\omega^{-1}\Gamma_{\omega}^{\frac{2}{\psi}-1}\left\langle y^{2}\right\rangle ,\\
\left.\left\langle l^{4}\right\rangle \right|_{\omega} & =\omega^{-1}\Gamma_{\omega}^{\frac{4}{\psi}-1}\left\langle y^{4}\right\rangle .
\end{align}
Keeping only the leading terms in Eqs.~\eqref{eq:sigma-avg} and
\eqref{eq:sigma-variance},
\begin{align}
\left\langle \sigma\right\rangle  & \sim\Gamma_{\omega}^{\frac{1}{\psi}-1},\\
\text{Var}\,\,\sigma & \sim\frac{\omega}{L}\delta\left(0\right)\Gamma_{\omega}^{\frac{3}{\psi}-1}. \label{eq:varleading}
\end{align}

We will later compare these results with numerical calculations. For that, scaling with different system sizes will be important. From the general form of activated dynamical scaling $\Gamma_\omega \sim L^\psi$, we find that the average conductivity obeys the following scaling
\begin{equation}
\label{eq:sigmascaling}
\frac{\left\langle \sigma\right\rangle}{L^{1-\psi}}\sim\left(\frac{\Gamma_{\omega}}{L^{\psi}}\right)^{\frac{1}{\psi}-1},
\end{equation}
which, for $\psi=1/2$, agrees with Ref.~\cite{MotrunichPRB2001}.

We now need to discuss what choice we make for $\delta(0)\sim 1/b$, where $b$ is the energy scale characterizing the broadening of the localized states, in order to make sense of Eq.~\eqref{eq:varleading}. We are going to choose $b \sim \omega$ or  $\omega \delta(0)\sim \mathrm{const.}$ Note that with this choice, a finite factor $\omega \delta(0)\sim \mathrm{const.}$ will appear in all higher moments of the distributions, a fact which can be traced back to the argument of the exponential in Eq.~\eqref{eq:R_lambda-1}. If the extrinsic processes lead to a different broadening behavior, the regularization has to be changed accordingly, but, as we argue next, this would not be achieved by simply choosing $s\ne1$ in our treatment.
If we choose $b\sim \omega^{s}$ with $s<1$, the contribution in Eq.~\eqref{eq:varleading} would be dominated by the second term on the r.h.s of Eq.~\eqref{eq:sigma-variance} at low frequencies. This is inconsistent, however, since it would lead to a negative variance. Obviously, this would simply reflect an improper choice of the broadening scheme. On the other hand, if we use $b\sim \omega^{s}$, with $s>1$, the variance will diverge as a power law of the frequency which, through the scaling with length $L \sim \Gamma_{\omega}^{\frac{1}{\psi}}$, leads to a conductivity variance exponentially divergent with size, which is non-physical. Again, the phenomenological delta function regularization would have to be changed. It should be noted that a line width that is linear in frequency is common in metals (Landau damping) and has been used before in studies of metallic dissipation of disordered systems~\cite{Hoyos_Vojta_PRL_2007,Sachdev_2008_PRL}.

With this choice for $b$, the ratio of the standard deviation to the average is
\begin{equation}
\frac{\text{std}\left(\sigma\right)}{\left\langle \sigma\right\rangle }\sim \frac{\Gamma_{\omega}^{\frac{1}{2}\left(\frac{1}{\psi}+1\right)}}{\sqrt{L}}.
\end{equation}
This ratio diverges when $\omega\rightarrow0$ $\left(\Gamma_{\omega}\rightarrow\infty\right)$
indicating that the distributions become infinitely broad and the average behavior may not be the best indicator of transport properties.

One argument in favor of the above choice for $b$ is the following. In the XXZ case, where $\psi=1/2$, we find
\begin{align}
\left\langle \sigma\right\rangle  & \sim\Gamma_{\omega},\\
\text{std}\,\,\sigma & \sim\frac{\Gamma_{\omega}^{5/2}}{\sqrt{L}}. \label{eq:xxzcond}
\end{align}
If we now use the known energy-length scaling $\Gamma_\omega \sim \sqrt{L}$, we recover the correct DC behavior obtained from the Landauer formalism mentioned in the Introduction~\cite{SoukoulisEconomou_PRB_1981,MudryFurusaki_PRB_2000}
\begin{align}
\left\langle \sigma\right\rangle  & \sim\sqrt{L},\\
\text{std}\,\,\sigma & \sim L^{3/4}. \label{eq:xxzcond2}
\end{align}
We mention that both of these results can be straightforwardly extracted from a previous SDRG treatment of this problem~\cite{Mard_PRB_2014}. Although this is no rigorous proof, it strongly suggests the appropriateness of our choice of broadening Ansatz.



In the next section, we calculate the conductivity distribution numerically and show that, indeed, the average value is much larger than the typical one in the thermodynamic limit: whereas the former points to a metallic behavior~\cite{MotrunichPRB2001}, the latter predicts that the system is actually an insulator.

\section{Numerical Results~\label{sec:Numerical_results}}

In this section, we numerically implement the SDRG procedure for both spin-1/2 and spin-1 cases and compare it with the exact numerical diagonalization of the XX model. The SDRG steps are performed numerically
and the energy cutoff $\Omega$ is progressively lowered, until it reaches the frequency $\omega$ of interest. From Eq.~\eqref{eq:sigma2}, we see that the conductivity at frequency $\omega$ receives contributions from excitation energies of order $\omega$. These correspond to breaking bonds decimated precisely when the cutoff $\Omega=\omega$. To obtain those contributions we compute the corresponding term in Eq.~\eqref{eq:sigma2}. After appropriately discretizing the frequency $\omega$ axis, we build a histogram of $\sigma(\omega)$ from several disorder realizations. From this we get the average and typical values of the conductivity $\sigma(\omega)$, as well as its full distribution, at each frequency $\omega$. 

In order to implement this procedure, a choice must be made of how to regularize the delta function in Eq.~\eqref{eq:sigma2} (the ``inside'' delta function), in close analogy to what was done in the analytical calculations. The simplest possible choice, a uniform box, generates a finite fraction of $\sigma$'s that are identically zero if the sum over delta functions has zero support in some frequency range (see Fig.~\ref{fig:delta_schemes} and the discussion there). As we will see, this inevitably happens at sufficiently low energy scales.
This is inconvenient for, e.g., the calculation of the geometric average,  which is then identically zero. This can be avoided, however, with the use of a smooth function, e.g., a Lorentzian. 
Thus, when the local gap is equal to the running cutoff  in Eq.~\eqref{eq:sigma2}, $\alpha_{\tilde{S},k}K_j=\Omega$, we could use
\begin{equation}\label{eq:simpledelta}
\delta(\omega-\Omega) = \frac{b/\pi}{(\omega-\Omega)^2+b^2},
\end{equation}
where, as discussed in Section~\ref{sec:Conductivity-distribution}, $b= \Lambda \omega$, where $\Lambda$ is a constant. However, due to the multiplicative structure of renormalizations in second-order perturbation theory, it is much more natural to work with a logarithmic running energy scale
$\Gamma_{\omega}=\ln\left(\frac{\Omega_{0}}{\omega}\right)$, rather than with $\omega$ itself~\cite{fisher94-xxz}. This leads us to consider using instead
\begin{equation}
\delta\left(\Gamma_{\omega}-\Gamma_{\Omega}\right)=\frac{\Lambda/\pi}{\left(\Gamma_{\omega}-\Gamma_{\Omega}\right)^{2}+\Lambda^{2}}, \label{eq:Lorentzian-function2}
\end{equation}
where $\Gamma_{\Omega}=\ln\left(\frac{\Omega_{0}}{\Omega}\right)$.
To see how to switch from Eq.~\eqref{eq:simpledelta} to Eq.~\eqref{eq:Lorentzian-function2}, note that, for any representation of the delta function,
$\delta\left(\Gamma_{\omega}-\Gamma_{\Omega}\right)=\omega \delta(\omega-\Omega).$
Now, for a sufficiently small $\Lambda$, only the region $\Gamma_{\omega}\approx\Gamma_{\Omega}$ matters. In that case, 
\begin{equation}
|\Gamma_{\omega}-\Gamma_{\Omega}|=\left|\ln\left(\frac{\Omega}{\omega}\right)\right|\approx \frac{|\Omega-\omega|}{\omega},
\end{equation}
Starting from Eq.~\eqref{eq:simpledelta}, changing the representation of the delta function from $\omega$ to $\Gamma_{\omega}$, and using this approximation, we arrive at Eq.~\eqref{eq:Lorentzian-function2}. This shows that the choice of the logarithmic scale delta function \emph{  with a constant, frequency-independent $\Lambda$}, is fully compatible with the ``inside'' delta function of Sec.~\ref{sec:Conductivity-distribution}.

An inconvenience of a symmetric Lorentzian, however, is the fact that it has finite support at negative values of the strictly positive argument $\Gamma_\omega$. We therefore used a modified Lorentzian form $\delta^{mL}\left(\Gamma_{\omega}-\Gamma_{\Omega}\right)$
which, while having all the required properties of a delta function representation, has zero support for negative values of its argument. The explicit form of this modified Lorentzian is given in Appendix~\ref{sec:EDdelta}.

In our numerical simulations, we verified that the results do not change for sufficiently small values of $\Lambda$. We indicate the value of $\Lambda$ chosen for each plot in its respective caption.

\subsection{XXZ chains}

We start by discussing the systems described by Eq.~\eqref{eq:spin-1/2-H}, the XXZ chains~\cite{MotrunichPRB2001,DamleMotrunichHusePRL}. We studied systems with sizes ranging from $L=2^{9}=512$ to $L=2^{14}=16384$ with the SDRG method. We will also show, for comparison, exact diagonalization results for $L=2^{7}=128$. The random couplings $J_{j}^{\perp}$ were drawn from a uniform distribution ranging from 0 to $\Omega_0=1$. We chose $\Delta_{j}=0$ because this is an irrelevant coupling for $-1/2<\Delta_{j}<1$~\cite{fisher94-xxz}. Indeed, we verified that, starting with finite $0<\Delta_{j}<1$ values, the anisotropy does flow to zero, and our results are the same asymptotically. For each $L$, we averaged over $1.2\times10^{5}$ disorder realizations. This yields very small error bars, which are not shown for clarity.

\begin{figure}
\includegraphics[width=1\columnwidth]{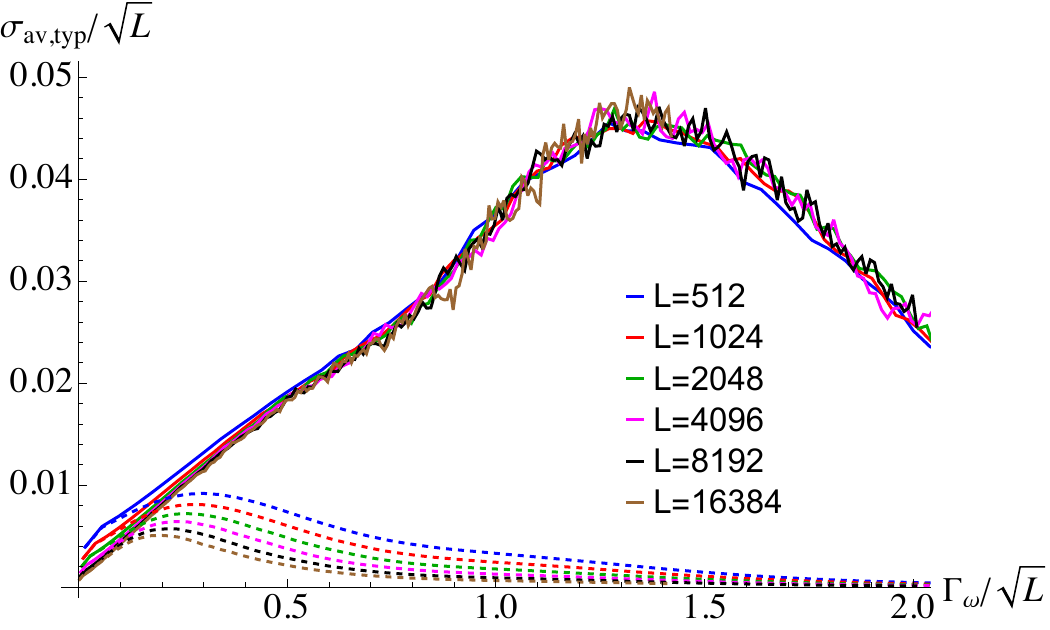}

\caption{Average [solid lines, $\sigma_{\text{av}}(\omega)$] and typical [dashed lines, $\sigma_{\text{typ}}(\omega)$] conductivities of the spin-1/2 XX chain for different system sizes $L$ as functions of $\Gamma_\omega=\ln\frac{\Omega_{0}}{\omega}$. Both axes are rescaled by $L^{-1/2}$ so that $\sigma_{\text{av}}(\omega)$ collapses onto a universal curve, as discussed in the text.
The behavior for $\Gamma_\omega/L^{1/2}\lessapprox1.4$, indicates that the average conductivity $\sigma_{\text{av}}(\omega)$ diverges at small frequencies $(\Gamma_\omega\rightarrow\infty)$, in the thermodynamic limit. The ratio of typical to average conductivities $\sigma_{\text{typ}}(\omega)/\sigma_{\text{av}}(\omega)$ for a fixed frequency, on the other hand, vanishes with increasing system size,
indicating an insulating
behavior. We chose the Lorentzian broadening parameter to be $\Lambda=0.4$; see Fig.~\ref{fig:delta_schemes}. \label{fig:XX_avg_typ}}
\end{figure}

The results are shown in Fig.~\ref{fig:XX_avg_typ}, after appropriate scaling with the system size [using $\psi=1/2$ in Eq.~\eqref{eq:sigmascaling}], in agreement with previous results~\cite{MotrunichPRB2001}.
As expected, for $\Gamma_\omega/L^{1/2}\lesssim 1.4$,
the scaled average conductivity $\sigma_{\text{av}}(\omega)/L^{1/2}$  increases linearly with $\Gamma_\omega/L^{1/2}$. This linear dependence implies a metallic conductivity in the thermodynamic limit (note that $\Gamma_\omega \to \infty$ as $\omega \to 0$), as noticed in Ref.~\cite{MotrunichPRB2001}. The downturn in the region $\Gamma_\omega/L^{1/2}\gtrsim 1.4$ represents the regime in which the finite size of the system begins to be ``felt''.
Note that the limits $\omega \rightarrow 0$ and   $ L\rightarrow \infty$ do not commute. The average value, however, is not a good indicator of the system's behavior, since the conductivity has an extremely broad distribution, as will be shown.
This is brought to light when we look at the typical value of the conductivity
\begin{equation}
\sigma_{\text{typ}}(\omega)=\sigma_{\text{geo}}(\omega)=\exp{\langle \ln{\sigma(\omega)}\rangle}.
\end{equation}
The numerical results for $\sigma_{\text{typ}}(\omega)$ are also shown in Fig.~\ref{fig:XX_avg_typ} for comparison, with the same rescaling of the average conductivity. It is clear that, for a fixed frequency $\omega$, the ratio $\sigma(\omega)_{\text{typ}}/\sigma(\omega)_{\text{av}} \to 0$ as $L \to \infty$. This behavior holds for frequencies well below the finite-size peak of $\sigma(\omega)_{\text{av}}$.
This leads us to conclude that the true conductivity of a typical sample is that of an insulator, not a metal.  As is common in situations where a physical quantity is strongly non-self-averaging, the apparent ``metallic'' behavior suggested by the average conductivity is a consequence of rare events in which a sample has atypically large values. The same phenomenon is seen, e.g., in the system's spin-spin correlation function, whose average value decays as a power law of the distance, whereas its typical value is a stretched exponential~\cite{fisher94-xxz}. 

\begin{figure}
\includegraphics[width=1\columnwidth]{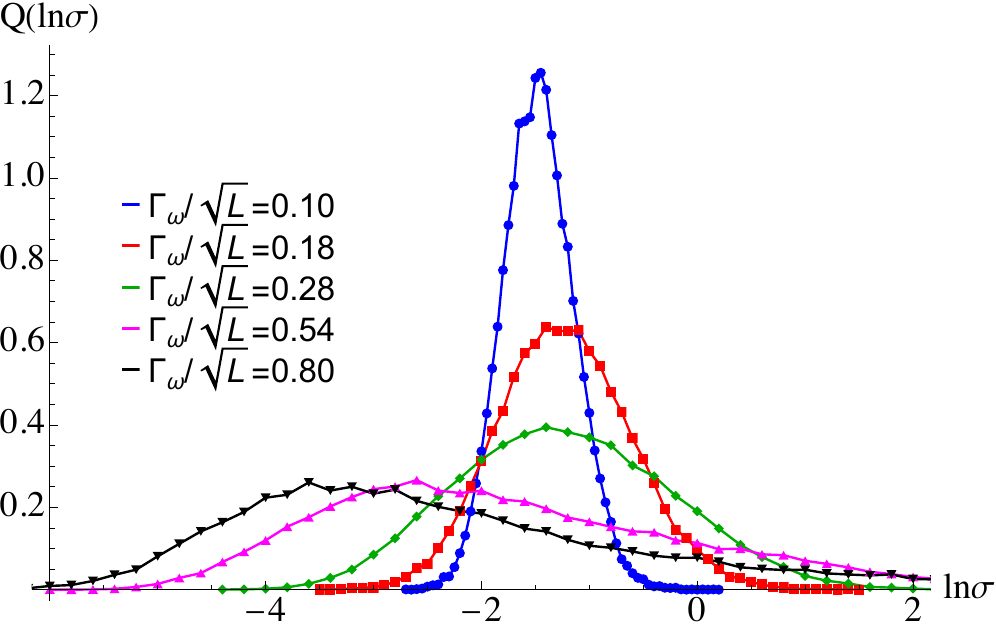}
\caption{
Conductivity distribution $Q\left(\ln\sigma\right)$ of the spin-1/2 XX chain for different frequency values, obtained by implementing the SDRG method numerically. We keep the system size fixed at $L=2048$. As the frequency $\omega$ decreases (and $\Gamma_\omega=\ln\frac{\Omega_{0}}{\omega}$ increases), the distribution becomes increasingly broader, explaining the large separation between the typical and the average conductivities. We chose the Lorentzian broadening parameter to be $\Lambda=0.4$; see Fig.~\ref{fig:delta_schemes}. \label{fig:dist_XX}}
\end{figure}

To corroborate this, we numerically obtained the full conductivity distribution. This is shown in Fig.~\ref{fig:dist_XX}, where we plot $Q\left(\log\sigma\right)$ for $L=2048$ and different frequency values. It is clear that, as the frequency decreases, the distributions become extremely broad on the log scale while their peaks eventually shift towards lower values of $\sigma(\omega)$. This explains the large separation between average and typical values. We conclude that the physically relevant quantity to determine the transport properties is, therefore, $\sigma_{\text{typ}}(\omega)$, not $\sigma_{\text{av}}(\omega)$.

We now compare the results of the SDRG method with exact diagonalization results. For the exact diagonalization, we make use of the Jordan-Wigner transformation~\cite{Lieb1961}, which maps the spin-1/2 XX chain into a noninteracting fermionic problem. We note that this exact mapping is only possible for the $S=1/2$ case and the fermions are non-interacting only when $\Delta_j=0$. We first express the Hamiltonian in Eq.~\eqref{eq:spin-1/2-H} in the fermionic basis obtained from this transformation and then obtain the corresponding eigenvectors using a numerical eigensolver.  Technical details and tricks to deal with numerical instabilities due to the exceedingly small finite-size gap can be found in Ref.~\cite{Getelina2020}. The eigenvectors are used to compute the conductivity, with the current operator of Eq.~\eqref{eq:Kubo-formula} written in terms of fermions.

\begin{figure}
\includegraphics[width=1\columnwidth]{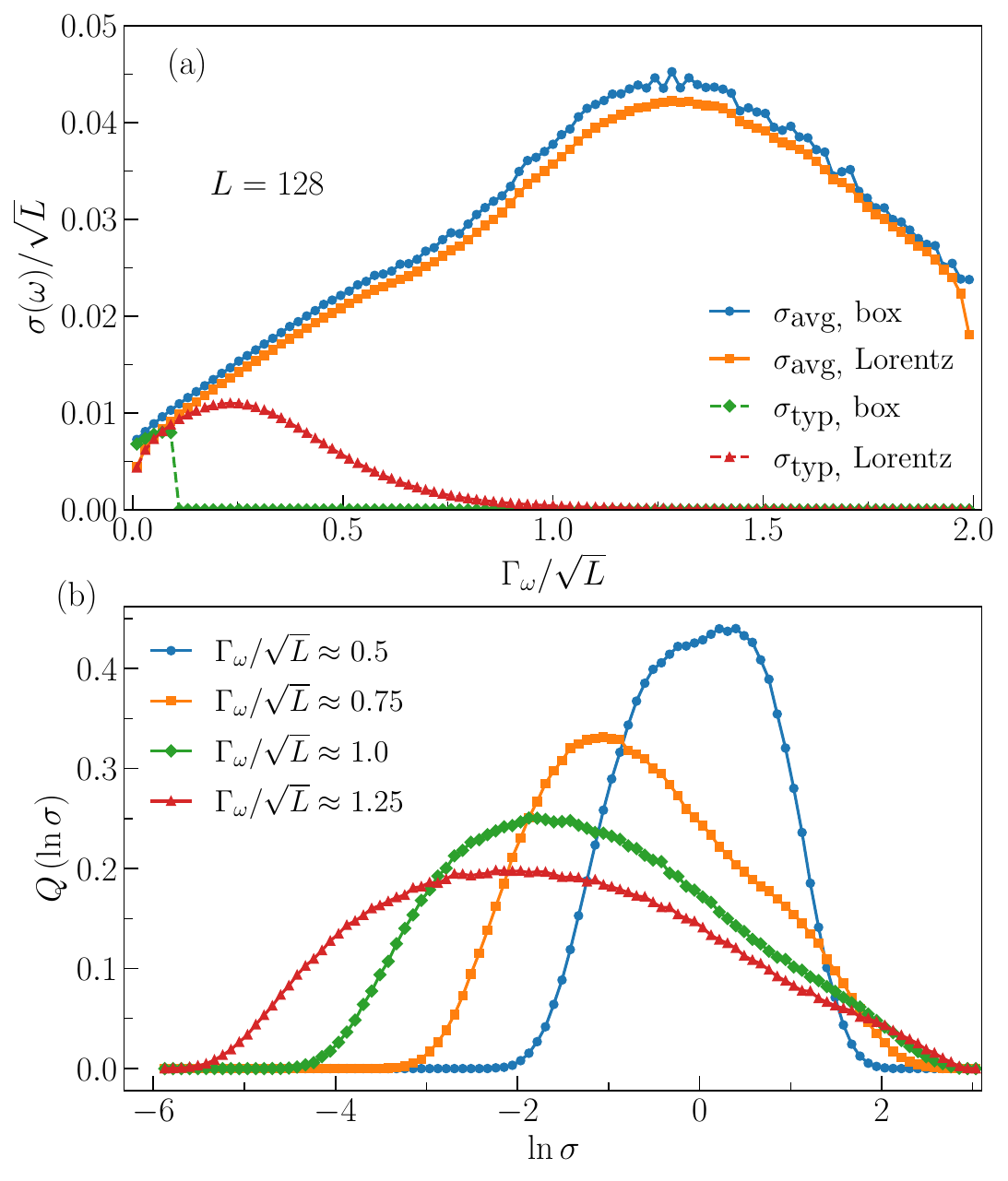}
\caption{Exact diagonalization results for the conductivities of the XX spin-1/2 chain, with system size $L=128$, averaged over $10^6$ different realizations. (a) Average [$\sigma_{\text{av}}(\omega)$] and typical [$\sigma_{\text{typ}}(\omega)$] conductivities (solid and dashed lines, respectively), obtained from the box-like and the Lorentzian representations of the delta function. The broadening parameter is $\Lambda=0.4$. The typical value of the former rapidly collapses to zero.  (b) Distributions of the conductivity $Q\left(\ln\sigma\right)$ for various values of frequency $\omega$. \label{fig:exact_diag}}
\end{figure}

The average and typical values of $\sigma(\omega)$ are shown in Fig.~\hyperref[fig:exact_diag]{\ref{fig:exact_diag}(a)} as a function of frequency for both the box-like and Lorentzian representations of the delta function. For the average, both choices of $\delta$ lead to very similar results, as expected from the discussion in section~\ref{sec:Conductivity-distribution}. The typical value collapses to zero for the box delta function as $\Gamma_{\omega}$ increases. This reflects the values of $\sigma(\omega)=0$ generated by this choice of the delta function representation (see the discussion following Fig.~\ref{fig:delta_schemes}). There is good agreement between the numerical results of Fig.~\hyperref[fig:exact_diag]{\ref{fig:exact_diag}(a)} and the SDRG results of Fig.~\ref{fig:XX_avg_typ}.
The distribution of $\sigma(\omega)$ is shown in Fig.~\hyperref[fig:exact_diag]{\ref{fig:exact_diag}(b)}. The distributions broaden and their peaks shift to the left in a fashion very similar to the SDRG results of Fig.~\ref{fig:dist_XX}.

\subsection{Spin-1 chains}

We also studied numerically, using the SDRG, the spin conductivity of the SU(2)-symmetric spin-1 chain, described by Eq.~\eqref{eq:spin-1-H}.
As demonstrated in reference~\cite{Quito_PhysRevLett.115.167201} and briefly reviewed in section~\ref{sec:Method}, this system exhibits two phases depending on the initial angle $\theta_j=\theta_{0},~\forall j$, taken to be uniform throughout the chain. We show two sets of results that are representative of the two phases. We choose  $\theta_{0}=\mp \pi/4$, corresponding to the mesonic and baryonic random-singlet phases, respectively. The corresponding SDRG flows were extensively studied in Ref.~\cite{Quito_PhysRevLett.115.167201} and shown to tend towards a universal IRFP. Here, for  $\theta_{0}=- \pi/4$, we use $P(r)\sim r^{-1/2}$, while for  $\theta_{0}=\pi/4$, we use $P(r)$ to be an uniform distribution in the range $0<r_j<\sqrt{2}$. In both cases, we use system sizes from $L=2^{9}=512$ to $L=2^{13}=8192$. We averaged over 120 thousand disorder realizations. We calculated the average and typical (geometric average) values of $\sigma(\omega)$.
Once again, the error bars are too small and are omitted for clarity.



\begin{figure}
\includegraphics[width=0.95\columnwidth]{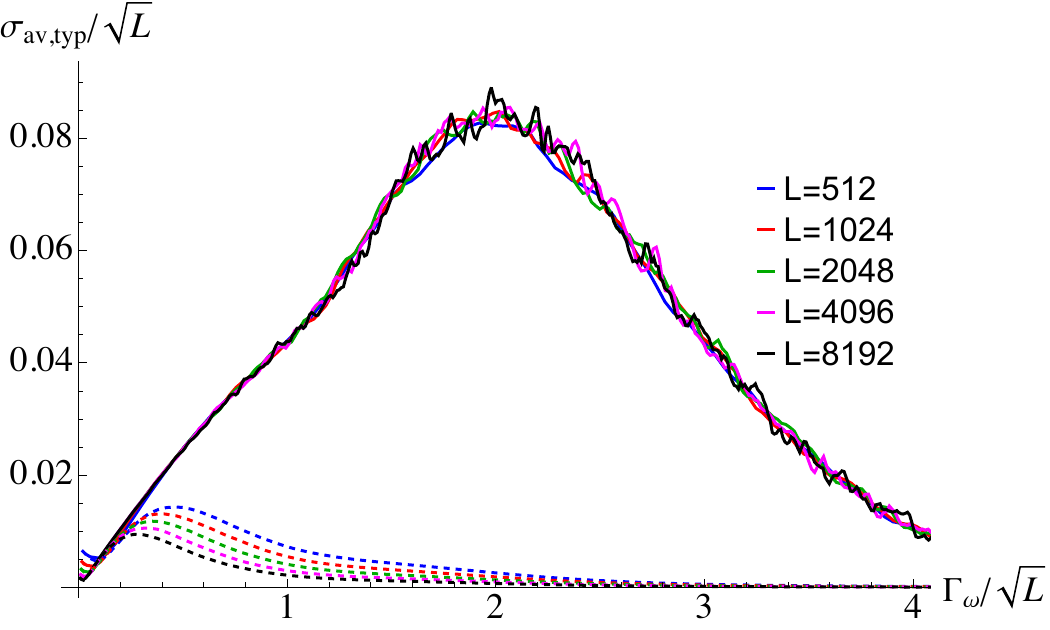}
\includegraphics[width=0.95\columnwidth]{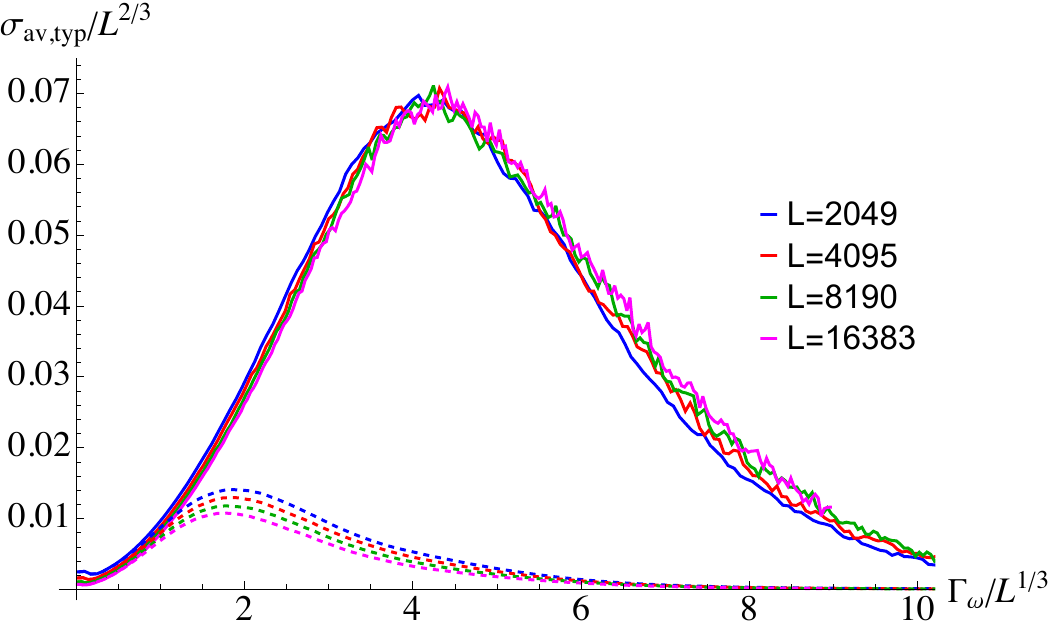}

\caption{Average [$\sigma_{\text{av}(\omega)},$ solid] and typical [$\sigma_{\text{typ}}(\omega)$, dashed] scaled spin conductivities of the $S=1$ model, obtained by a numerical implementation of the SDRG method. The initial distribution has a fixed angle $\theta_j=\theta_{0}$ and random $r_j$ (see text). (a) Mesonic phase: The initial angle is $\theta_{0}=-\pi/4$. The behavior is very similar to the spin 1/2 XXZ chain of Fig.~\ref{fig:XX_avg_typ}. (b) Baryonic phase: The initial angle is $\theta_{0}=\pi/4$. The scaling factors of $\left\langle \sigma\right\rangle$  and $\Gamma_\omega$
in the baryonic phase are $L^{2/3}$ and $L^{1/3}$, respectively, in contrast to $L^{1/2}$ in the mesonic phase. The broadening parameter is $\Lambda=0.4$.} \label{fig:spin1_avg_typ}
\end{figure}

The results for the average [$\sigma_{\text{av}}(\omega)$] and typical [$\sigma_{\text{typ}}(\omega)$] spin conductivities are shown for the mesonic phase in Fig.~\hyperref[fig:spin1_avg_typ]{\ref{fig:spin1_avg_typ}(a)}
and for the baryonic phase in Fig.~\hyperref[fig:spin1_avg_typ]{\ref{fig:spin1_avg_typ}(b)}. Note how we must use different scalings depending on the phase, in agreement with Eq.~\eqref{eq:sigmascaling}. Indeed, whereas the results in the mesonic phase, like in spin-1/2 XXZ chain, require scaling with $\psi=1/2$, in the baryonic phase $\psi=1/3$ must be used~\cite{Quito_PhysRevLett.115.167201}, as discussed in section~\ref{sec:Conductivity-distribution}. Once this is done, the average conductivity of both sets fall onto a universal curve.

The behavior in the mesonic phase is very similar to the XXZ chains. The linear dependence of $\sigma_{\text{av}}(\omega)$ with $\Gamma_\omega$ to the left of the finite-size peak points to a diverging spin conductivity as $\omega \to 0$ in the thermodynamic limit. The region to the right of the peak reflects the fact that the limits $L \rightarrow \infty$ and $\omega \rightarrow 0$ do not commute when computing $\sigma(\omega,L)$. Similar to the spin-1/2 case, the typical spin conductivity $\sigma_{\text{typ}}(\omega)$ points to an insulating behavior. For a fixed frequency $\omega$, the ratio $\sigma_{\text{typ}}(\omega)/\sigma_{\text{av}}(\omega)$ decreases with the system size, also in close analogy to the spin-1/2 case.

The results for the baryonic phase are qualitatively similar.
Note, however, the larger finite-size effects. This reflects the slower approach to the asymptotic regime of the baryonic phase, which can be understood as follows. Unlike in the mesonic phase, where only second-order decimations are present asymptotically, 
in the baryonic phase, both first- and second-order steps persist at all energy scales, see Eqs.~(\ref{eq:J14-order-2}-\ref{eq:D1-order1-triplet}). Whereas second-order steps are very effective at lowering the energy scale (due to its multiplicative form), first-order decimations are fairly ineffective, as they only renormalize the couplings by a pre-factor of order 1. As a result, the distribution of couplings flows more slowly toward the IRFP. This is closely tied to the smaller value of $\psi=1/3$, as compared to the mesonic phase $\psi=1/2$. As a consequence of this slow approach to asymptotics, the contrast between a  metallic $\sigma_{\text{av}}(\omega)$ and an insulating $\sigma_{\text{typ}}(\omega)$ is less marked than in the mesonic phase. Nevertheless, all the indications are that the same behavior also holds in the baryonic phase. 

\begin{figure}  

\includegraphics[width=0.95\columnwidth]{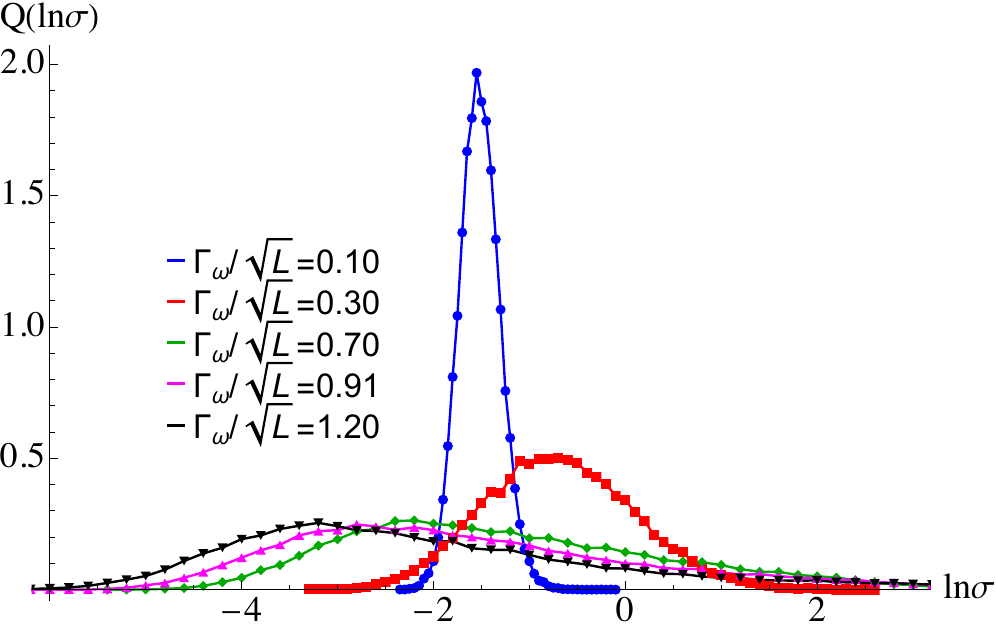}

\includegraphics[width=0.95\columnwidth]{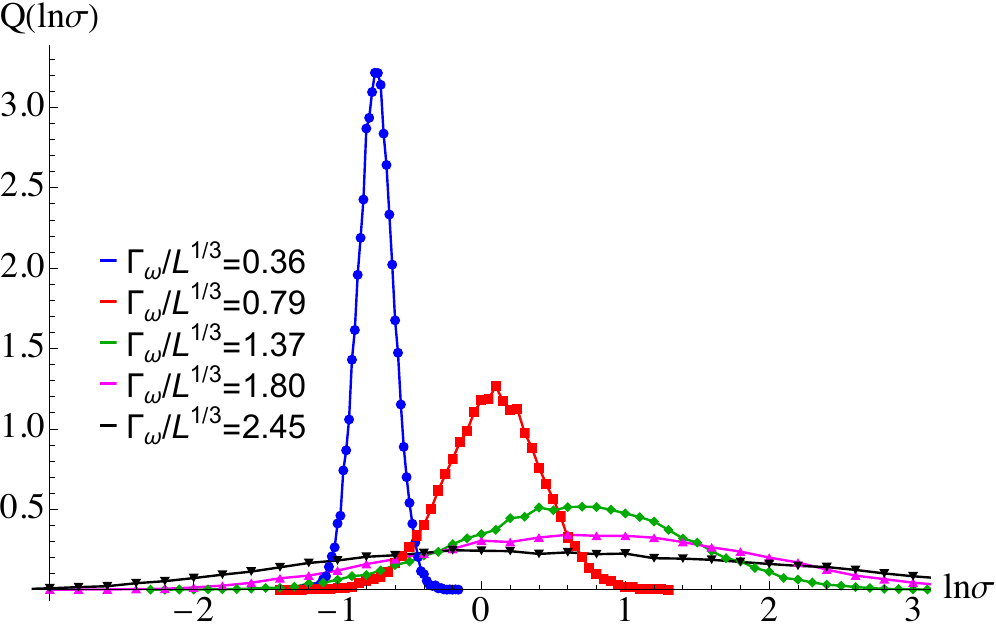}

\caption{Distribution of $\ln\sigma(\omega)$ for different frequency values $\omega$ ($\Gamma_\omega =\ln(\Omega_0/\omega)$, obtained by implementing the SDRG method numerically. In the (a) mesonic and (b) baryonic phases of random spin-1 chains. As the frequency decreases, the distribution broadens by several orders of magnitude. These wide distributions lead to very different values of the average and the typical spin conductivity. The broadening parameter is $\Lambda=0.4$. 
\label{fig:spin1_distribution}}
\end{figure}

We also looked at the conductivity distributions, as shown in Fig.~\ref{fig:spin1_distribution}. In both phases, the distributions become increasingly broader as the frequency decreases, although this tendency is less sharp in the baryonic phase. Again, this is to be expected from the slower approach to asymptotics in the latter case. Nevertheless, the clear tendency to ever broader distributions in both cases are consistent with the sharp distinction between the typical and average values.

We conclude that, both in the spin-1/2 and the spin-1 disordered chains studied here, the signature of the IRFP is conspicuously reflected in the conductivity behavior. As a result of this, the true transport characteristics are only captured by the geometric average or typical value of the conductivity. 

\section{Conclusions \label{sec:conclusion}}

By calculating the average conductivity of the XX chain, Eq.~\eqref{eq:average_sigma},
it was concluded in Ref.~\cite{MotrunichPRB2001} that this system exhibits metallic behavior. This was based on the average conductivity
having a logarithm divergence, $\left\langle \sigma\right\rangle \sim\ln\left(\frac{\Omega_{0}}{\omega}\right)$,
as the frequency goes to zero. In this work, we showed that the distribution of $\sigma$ broadens significantly, so that the average is no longer a good indicator of the transport properties, which led us to look at the typical conductivity instead. The value of the typical conductivity collapses to zero in the $\omega\rightarrow0$ limit, as
can be seen from Fig.~\ref{fig:XX_avg_typ}, which indicates an \emph{insulating}
behavior. This result agrees with the expectation since the XX model can
be mapped into a spinless free fermion chain, whose solution is known to be  localized, albeit anomalously~\cite{Mard_PRB_2014}.
The discrepancy between the average and typical behaviors can already be seen
at the early stages of the SDRG as the region in which the conductivity is self-averaging is very small. The $\sigma$ distributions plotted in log scale in Fig.~\ref{fig:dist_XX} clearly present a very wide broadening, with a width spanning several decades in linear scale. This feature is reflected in the behavior of quantities calculated in the small frequency limit, which is reached in later stages of the SDRG flow.

For the $S=1$ model, a similar analysis indicates that the typical value approaches zero as the system size is increased, also indicating an insulating behavior. The average conductivity is again not a good indicator of the phase, as the distribution of conductivity broadens without limit. A distinction between the spin-1/2 and spin-1 chains is that the latter exhibits two RSPs with distinct scaling behaviors~\cite{Quito_PhysRevLett.115.167201}. The different phases are tuned by the angle between bilinear and biquadratic exchange couplings, and their scaling behaviors are reflected in the scaling of the average conductivity, as given in Eq.~\eqref{eq:sigmascaling}, which is governed by the tunneling exponent: $\psi=1/2$ in the mesonic phase while $\psi=1/3$ in the baryonic one.

Even though the numerical results we showed are for $S=1/2$ and $S=1$ systems, the framework derived in this work can be readily implemented for generic SU(2) invariant spin-S chains in regions of the phase diagram in which the spin size does not grow at low energies. Based on our analytical results, we conclude that the RSPs of \emph{all} such chains are \emph{spin insulators.} We leave the case in which the spin sizes can also flow under the SDRG for future work. 

Although the SDRG procedure here was constructed explicitly for spin transport, the derivations of section~\ref{sec:Current-renormalization} can be readily generalized to other conserved quantities. In particular, the average conductivity was calculated in the metallic phase of the disordered metal-superconductor transition \cite{DelMaestro_PRL_2010}. Quite possibly, the same distinction between average and typical values applies to that case as well. We leave for future work the analysis of the behavior of other quantities, in particular, for the class of systems here considered, the energy current. An open
question is whether the distributions of all conserved quantities broaden similarly to the spin conductivity or whether their flow is significantly distinct. 

\section*{Acknowledgments}

LFCF acknowledges financial support from Fapesp through grant 2015/01966-1. VLQ acknowledges financial support from the National High Magnetic Field Laboratory through NSF Grant No. DMR-1157490, where this project was started, CNPq (grant 311565/2023-9), FAPESP via process number 2024/09202-0, and the hospitality of the Aspen Center for Physics, supported by National Science Foundation grant PHY-1607611. JCG acknowledges financial support from Coordenação de Aperfeiçoamento de Pessoal de Nível Superior - Brasil (CAPES) - Finance Code 001. JAH acknowledges financial support from FAPESP and CNPq. JAH thanks IIT Madras for a visiting position under the IoE program, which facilitated the completion of this research work. EM acknowledges financial support from CNPq (grant 309584/2021-3) and FAPESP, process number 2022/15453-0. VLQ and EM also acknowledge the FAPESP grant 2021/14335-0 for support in May and June 2023.

\section*{Data availability}
The data that support the findings of this article are not publicly available. The data are available from the authors upon reasonable request.

\appendix

\section{Derivation of the renormalization of the current operator \label{sec:current-operator-derivation}}

In this Appendix, we give a proof of the expressions used for the
renormalized spin current. Some of these expressions have appeared
before for small spin sizes~\cite{DamleMotrunichHusePRL,MotrunichPRB2001} but our results
generalize them to any value of spin-$S$. We assume only the SO(2) symmetry of rotations around one axis, here taken to be the $z$-axis, see Eqs.~\eqref{eq:generic-SU2-hamilt}
and \eqref{eq:spin-1/2-H}. For this, the commutator form of the
current operator in Eq.~\eqref{eq:tau-def} is crucial, as are the
techniques developed for generic spin-$S$ in reference~\cite{quitoprb2016}.

The process of computing the SDRG-based perturbative effects to the current operator
starts from correcting the eigenstates of the pair of spins to be
decimated (in this case, spins on sites 2 and 3). Let us call $S_{G}$
the spin of the ground state multiplet, which can be a singlet, $S_{G}=0$,
or a higher angular momentum $S_{G}\ne0$. In what follows, we represent
the states as $\left|S_{G},M\right\rangle $, corresponding to a shorthand
notation, since the problem to be solved involves actually the four sites 1, 2, 3, and 4. For example, the complete description of a state
$\left|\psi^{\left(0\right)}\right\rangle $ involves four quantum
numbers, $\left|\psi^{\left(0\right)}\right\rangle =\left|m_{1},m_{4};S_{G},M\right\rangle $
where $m_{1},m_{4}$ are the quantum numbers coming from the eigenvalues of $S_{1}^{z}$
and $S_{4}^{z}$, respectively. Up to second order in $V$, defined in Eq.~\eqref{eq:H_broken},
the corrected eigenstate of $H=H_{0}+V$ reads~\cite{cohen1977quantum}
\begin{equation}
\left|S_{G}M\right\rangle _{cor}=\left|\psi^{\left(0\right)}\right\rangle +\left|\psi^{\left(1\right)}\right\rangle +\left|\psi^{\left(2\right)}\right\rangle ,\label{eq:corrected_eigenstate}
\end{equation}
where 
\begin{align}
\left|\psi^{\left(0\right)}\right\rangle  & =\left|S_{G},M\right\rangle \label{eq:eigen_order_0}\\
\left|\psi^{\left(1\right)}\right\rangle  & =\sum_{S^{\prime}\ne S_{G},M^{\prime}}\frac{\left\langle S^{\prime},M^{\prime}\left|V\right|S_{G},M\right\rangle }{\Delta E\left(S_{G},S^{\prime}\right)}\left|S^{\prime},M^{\prime}\right\rangle \label{eq:eigen_order_1}\\
\left|\psi^{\left(2\right)}\right\rangle  & =\sum_{S^{\prime}\ne S_{G},S^{\prime\prime}\ne S_{G},M^{\prime},M^{\prime\prime}}\frac{\left\langle S^{\prime\prime},M^{\prime\prime}\left|V\right|S^{\prime},M^{\prime}\right\rangle }{\Delta E\left(S_{G},S^{\prime\prime}\right)\Delta E\left(S_{G},S^{\prime}\right)}\times\nonumber \\
 & \times\left\langle S^{\prime},M^{\prime}\left|V\right|S_{G},M\right\rangle \left|S^{\prime\prime},M^{\prime\prime}\right\rangle+\ldots .\label{eq:eigen_order_2}
\end{align}

We have defined $\Delta E\left(S_{G},S\right)=E_{S_{G}}-E_{S}$, the unperturbed energy difference between the multiplets of total angular
momentum $S_{G}$ and $S$. The unperturbed energies come
from diagonalizing $H_{0}$ of Eq.~\eqref{eq:H_broken}. Besides
the second order correction shown in Eq.~\eqref{eq:eigen_order_2}, there
are two more terms, but they do not contribute to the final result, as
shown in Appendix~\ref{sec:Other-second-order-corrections}. Higher-order
corrections $\left|\psi^{\left(n>2\right)}\right\rangle $ to the state lead to sub-leading contributions and can be neglected. 

It is significantly more convenient, in many of the derivations, to work with operators, instead of states. We define the following resolvent operator~\cite{lindgren1974BW}
\begin{equation}
R=\sum_{S^{\prime} \ne S_{G},M^{\prime}}\frac{\left|S^{\prime},M^{\prime}\right\rangle \left\langle S^{\prime},M^{\prime}\right|}{\Delta E\left(S_{G},S^{\prime}\right)}
=\sum_{S^{\prime} \ne S_{G}}\frac{P_{S^\prime}}{\Delta E\left(S_{G},S^{\prime}\right)},\label{eq:resolvent}
\end{equation}
where $P_S$ is a projector onto the multiplet of total angular momentum $S$ of the pair of spins 2 and 3.
We emphasize that $R$ involves \emph{only sites 2 and 3} and therefore commutes with any operator acting on sites 1 and 4.
The corrected states can be written in a compact way as~\cite{lindgren1974BW}
\begin{align}
\left|\psi^{\left(0\right)}\right\rangle  & =\left|S_{G},M\right\rangle \label{eq:eigen_order_0-1}\\
\left|\psi^{\left(1\right)}\right\rangle  & =RV\left|S_{G},M\right\rangle \label{eq:eigen_order_1-1}\\
\left|\psi^{\left(2\right)}\right\rangle  & =RVRV\left|S_{G},M\right\rangle \label{eq:eigen_order_2-2}
\end{align}

The various contributions to the renormalized currents come from matrix elements of $\tau_i\ (i=1,2,3)$ between various combinations of the states in Eqs.~\eqref{eq:eigen_order_0-1}, \eqref{eq:eigen_order_1-1}, and \eqref{eq:eigen_order_2-2}. The latter are given by operators acting on $\left|S_{G},M\right\rangle$. In order to manipulate only operators, we will, in what follows, bracket all of them with $P_{S_G}$, the projector onto the sup-space of $\left|S_{G},M\right\rangle$.

We now determine the leading power of the local gap $\Omega$ of each contribution to the renormalized spin currents. First, note that $R\sim \mathcal{O}(\Omega^{-1})$, $\tau_{2} \sim \mathcal{O}(\Omega^{1})$, and $\tau_{1,3} \sim \mathcal{O}(\Omega^{0})$.
Using this and depending on whether $S_G=0$ or $S_G\neq 0$, we can construct Table~\ref{tab:Summary-renorm}. This Table will guide the calculations in what follows.


\begin{table}
\begin{centering}
\begin{tabular}{|c|c|}
\hline 
\multicolumn{1}{|c|}{} & $S_{G}=0$\tabularnewline
\hline 
\hline 
$\tilde{\tau}_{1,3}$  & $\left\langle \psi^{\left(1\right)}\left|\tau_{1,3}\right|\psi^{\left(0\right)}\right\rangle +\left\langle \psi^{\left(0\right)}\left|\tau_{1,3}\right|\psi^{\left(1\right)}\right\rangle \sim\mathcal{O}\left(\frac{1}{\Omega}\right)$\tabularnewline
\hline 
$ \tilde{\tau}_{2}$  & $\left\langle \psi^{\left(2\right)}\left|\tau_{2}\right|\psi^{\left(0\right)}\right\rangle +\left\langle \psi^{\left(0\right)}\left|\tau_{2}\right|\psi^{\left(2\right)}\right\rangle \sim\mathcal{O}\left(\frac{1}{\Omega}\right)=\tilde{\tau}_{1,3}$\tabularnewline
\hline 
\multicolumn{1}{|c|}{} & $S_{G}\ne0$\tabularnewline
\hline 
$\tilde{\tau}_{1,3}$  & $\left\langle \psi^{\left(0\right)}\left|\tau_{1,3}\right|\psi^{\left(0\right)}\right\rangle \sim\mathcal{O}\left(\Omega^{0}\right)$\tabularnewline
\hline 
$\tilde{\tau}_{2}$  & $\left\langle \psi^{\left(1\right)}\left|\tau_{2}\right|\psi^{\left(0\right)}\right\rangle +\left\langle \psi^{\left(1\right)}\left|\tau_{2}\right|\psi^{\left(0\right)}\right\rangle \sim\mathcal{O}\left(\Omega^{0}\right)=\tilde{\tau}_{1,3}$\tabularnewline
\hline 
\end{tabular}
\par\end{centering}
\caption{Summary of the current renormalizations of a given SDRG step, listing
the leading non-zero contributions for each term. When the ground state
of the pair of spins at sites 2 and 3 is a singlet ($S_{G}=0$), the
first finite corrections leading to $\tilde{\tau}_{1,3}$ and $\tilde{\tau}_{2}$
go as $\frac{1}{\Omega}$, and are, in fact, all equal. For $S_{G}\protect\ne0$,
the corrections $\tilde{\tau}_{1,2,3}\sim\mathcal{O}\left(\Omega^{0}\right)$.
Even though $\tilde{\tau}_{2}$ corrections come from correcting the
state in higher order when compared to $\tau_{1,3}$, the contributions
are of the same order since $\tau_{2}$ itself is proportional to
$\Omega$. ~\label{tab:Summary-renorm}}
\end{table}

\subsection{Renormalization of $\tau_{1,3}$}

Let us focus on $\tau_{1}$, keeping in mind that the renormalization
of $\tau_{3}$ follows from similar steps by symmetry. 
To zeroth order [see Eq.~\eqref{eq:eigen_order_0-1}], we need
\begin{align}
\tilde{\tau}_{1} & =P_{S_{G}}\tau_{1}P_{S_{G}} \nonumber \\
& = i P_{S_{G}} [S_1^z,H_{12}] P_{S_{G}} \nonumber \\
& = i [S_1^z,P_{S_{G}} H_{12}P_{S_{G}} ],
\label{eq:tau-1-zero}
\end{align}
where we used the fact that $P_{S_{G}}$ acts only on sites 2 and 3 and therefore commutes with $\mathbf{S}_1$ and $\mathbf{S}_4$. In general, in all the following derivations, it will be very convenient to write the local current operators in terms of commutators.

If the ground state is a singlet ($S_G=0$), $P_0 H_{12}P_0$ is zero. This can be seen most easily by writing the Hamiltonian in terms of irreducible spherical tensors \cite{quitoprb2016}. Then, $H_{1,2}$ and $H_{3,4}$ involve irreducible spherical tensors of rank 1 or above
of spins $\mathbf{S}_{2}$ and $\mathbf{S}_{3}$, respectively. In fact, when the Hamiltonians are just bi-linear, like the XXZ or the Heisenberg cases, 
the irreducible spherical operators are the spin components themselves, which are of rank 1. From the Wigner-Eckart theorem, the projection of tensors of rank 1 or above onto a singlet state is zero \cite{cohen1977quantum}. Therefore, this contribution vanishes.

If the local ground state carries finite angular momentum $S_{G}\ne0$, $P_{S_{G}} H_{12}P_{S_{G}}$ is the renormalized Hamiltonian (obtained in first-order perturbation theory) that connects the spin 1 to the new effective spin $\mathbf{S}_{(23)}$ representing the ground pair multiplet $S_G$
\begin{equation}
\tilde{H}_{1,(23)}\equiv P_{S_{G}}H_{1,2}P_{S_{G}}.\label{eq:h123}
\end{equation}
Thus,
\begin{equation}
\tilde{\tau}_{1}=i\left[S_{1}^{z},\tilde{H}_{1,(23)}\right].\label{eq:multirentau1}
\end{equation}
For ground states of finite angular momentum, this is the final result. The
renormalization of $\tau_{3}$ follows from similar steps, 
\begin{equation}
\tilde{\tau}_{3}=-i\left[S_{4}^{z},\tilde{H}_{(23),4}\right].\label{eq:multirentau3}
\end{equation}

If the local ground state of sites 2 and 3 is a singlet we need to go to higher order. The leading
order terms come from matrix elements between either the bra or the ket corrected to first
order in perturbation theory, Eq.~\eqref{eq:eigen_order_1-1}, and the zeroth order multiplet states, Eq.~\eqref{eq:eigen_order_0-1} (for guidance, see the first line of Table \ref{tab:Summary-renorm}). Thus, we need
\begin{align}
\tilde{\tau}_{1} & =P_{0}VR\tau_{1}P_{0}+P_{0}\tau_{1}RVP_{0}.
\label{eq:tau1first}
\end{align}
Since $\left[S_{1}^{z},H_{3,4}\right]=0$ we rewrite $\tau_{1}$ as
\begin{equation}
\tau_{1}=i\left[S_{1}^{z},V\right],
\end{equation}
as $V=H_{1,2}+H_{3,4}$. We then get
\begin{align}
\tilde{\tau}_{1} & =i\left(P_{0}VR\left[S_{1}^z,V\right]P_{0}+P_{0}\left[S_{1}^z,V\right]RVP_{0}\right),\nonumber \\
 & =iP_{0}V\left[R,S_{1}^z\right]VP_{0}+iP_{0}\left[S_{1}^z,VRV\right]P_{0}.
\end{align}
Since $\left[R,S_{1}^z\right]=0$,
\begin{align}
\tilde{\tau}_{1} & =i\left[S_{1}^z,P_{0}VRVP_{0}\right],\nonumber \\
 & =i\left[S_{1}^z,\tilde{H}_{1,4}\right],
\end{align}
where 
\begin{eqnarray}
\tilde{H}_{1,4} & = & P_{0}VRVP_{0}
\end{eqnarray}
is a closed-form generic expression for the renormalized Hamiltonian and its couplings constants after decimating a singlet pair, valid for any Hamiltonian, and extensively used before in studying disordered spin systems. The renormalization rules for the coupling constants in the case of any SU(2)-symmetric spin-$S$ Hamiltonian were given in Ref.~\cite{quitoprb2016}. Finally, we find 
\begin{equation}
\tilde{\tau}_{1}=i\left[S_{1}^{z},\tilde{H}_{1,4}\right].\label{eq:tau-1-final}
\end{equation}
The calculation of the renormalized $\tilde{\tau}_{3}$ for singlet ground states follows similarly, resulting in 
\begin{equation}
\tilde{\tau}_{3}=-i\left[S_{4}^{z},\tilde{H}_{1,4}\right].
\end{equation}
Since the SDRG step keeps all the underlying symmetries [in this case,  SO(2)]~\footnote{In fact, SU(2) invariance is kept for the class of Hamiltonians invariant under any rotation, but the SO(2) symmetry is sufficient for this calculation}, $\left[S_{1}^{z}+S_{4}^{z},\tilde{H}_{1,4}\right]=0$, and we conclude
that, for singlet ground states, 
\begin{align}
\tilde{\tau}_{3} & =-i\left[S_{4}^{z},\tilde{H}_{1,4}\right]=i\left[S_{1}^{z},\tilde{H}_{1,4}\right]=\tilde{\tau}_{1}.\label{eq:singletrentau13}
\end{align}

Eqs.~\eqref{eq:multirentau1}, \eqref{eq:multirentau3}
and \eqref{eq:singletrentau13} give two of the three contributions to the renormalized spin
current, for any value of $S_{G}$. The route we followed in the derivation makes manifest an important result:
the current \emph{retains its form, but with coupling constants which
are precisely those in the renormalized Hamiltonian}. This will be shown
to hold also in the case of the last contribution, $\tau_{2}$, in
the next subsection. We stress how the commutator form of the current
operator in Eq.~\eqref{eq:tau-def} was central to this proof.

A byproduct of this result is that we can now read the leading order in ${\Omega}$ of the renormalized current off the order of the renormalized Hamiltonian.
In particular, note the following contrast between the renormalized
current from decimations of singlets and from finite angular momentum multiplets.
In a singlet decimation, the couplings in the effective Hamiltonian
$\tilde{H}$ are of order ${\Omega^{-1}}$, and so is the
renormalized current (as expected since the correction of the bra
or ket is in first-order perturbation theory). For decimations with $S_{G}\ne0$ ground
multiplets, the effective couplings are $\mathcal{O}(\Omega^{0})$, which is also reflected in the renormalized
currents (see Table \ref{tab:Summary-renorm}).

\subsection{Renormalization of $\tau_{2}$}

The aim of this subsection is to find the renormalization of $\tau_{2}$.
The operator $\tau_{2}$ is given in Eq.~\eqref{eq:tau-def}, 
\begin{equation}
\tau_{2}=i\left[S_{2}^{z},H_{2,3}\right].\label{eq:tau2-def}
\end{equation}
In general, the matrix elements of the operator $\tau_{2}$, which
acts only on the spins 2 and 3, can be easily calculated in the basis
of their total angular momentum. When sandwiched between any two projectors on $S$,$S^\prime$ multiplets,
it yields 
\begin{align}
P_{S}\tau_{2}P_{S^{\prime}} & =iP_{S}\left[S_{2z},H_{23}\right]P_{S^{\prime}}\nonumber \\
 & =i\left(E_{S^{\prime}}-E_{S}\right)P_{S}S_{2z}P_{S^{\prime}}\nonumber \\
 & \equiv -i\,\Delta E\left(S,S^{\prime}\right)P_{S}S_{2z}P_{S^{\prime}}.\label{eq:tau2-matrix-element}
\end{align}
Notice that, when
computed between  ground and excited states, the matrix element
is proportional to $\Omega$.

The task now is to find the renormalized $\tau_{2}$. We will keep the discussion generic, valid for both singlet and non-singlet ground states of the decimated pair, and make the proper distinctions when necessary.

The zeroth order correction is always zero, since 
\begin{align}
\tilde{\tau}_{2}^{\left(0\right)} & =P_{S_{G}}\tau_{2}P_{S_{G}}=0,\label{eq:tau2-order0}
\end{align}
since, from Eq.~\eqref{eq:tau2-matrix-element}, $\Delta E=0$ when $S=S^\prime=S_{G}$.

The contribution to next order comes from correcting the bra or ket to first order [see Eq.~\eqref{eq:eigen_order_1-1}], that is,
\begin{align}
\tilde{\tau}_{2}^{\left(1\right)} & =P_{S_{G}}VR\tau_{2}P_{S_{G}}+P_{S_{G}}\tau_{2}RVP_{S_{G}}.\label{eq:tau-2-1-1}
\end{align}
The following operator identities are useful
\begin{align}
R\tau_{2}P_{S_{G}} & =i\left(1-P_{S_{G}}\right)S_{2z}P_{S_{G}},\label{eq:tau2-id-1}\\
P_{S_{G}}\tau_{2}R & =-iP_{S_{G}}S_{2z}\left(1-P_{S_{G}}\right).\label{eq:tau2-id-2}
\end{align}
They can be easily proven by using the explicit form of the resolvent operator, Eq.~\eqref{eq:resolvent}, and Eq.~\eqref{eq:tau2-matrix-element}.
Using these identities, $\tilde{\tau}_{2}^{\left(1\right)}$ becomes
\begin{align}
\tilde{\tau}_{2}^{\left(1\right)}= & iP_{S_{G}}V\left(1-P_{S_{G}}\right)S_{2z}P_{S_{G}} \nonumber
\\
&-iP_{S_{G}}S_{2z}\left(1-P_{S_{G}}\right)VP_{S_{G}},\nonumber \\
= & i\left(P_{S_{G}}\left[V,S_{2}^{z}\right]P_{S_{G}}+\left[P_{S_{G}}S_{2}^{z}P_{S_{G}},P_{S_{G}}VP_{S_{G}}\right]\right).\label{eq:tau2-interm}
\end{align}
where we have used $P_{S_{G}}^{2}=P_{S_{G}}$. 

If the pair ground state is
a singlet, $P_{0}VP_{0}=0$ and the second commutator is zero. As
for the first one, 
\begin{eqnarray}
P_{0}\left[V,S_{2}^{z}\right]P_{0} & = & P_{0}\left[H_{1,2},S_{2}^{z}\right]P_{0}\nonumber \\
 & = & -P_{0}\left[H_{1,2},S_{1}^{z}\right]P_{0}\nonumber \\
 & = & -\left[P_{0}H_{1,2}P_{0},S_{1}^{z}\right]=0,\label{eq:1stcomm}
\end{eqnarray}
where we used, at each step, respectively, $\left[H_{3,4},S_{2}^{z}\right]=0$,
$\left[H_{1,2},S_{1}^{z}+S_{2}^{z}\right]=0$, $\left[S_{1}^{z},P_{0}\right]=0$,
and finally that the projection of $H_{1,2}$ (and $H_{3,4}$) onto
a singlet ground state is zero, as already discussed in the previous subsection. We conclude, therefore, that
for singlet ground states, $\tilde{\tau}_{2}$ vanishes.

As for non-singlet ground states, we can repeat the steps in Eq.~\eqref{eq:1stcomm}
except for the last one and write for the first commutator of Eq.~\eqref{eq:tau2-interm}, using Eq.~\eqref{eq:h123},
\begin{align}
iP_{S_{G}}\left[V,S_{2}^{z}\right]P_{S_{G}} & =-i\left[P_{S_{G}}H_{1,2}P_{S_{G}},S_{1}^{z}\right],\nonumber \\
 & =-i\left[\tilde{H}_{1,(23)},S_{1}^{z}\right].\label{eq:tau-final-order-1}
\end{align}
The second term of Eq.~\eqref{eq:tau2-interm} involves $P_{S_{G}}S_{2}^{z}P_{S_{G}}$,
the projection of the $S_{2}^{z}$ spin onto the ground-state multiplet.
With the help of the projection theorem~\cite{cohen1977quantum}, a corollary of the Wigner-Eckart theorem~\cite{Westerberg1997}, this can be written as 
\begin{equation}
P_{S_{G}}S_{2}^{z}P_{S_{G}}=g\tilde{S}_{\left(23\right)}^{z},
\end{equation}
with the Land\'e $g$-factor
\begin{equation}
g=\frac{S_{G}\left(S_{G}+1\right)+S_{2}\left(S_{2}+1\right)-S_{3}\left(S_{3}+1\right)}{2S_{G}\left(S_{G}+1\right)}.\label{eq:Lande}
\end{equation}
Note that when $S_{2}=S_{3}$, $g=1/2$, as expected by symmetry.
The second term of Eq.~\eqref{eq:tau2-interm} then becomes 
\begin{align}
&i\left[P_{S_{G}}S_{2}^{z}P_{S_{G}},P_{S_{G}}VP_{S_{G}}\right] \nonumber \\
&=ig\left[\tilde{S}_{(23)}^{z},\tilde{H}_{1,(23)}+\tilde{H}_{(23),4}\right]\nonumber \\
 & =ig\left[\tilde{S}_{(23)}^{z},\tilde{H}_{1,(23)}\right]+ig\left[\tilde{S}_{(23)}^{z},\tilde{H}_{(23),4}\right]\nonumber \\
 & =-ig\left[S_{1}^{z},\tilde{H}_{1,(23)}\right]+ig\left[\tilde{S}_{(23)}^{z},\tilde{H}_{(23),4}\right].\label{eq:tau-finalB-order-1}
\end{align}
Adding the contributions from Eqs.~\eqref{eq:tau-final-order-1}
and \eqref{eq:tau-finalB-order-1}, 
\begin{equation}
\tilde{\tau}_{2}=i(1-g)\left[S_{1}^{z},\tilde{H}_{1,(23)}\right]-ig\left[S_{4}^{z},\tilde{H}_{(23),4}\right].
\end{equation}
This is the final result for non-singlet ground states. The contributions of different terms in the Hamiltonian are weighted by the corresponding Land\'e factors.

For singlet ground states, the leading finite term requires second order corrections. Thus, we must consider 
\begin{align}
\tilde{\tau}^{(2)}_{2,a} & =P_{0}VR\tau_{2}RVP_{0},\label{eq:tau2_1_1_ren}
\end{align}
which comes from both bra and ket corrections to first order [see Eq.~\eqref{eq:eigen_order_1-1}] and
\begin{align}
\tilde{\tau}^{(2)}_{2,b} & =P_{0}\tau_{2}RVRVP_{0},\label{eq:tau2_0_2_ren}\\
\tilde{\tau}^{(2)}_{2,c} & =P_{0}VRVR\tau_{2}P_{0},\label{eq:tau2_2_0_ren}
\end{align}
which correspond to a matrix element between a ket or a bra correction to second order [Eq.~\eqref{eq:eigen_order_2-2}] and an unperturbed state, respectively.
The contribution $\tilde{\tau}^{(2)}_{2,a}$ can be written in terms of the commutator
form of $\tau_{2}$ as 
\begin{align}
\tilde{\tau}^{(2)}_{2,a} & =iP_{0}\left(VR\left[S_{2}^{z},H_{2,3}-E_{0}\right]RV\right)P_{0}.
\end{align}
We introduced the constant $E_{0}$ in the commutator for later convenience. Expanding the
commutator and using $\left(H_{2,3}-E_{0}\right)R=R\left(H_{23}-E_{0}\right)=P_{0}-1$,
\begin{align}
\tilde{\tau}^{(2)}_{2,a}= & iP_{0}\left[VRS_{2}^{z}\left(P_{0}-1\right)V-V\left(P_{0}-1\right)S_{2}^{z}RV\right]P_{0}.
\end{align}
Recalling that $P_{0}VP_{0}=0$, $\tilde{\tau}^{(2)}_{2,a}$ is further
simplified to 
\begin{align}
\tilde{\tau}^{(2)}_{2,a} & =-iP_{0}\left(VRS_{2}^{z}V-VS_{2}^{z}RV\right)P_{0},\nonumber \\
 & =-iP_{0}V\left[R,S_{2}^{z}\right]VP_{0}.\label{eq:tau2c}
\end{align}
We leave this result for now and focus on the contributions $\tilde{\tau}^{(2)}_{2,b}$
and $\tilde{\tau}^{(2)}_{2,c}$ from Eqs.~\eqref{eq:tau2_0_2_ren} and \eqref{eq:tau2_2_0_ren}.
We now particularize
Eqs.~\eqref{eq:tau2-id-1} and \eqref{eq:tau2-id-2}
to the case when ${S_G}=0$. Using $P_{0}S_{2z}P_{0}=0$,
a consequence of the Wigner-Eckart theorem, those identities become
\begin{align}
R\tau_{2}P_{0} & =iS_{2z}P_{0},\label{eq:tau2-id-singlets-1}\\
P_{0}\tau_{2}R & =-iP_{0}S_{2z}.\label{eq:tau2-ide-singlets-2}
\end{align}
Then, summing Eqs.~\eqref{eq:tau2_0_2_ren} and \eqref{eq:tau2_2_0_ren},
\begin{align}
\tilde{\tau}^{(2)}_{2,b}+\tilde{\tau}^{(2)}_{2,c} & =\left(P_{0}\tau_{2}R\right)VRVP_{0}+P_{0}VRV\left(R\tau_{2}P_{0}\right),\nonumber \\
 & =-i\left(P_{0}S_{2}^{z}VRVP_{0}-P_{0}VRVS_{2}^{z}P_{0}\right),\nonumber \\
 & =-iP_{0}\left[S_{2}^{z},VRV\right]P_{0}.
\end{align}
The commutator can be decomposed as
\begin{equation}
\left[S_{2}^{z},VRV\right]=\left[S_{2}^{z},V\right]RV+VR\left[S_{2}^{z},V\right]+V\left[S_{2}^{z},R\right]V.\label{eq:comm_identity}
\end{equation}
When sandwiched with $P_0$, the third term on the
r.h.s. cancels out with the $\tilde{\tau}^{(2)}_{2,a}$ from Eq.~\eqref{eq:tau2c}. Using $\left[S_{2}^{z},V\right]=i\tau_{1}$, the first two
terms give 
\begin{align}
\tilde{\tau}^{(2)}_{2,b}+\tilde{\tau}^{(2)}_{2,c} & =P_{0}\left(\tau_{1}RV+VR\tau_{1}\right)P_{0} =\tilde{\tau}_{1},
\end{align}
where we used Eq.~\eqref{eq:tau1first}. In summary, the contribution
$\tilde{\tau}_{2}$ has been shown to be the same as that of $\tilde{\tau}_{1}$.
The explicit form of $\tilde{\tau}_{1}$ was found in the previous subsection.

The summary of the results of this Appendix is the following. We showed
that when the local ground state of the two-site problem is a singlet,
the renormalizations of $\tau_{1,2,3}$ are all identical and given by
\begin{equation}
\tilde{\tau}_{1}=\tilde{\tau}_{2}=\tilde{\tau}_{3}=i\left[S_{1}^{z},\tilde{H}_{1,4}\right],\label{eq:current-step-final}
\end{equation}
with $\tilde{H}_{1,4}$ connecting sites 1 to 4, after sites 2 and
3 have been removed. When the pair ground state is, instead, a
multiplet of finite angular momentum, sites 1 and 4 are connected
to a new effective spin $\mathbf{S}_{(23)}$ that replaces spins 2 and 3 and mimics the
ground state multiplet of the pair. The renormalized currents are
\begin{align}
\tilde{\tau}_{1} & =i\left[S_{1}^{z},\tilde{H}_{1,(23)}\right],\\
\tilde{\tau}_{2} & =i(1-g)\left[S_{1}^{z},\tilde{H}_{1,(23)}\right]-ig\left[S_{4}^{z},\tilde{H}_{(23),4}\right],\\
\tilde{\tau}_{3} & =-i\left[S_{4}^{z},\tilde{H}_{(23),4}\right].
\end{align}
Recall that $g$ is defined in Eq.~\eqref{eq:Lande}. This is the net result of a single SDRG step. The totality of these results confirm the remark already made about the renormalized currents: they retain their form, but with the same coupling constants as in the renormalized Hamiltonian, as the commutator form makes explicit. The pre-factors $g$ and $1-g$, which appear to modify the coupling constants, can actually be ascribed to a change of the effective length of the new bonds
\begin{align}
\tilde{l}_{1,\left(2,3\right)} & =l_{1}+\left(1-g\right)l_{2},\\
\tilde{l}_{\left(2,3\right),4} & =gl_{2}+l_{3}.
\end{align}
so that the total current
\begin{eqnarray}
\tau = \sum_i l_{i}\tau_{i}
\end{eqnarray}
gets contributions from the current itself and the lengths. This choice is immaterial in one dimension, but less so in higher dimensions, due to the less trivial connectivity. Finally, as already mentioned, $g=1/2$ in the particular cases we focused on in this paper, but it need not be so in general.

\section{Vanishing second-order corrections to the states~\label{sec:Other-second-order-corrections}}

In this Appendix, we list the second-order perturbation contributions to the states that were not written in Eq.~\eqref{eq:eigen_order_2} and show that they give vanishing contributions to
the renormalization of the current operator when $S_G=0$. Recall that these second-order corrections are required only in the current
renormalization of $\tau_{2}$ and only when the ground state of the decimated pair is a singlet, $S_{G}=0$ {[}see Table~\eqref{tab:Summary-renorm}{]}.
Besides the contributions
in Eq.~\eqref{eq:eigen_order_2}, there exist also the following
terms to second-order in $V$:
\begin{align}
\left|\psi_{a}^{\left(2\right)}\right\rangle  & =-R^{2}VP_{0}VP_{0}\left|S_{G}=0,M=0\right\rangle ,\label{eq:eigen_order_2-1-1}\\
\left|\psi_{b}^{\left(2\right)}\right\rangle  & =-\frac{1}{2}P_{0}VR^{2}V\left|S_{G}=0,M=0\right\rangle .
\end{align}

The state $\left|\psi_{a}^{\left(2\right)}\right\rangle $
is identically zero because, as we have seen, $P_{0}VP_{0}=0$. The state $\left|\psi_{b}^{\left(2\right)}\right\rangle $ is not zero, but it does not contribute to the current renormalization. This state enters into the matrix elements of $\tau_{2}$ between $\left|\psi_{b}^{\left(2\right)}\right\rangle $ and the unperturbed states in \eqref{eq:eigen_order_0}. Let us call this extra contribution
$\delta\tau_{2}$, 
\begin{equation}
\delta\tau_{2}=-\frac{1}{2}\left[P_{0}\tau_{2}\left(P_{0}VR^{2}VP_{0}\right)+\left(P_{0}VR^{2}VP_{0}\right)\tau_{2}P_{0}\right].
\end{equation}
From Eq.~(\ref{eq:tau2-order0}), $P_{0}\tau_{2}P_{0}=0$, 
and this operator vanishes.

\section{Matrix elements of the current operator~\label{sec:Matrix-elements-current}}

In this Appendix, we calculate the matrix elements of the spin current
operator $\tau_{2}$. We will use the
variables $S$ and $S^{\prime}$ to denote the total angular momentum
values of the decimated spin pair $S_{2}$ and $S_{3}$. Generically, the matrix
elements read
\begin{align}
\left\langle S^{\prime}M^{\prime}\left|\tau_{2}\right|S,M\right\rangle  & =i\left\langle S^{\prime},M^{\prime}\left|\left[S_{2}^{z},H_{2,3}\right]\right|S,M\right\rangle ,\nonumber \\
 & =i\Delta E\left(S,S^{\prime}\right)\left\langle S^{\prime},M^{\prime}\left|S_{2}^{z}\right|S,M\right\rangle ,
\end{align}
with $\Delta E\left(S,S^{\prime}\right)$ the energy difference between
the states of total angular momentum $S$ and $S^{\prime}$.

The remaining task is to calculate $\left\langle S^{\prime},M^{\prime}\left|S_{2}^{z}\right|S,M\right\rangle $.
For completeness, this will be done in two distinct ways. First, for the spin-1 chain, we use the eigenstates of the spin-1 problem to derive the matrix elements explicitly, as this is the case of main interest in this manuscript (the spin 1/2 case has been derived before \cite{MotrunichPRB2001}). Then, we derive expressions for generic SU(2)-symmetric chains, using the tools developed in Ref.~\cite{quitoprb2016}.

\subsection{Matrix elements for the spin-1 case}

In this subsection, we derive the matrix elements using the eigenstates
of the total angular momentum operators of the two-site problem. We
denote by $S$ the total angular momentum and by $M$ its $z$ component.
The nine eigenstates of the two-site problem written in terms of eigenvalues of the $z$-components of the individual spins are the following
\begin{align}
\left|S=0,M=0\right\rangle  & =\frac{1}{\sqrt{3}}\left(\left|1,-1\right\rangle -\left|0,0\right\rangle +\left|-1,1\right\rangle \right),\\
\left|S=1,M=1\right\rangle  & =\frac{1}{\sqrt{2}}\left(\left|0,1\right\rangle -\left|1,0\right\rangle \right),\\
\left|S=1,M=0\right\rangle  & =\frac{1}{\sqrt{2}}\left(\left|-1,1\right\rangle -\left|1,-1\right\rangle \right),\\
\left|S=1,M=-1\right\rangle  & =\frac{1}{\sqrt{2}}\left(\left|-1,0\right\rangle -\left|0,-1\right\rangle \right),\\
\left|S=2,M=2\right\rangle  & =\left|1,1\right\rangle, \\
\left|S=2,M=1\right\rangle  & =\frac{1}{\sqrt{2}}\left(\left|0,1\right\rangle +\left|1,0\right\rangle \right),\\
\left|S=2,M=0\right\rangle  & =\frac{1}{\sqrt{6}}\left(\left|-1,1\right\rangle +2\left|0,0\right\rangle +\left|1,-1\right\rangle \right),\\
\left|S=2,M=-1\right\rangle  & =\frac{1}{\sqrt{2}}\left(\left|0,-1\right\rangle -\left|-1,0\right\rangle \right),\\
\left|S=2,M=-2\right\rangle  & =\left|-1,-1\right\rangle
\end{align}
The matrix elements of $S_{2}^{z}$ and $S_{3}^{z}$ follow immediately from these states. Using the notation from the main text, $\beta_{\tilde{S},\pm,M}=\left\langle \tilde{S},M\left|S_{2}^{z}\right|\tilde{S}\pm1,M\right\rangle $,
\begin{align}
\beta_{1,-,0}=\beta_{0,+,0}& =\left\langle 1,0\left|S_{2}^{z}\right|0,0\right\rangle =\sqrt{\frac{2}{3}}, \label{eq:beta1-spin-1} \\
\beta_{1,+,0} & =\left\langle 1,0\left|S_{2}^{z}\right|2,0\right\rangle =\frac{1}{\sqrt{3}}, \label{eq:beta2-spin-1} \\
\beta_{1,+,1} & =\left\langle 1,1\left|S_{2}^{z}\right|2,1\right\rangle =\frac{1}{2}, \label{eq:beta3-spin-1} \\
\beta_{1,+,-1} & =\left\langle 1,-1\left|S_{2}^{z}\right|2,-1\right\rangle =\frac{1}{2}. \label{eq:beta4-spin-1}
\end{align}

\subsection{Matrix elements for generic SU(2) Hamiltonians}

In the generic SU(2)-symmetric case, we first define
\begin{equation}
 f\left(S_{2}S_{3};SS^{\prime},MM^{\prime}\right) =\left\langle S_{2}S_{3};S^{\prime},M^{\prime}\left|S_{2}^{z}\right|S_{2}S_{3};S,M\right\rangle.
\end{equation}

Since $S_{2}^{z}$
is the zeroth component of a rank-1 irreducible spherical tensor, it follows from the Wigner-Eckart theorem that $M^{\prime}=M$, $S^{\prime}=S\pm1$, and
\begin{align}
 & f\left(S_{2}S_{3};SS^{\prime},MM^{\prime}\right)\nonumber \\
 & =\delta_{M,M^{\prime}}\left(-1\right)^{1-S+S^{\prime}}\frac{\left\langle 1,0;S,M\left|1,S;S^{\prime},M\right.\right\rangle }{\sqrt{2S^{\prime}+1}}\times\nonumber \\
\times & \left\langle S_{2}S_{3},S^{\prime}\left|\left|S_{2}^{z}\right|\right|S_{2}S_{3},S\right\rangle,\label{eq:mat_elem_Sz}
\end{align}
where $\left\langle 1,0;S,M\left|1,S;S^{\prime},M\right.\right\rangle$ is a Clebsch-Gordan coefficient and $\left\langle S_{2}S_{3},S^{\prime}\left|\left|S_{2}^{z}\right|\right|S_{2}S_{3},S\right\rangle $ is the so-called reduced matrix element \cite{Edmondsbook}, which is independent of $M$ and $M^{\prime}$. We are using the notation of ref.~\cite{Edmondsbook} for the Clebsch-Gordan coefficient, according to which
$\langle J_{1},M_{1};J_{2},M_{2} | J_{1},J_{2};J,M \rangle$ denotes the sum of angular momenta $J_1$ and $J_2$ to form the total angular momentum $J$, with the accompanying $z$-components $M_1$, $M_2$, and $M$, respectively. The 2-site reduced matrix element can be recasted
in terms of the reduced matrix element $\left\langle S_{2}\left|\left|S_{2}^{z}\right|\right|S_{2}\right\rangle $ that involves only the site 2~\cite{Edmondsbook},
\begin{align}
\left\langle S_{2}S_{3},S^{\prime}\left|\left|S_{2}^{z}\right|\right|S_{2}S_{3},S\right\rangle  & =\left(-1\right)^{S_{2}+S_{3}+S+1}\times\nonumber \\
 & \times\left[\left(2S+1\right)\left(2S^{\prime}+1\right)\right]^{1/2}\times\nonumber \\
 & \times\left\{ \begin{array}{ccc}
S_{2} & S^{\prime} & S_{3}\\
S & S_{2} & 1
\end{array}\right\} \left\langle S_{2}\left|\left|S_{2}^{z}\right|\right|S_{2}\right\rangle .\label{eq:red_mat_elem_Sz}
\end{align}
The object within braces is a Wigner $6j$ symbol \cite{Edmondsbook}. The remaining
reduced matrix element $\left\langle S_{2}\left|\left|S_{2}^{z}\right|\right|S_{2}\right\rangle $
is simply evaluated by using the Wigner-Eckart theorem to compute the expectation value of $S_{2}^{z}$ in $\left|S_{2},M_{2}\right\rangle$ and solving for $\left\langle S_{2}\left|\left|S_{2}^{z}\right|\right|S_{2}\right\rangle $,
\begin{align}
\left\langle S_{2}\left|\left|S_{2}^{z}\right|\right|S_{2}\right\rangle  & =-\frac{\sqrt{2S_{2}+1}\left\langle S_{2},M_{2}\left|S_{2}^{z}\right|S_{2},M_{2}\right\rangle }{\left\langle 1,0;S_{2},M_{2}\left|1,S_{2};S_{2},M_{2}\right.\right\rangle },\nonumber \\
 & =\sqrt{S_{2}\left(S_{2}+1\right)\left(2S_{2}+1\right)},\label{eq:red_S2z_2}
\end{align}
where we used
\begin{align}
\left\langle 1,0;S_{2},M_{2}\left|1,S_{2};S_{2},M_{2}\right.\right\rangle  & =-\frac{M_{2}}{\sqrt{S_{2}\left(S_{2}+1\right)}},\\
\left\langle S_{2},M_{2}\left|S_{2}^{z}\right|S_{2},M_{2}\right\rangle  & =M_{2}.
\end{align}
Plugging \eqref{eq:red_S2z_2} into \eqref{eq:red_mat_elem_Sz} and
taking the result into \eqref{eq:mat_elem_Sz},
\begin{align}
f\left(S_{2}S_{3},SS^{\prime},MM^{\prime}\right) & =\delta_{M,M^{\prime}}\left(-1\right)^{S_{2}+S_{3}+S+1}\times\nonumber \\
 & \left[S_{2}\left(S_{2}+1\right)\left(2S_{2}+1\right)\right]^{1/2}\times\nonumber \\
 & \times g\left(S,S^{\prime},M\right)\left\{ \begin{array}{ccc}
S_{2} & S^{\prime} & S_{3}\\
S & S_{2} & 1
\end{array}\right\} ,\label{eq:f-matrix-element}
\end{align}
where we defined
\begin{align}
g\left(S,S^{\prime},M\right) & =\sqrt{2S+1}\left\langle 1,0;S,M\left|1,S;S^{\prime},M\right.\right\rangle \nonumber \\
 & =\begin{cases}
\sqrt{S+1-\frac{M^{2}}{S+1}}, & S^{\prime}=S+1,\\
-\sqrt{S-\frac{M^{2}}{S}}, & S^{\prime}=S-1.
\end{cases}\label{eq:g-function}
\end{align}
Equation \eqref{eq:f-matrix-element} is the final result for generic spin-$S$ chains. The coefficient $\beta$ defined in the main text in Eq.~\eqref{eq:betadef} depends, in the general case, on the spins $S_2$ and $S_3$ that are decimated. This is why, in this Appendix, we will switch to an unambiguous notation and write it as $\beta_{S,\pm,M}\left(S_{2},S_{3}\right)$. We keep the previous lighter notation in the rest of the paper in each of the particular cases of $s=1/2$ or $s=1$ because it gives rise to no ambiguity. It is then given by
\begin{equation}
\beta_{S,\pm,M}\left(S_{2},S_{3}\right)=f\left(S_{2}S_{3};S\pm1S,MM\right).
\end{equation}
It is straightforward to show that the particular cases of $s=1/2$ and $s=1$ given in of Eqs.~\eqref{eq:beta1-spin-1}-\eqref{eq:beta4-spin-1} are recovered from these general formulas. 

\section{Treating powers of the broadened delta function~\label{sec:delta_n}}
The goal of this Appendix is to justify the treatment of terms involving powers of the $\delta$-function. We will show that the choice of keeping one of the deltas as ``delta sharp'' while broadening the other ones produces the leading term in $\omega$ for small $\omega$, which justifies our treatment. The reasoning is analogous to the Sommerfeld expansion for fermionic systems, in which the derivative of the Fermi-Dirac distribution becomes concentrated around the Fermi surface at low temperatures. The idea is to treat powers of the Lorentzian as  kernels sharply concentrated around $x=a$, with a  width of order $b$. The $n$-th power of the Lorentzian is thus 
\begin{equation}
K_{n,b}(x-a)
=
\frac{b^n}{\pi^n}
\frac{1}{\left[(x-a)^2+b^2\right]^n}.
\end{equation}
We will focus on the small-$b$ limit. We assume that
\begin{equation}
a>0, \qquad a\gg b>0.
\end{equation}
The peak at $x=a$ is, therefore, far from the boundary at $x=0$.
For concreteness, we will first consider the case of $n=2$ and later generalize to  $n>2$.
\subsection{The $n=2$ case}
Consider the integral
\begin{equation}
I_{2}=\int_{0}^{\infty}dx\,f(x)K_{2,b}\left(x-a\right),
\end{equation}
where $f(x)$ is a test function.
Making the change of variables
$x=a+bt$, and using $a/b\gg 1$, we can replace the lower limit by $-\infty$
\begin{equation}
I_2=\frac{1}{\pi^2 b}
\int_{-\infty}^{\infty}
dt\,
\frac{f(a+bt)}{(1+t^2)^2}.
\end{equation}
\\
Now we expand $f$ in powers of $bt$
\begin{equation}
f(a+bt)
= f(a) + bt f'(a) + \frac{b^2t^2}{2}f''(a) + \cdots .
\end{equation}
Since the kernel is even, the term proportional to $bt$ vanishes. Moreover,
\begin{equation}
\int_{-\infty}^{\infty}
\frac{dt}{(1+t^2)^2}
= \int_{-\infty}^{\infty}
\frac{t^2\,dt}{(1+t^2)^2} =
\frac{\pi}{2}.
\end{equation}
Thus,
\begin{equation}
I_2
\simeq
\frac{f(a)}{2\pi b}
+
\frac{b}{4\pi}f''(a)
+
\cdots.
\end{equation}
Equivalently, we can write
\begin{equation}
K_{2,b}(x-a) \sim \frac{1}{2\pi b}\delta(x-a)+\frac{b}{4\pi}\delta''(x-a)+\cdots . \label{eq:K2b}
\end{equation}
The main features of this result are: (1) the leading term is proportional to $1/b$ as stated in the main text; (2) the other contributions in powers of $b$ involve derivatives of $f(x)$. In the main text, $f(x)=e^{-3x}x^{3/\psi-1}$, with $x=\Gamma_{\Omega}$ and $a=\Gamma_{\omega}$. These derivative terms are sub-leading in the small-$b$ expansion and do not contribute at small $\omega$. (3) By comparing~\eqref{eq:K2b} with the replacement $C\delta(0)\delta(x-a)$ made in the main text, we conclude that $C=1/2$. This prefactor is non-universal and depends on the choice of $K_{2,b}$. For the box representation, for instance, a simple calculation shows that $C=1$.

\subsection{Generalization to $n>2$}
Proceeding with the same steps as before, we define
\begin{equation}
I_n
\simeq
\frac{1}{\pi^n b^{n-1}}
\int_{-\infty}^{\infty}
dt\,
\frac{f(a+bt)}{(1+t^2)^n}.
\end{equation}
Again, expanding and using that the kernel is even, all odd powers of $t$ integrate to zero. 
The small-$b$ expansion has the leading terms
\begin{equation}
I_n\simeq\frac{\Gamma\left(n-\frac{1}{2}\right)}{\pi^{n-\frac{1}{2}}\Gamma(n)}\frac{1}{b^{n-1}}\left[f(a)+\frac{b^2}{2(2n-3)}f''(a)+\cdots\right].
\end{equation}
This translates into the following form of the kernel
\begin{equation}
\begin{aligned}
K_{n,b}(x-a)
&\sim
\frac{ \Gamma\left(n-\frac{1}{2}\right) }{ \pi^{n-\frac{1}{2}}\Gamma(n) } \frac{1}{b^{n-1}}\delta(x-a)+
\\
&\quad+ \frac{
\Gamma\left(n-\frac{3}{2}\right)}{
4\pi^{n-\frac{1}{2}}\Gamma(n)} \frac{1}{b^{n-3}}\delta''(x-a)+\cdots .
\end{aligned}
\end{equation}
We find that the leading term is proportional to $1/b^{n-1}$, justifying the treatment used in the main text.

\section{A modified Lorentzian representation of the delta function ~\label{sec:EDdelta}}

\begin{figure}
\includegraphics[width=0.6\columnwidth]{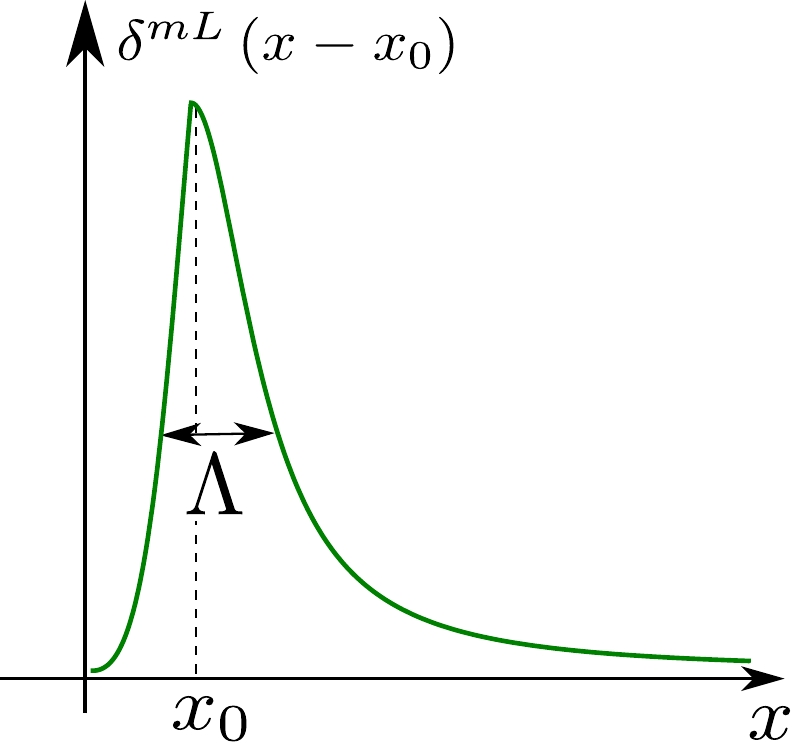}
\caption{The modified Lorentzian $\delta^{mL}\left(x-x_0\right)$ used as the delta function representation in the numerical results, Eq.~\eqref{eq:delta_approx}. For $x<x_0$, it grows as $(x/x_{0})^2$, starting from zero. The decay for $x>x_0$ is the same as that of an ordinary Lorentzian.
\label{fig:delta_schemes2}}
\end{figure}

In this Appendix, we give further details about the choice of the delta function representation. We showed in the main text the convenience of using the logarithmic scale Lorentzian representation of the delta function of Eq.~\eqref{eq:Lorentzian-function2}. In that form, it has the unwanted feature of having finite support for negative values of the strictly positive quantity $\Gamma_\omega=\ln(\Omega_0/\omega)$. We thus elected to use a modified Lorentzian that avoids this problem.

For the modified Lorentzian, we choose a continuous function of $x$ peaked at $x_0$ and identical to a Lorentzian for $x>x_0$, which, however, tends to zero when $x\rightarrow0$, thus having support \emph{only for positive arguments}:
\begin{eqnarray}
\delta^{mL}\left(x-x_0\right)&=&\frac{\kappa}{\Lambda^{2}+\left(x-x_0\right)^{2}}f(x,x_0), \label{eq:delta_approx} \\
f(x,x_0)&=&
1+\left[\left(\frac{x}{x_0}\right)^{2}-1\right]\Theta\left(x_0-x\right), \label{eq:delta-approx-w}
\end{eqnarray}
where $\Lambda$ is the broadening and
$\kappa$ guarantees that it is normalized to one. Integrating Eq.~\eqref{eq:delta_approx} in $x$ from zero to infinity we obtain

\begin{eqnarray}
1/\kappa=-\frac{\ln \left(\Lambda ^2+x_0 ^2\right)}{x_0 }-\frac{\Lambda  \tan ^{-1}\left(\frac{x_0 }{\Lambda }\right)}{x_0 ^2}+ \nonumber \\ \frac{2 \ln \Lambda }{x_0}+\frac{\tan ^{-1}\left(\frac{x_0 }{\Lambda }\right)}{\Lambda }+\frac{\pi }{2 \Lambda }+\frac{1}{x_0 }.
\end{eqnarray}

This modified Lorentzian is shown schematically in Fig.~\ref{fig:delta_schemes2}. Evidently, this function satisfies the usual requirements for a delta function representation in the limit $\Lambda \to 0$. Note that any symmetric form, such as an ordinary Lorentzian or a Gaussian, will inevitably produce weight in the negative argument region, which is undesirable. It is in order to avoid this that we choose to use this modified Lorentzian.


\begin{thebibliography}{41}%
\makeatletter
\providecommand \@ifxundefined [1]{%
 \@ifx{#1\undefined}
}%
\providecommand \@ifnum [1]{%
 \ifnum #1\expandafter \@firstoftwo
 \else \expandafter \@secondoftwo
 \fi
}%
\providecommand \@ifx [1]{%
 \ifx #1\expandafter \@firstoftwo
 \else \expandafter \@secondoftwo
 \fi
}%
\providecommand \natexlab [1]{#1}%
\providecommand \enquote  [1]{``#1''}%
\providecommand \bibnamefont  [1]{#1}%
\providecommand \bibfnamefont [1]{#1}%
\providecommand \citenamefont [1]{#1}%
\providecommand \href@noop [0]{\@secondoftwo}%
\providecommand \href [0]{\begingroup \@sanitize@url \@href}%
\providecommand \@href[1]{\@@startlink{#1}\@@href}%
\providecommand \@@href[1]{\endgroup#1\@@endlink}%
\providecommand \@sanitize@url [0]{\catcode `\\12\catcode `\$12\catcode
  `\&12\catcode `\#12\catcode `\^12\catcode `\_12\catcode `\%12\relax}%
\providecommand \@@startlink[1]{}%
\providecommand \@@endlink[0]{}%
\providecommand \url  [0]{\begingroup\@sanitize@url \@url }%
\providecommand \@url [1]{\endgroup\@href {#1}{\urlprefix }}%
\providecommand \urlprefix  [0]{URL }%
\providecommand \Eprint [0]{\href }%
\providecommand \doibase [0]{https://doi.org/}%
\providecommand \selectlanguage [0]{\@gobble}%
\providecommand \bibinfo  [0]{\@secondoftwo}%
\providecommand \bibfield  [0]{\@secondoftwo}%
\providecommand \translation [1]{[#1]}%
\providecommand \BibitemOpen [0]{}%
\providecommand \bibitemStop [0]{}%
\providecommand \bibitemNoStop [0]{.\EOS\space}%
\providecommand \EOS [0]{\spacefactor3000\relax}%
\providecommand \BibitemShut  [1]{\csname bibitem#1\endcsname}%
\let\auto@bib@innerbib\@empty
\bibitem [{\citenamefont {Igl\'oi}\ and\ \citenamefont
  {Monthus}(2005)}]{igloi-review}%
  \BibitemOpen
  \bibfield  {author} {\bibinfo {author} {\bibfnamefont {F.}~\bibnamefont
  {Igl\'oi}}\ and\ \bibinfo {author} {\bibfnamefont {C.}~\bibnamefont
  {Monthus}},\ }\bibfield  {title} {\bibinfo {title} {Strong disorder {RG}
  approach of random systems},\ }\href
  {https://doi.org/10.1016/j.physrep.2005.02.006} {\bibfield  {journal}
  {\bibinfo  {journal} {Phys. Rep.}\ }\textbf {\bibinfo {volume} {412}},\
  \bibinfo {pages} {277} (\bibinfo {year} {2005})}\BibitemShut {NoStop}%
\bibitem [{\citenamefont {Igl{\'o}i}\ and\ \citenamefont
  {Monthus}(2018)}]{Igloi2018}%
  \BibitemOpen
  \bibfield  {author} {\bibinfo {author} {\bibfnamefont {F.}~\bibnamefont
  {Igl{\'o}i}}\ and\ \bibinfo {author} {\bibfnamefont {C.}~\bibnamefont
  {Monthus}},\ }\bibfield  {title} {\bibinfo {title} {Strong disorder rg
  approach - a short review of recent developments},\ }\href
  {https://doi.org/10.1140/epjb/e2018-90434-8} {\bibfield  {journal} {\bibinfo
  {journal} {The European Physical Journal B}\ }\textbf {\bibinfo {volume}
  {91}},\ \bibinfo {pages} {290} (\bibinfo {year} {2018})}\BibitemShut
  {NoStop}%
\bibitem [{\citenamefont {Vojta}(2006)}]{Vojtareview2006}%
  \BibitemOpen
  \bibfield  {author} {\bibinfo {author} {\bibfnamefont {T.}~\bibnamefont
  {Vojta}},\ }\bibfield  {title} {\bibinfo {title} {Rare region effects at
  classical, quantum and nonequilibrium phase transitions},\ }\href
  {http://stacks.iop.org/0305-4470/39/i=22/a=R01} {\bibfield  {journal}
  {\bibinfo  {journal} {Journal of Physics A: Mathematical and General}\
  }\textbf {\bibinfo {volume} {39}},\ \bibinfo {pages} {R143} (\bibinfo {year}
  {2006})}\BibitemShut {NoStop}%
\bibitem [{\citenamefont {Fisher}(1994)}]{fisher94-xxz}%
  \BibitemOpen
  \bibfield  {author} {\bibinfo {author} {\bibfnamefont {D.~S.}\ \bibnamefont
  {Fisher}},\ }\bibfield  {title} {\bibinfo {title} {Random antiferromagnetic
  quantum spin chains},\ }\href {https://doi.org/10.1103/PhysRevB.50.3799}
  {\bibfield  {journal} {\bibinfo  {journal} {Phys. Rev. B}\ }\textbf {\bibinfo
  {volume} {50}},\ \bibinfo {pages} {3799} (\bibinfo {year}
  {1994})}\BibitemShut {NoStop}%
\bibitem [{\citenamefont {Getelina}\ and\ \citenamefont
  {Hoyos}(2020)}]{Getelina2020}%
  \BibitemOpen
  \bibfield  {author} {\bibinfo {author} {\bibfnamefont {J.~C.}\ \bibnamefont
  {Getelina}}\ and\ \bibinfo {author} {\bibfnamefont {J.~A.}\ \bibnamefont
  {Hoyos}},\ }\bibfield  {title} {\bibinfo {title} {The correlation functions
  of certain random antiferromagnetic spin-1/2 critical chains},\ }\href
  {https://doi.org/10.1140/epjb/e2019-100472-7} {\bibfield  {journal} {\bibinfo
   {journal} {The European Physical Journal B}\ }\textbf {\bibinfo {volume}
  {93}},\ \bibinfo {pages} {2} (\bibinfo {year} {2020})}\BibitemShut {NoStop}%
\bibitem [{\citenamefont {Xavier}\ \emph {et~al.}(2018)\citenamefont {Xavier},
  \citenamefont {Hoyos},\ and\ \citenamefont
  {Miranda}}]{XavierHoyosMiranda_PRB_2018}%
  \BibitemOpen
  \bibfield  {author} {\bibinfo {author} {\bibfnamefont {J.~C.}\ \bibnamefont
  {Xavier}}, \bibinfo {author} {\bibfnamefont {J.~A.}\ \bibnamefont {Hoyos}},\
  and\ \bibinfo {author} {\bibfnamefont {E.}~\bibnamefont {Miranda}},\
  }\bibfield  {title} {\bibinfo {title} {Adaptive density matrix
  renormalization group for disordered systems},\ }\href
  {https://doi.org/10.1103/PhysRevB.98.195115} {\bibfield  {journal} {\bibinfo
  {journal} {Phys. Rev. B}\ }\textbf {\bibinfo {volume} {98}},\ \bibinfo
  {pages} {195115} (\bibinfo {year} {2018})}\BibitemShut {NoStop}%
\bibitem [{\citenamefont {Damle}\ \emph {et~al.}(2000)\citenamefont {Damle},
  \citenamefont {Motrunich},\ and\ \citenamefont
  {Huse}}]{DamleMotrunichHusePRL}%
  \BibitemOpen
  \bibfield  {author} {\bibinfo {author} {\bibfnamefont {K.}~\bibnamefont
  {Damle}}, \bibinfo {author} {\bibfnamefont {O.}~\bibnamefont {Motrunich}},\
  and\ \bibinfo {author} {\bibfnamefont {D.~A.}\ \bibnamefont {Huse}},\
  }\bibfield  {title} {\bibinfo {title} {Dynamics and transport in random
  antiferromagnetic spin chains},\ }\href
  {https://doi.org/10.1103/PhysRevLett.84.3434} {\bibfield  {journal} {\bibinfo
   {journal} {Phys. Rev. Lett.}\ }\textbf {\bibinfo {volume} {84}},\ \bibinfo
  {pages} {3434} (\bibinfo {year} {2000})}\BibitemShut {NoStop}%
\bibitem [{\citenamefont {Motrunich}\ \emph {et~al.}(2001)\citenamefont
  {Motrunich}, \citenamefont {Damle},\ and\ \citenamefont
  {Huse}}]{MotrunichPRB2001}%
  \BibitemOpen
  \bibfield  {author} {\bibinfo {author} {\bibfnamefont {O.}~\bibnamefont
  {Motrunich}}, \bibinfo {author} {\bibfnamefont {K.}~\bibnamefont {Damle}},\
  and\ \bibinfo {author} {\bibfnamefont {D.~A.}\ \bibnamefont {Huse}},\
  }\bibfield  {title} {\bibinfo {title} {Dynamics and transport in random
  quantum systems governed by strong-randomness fixed points},\ }\href
  {https://doi.org/10.1103/PhysRevB.63.134424} {\bibfield  {journal} {\bibinfo
  {journal} {Phys. Rev. B}\ }\textbf {\bibinfo {volume} {63}},\ \bibinfo
  {pages} {134424} (\bibinfo {year} {2001})}\BibitemShut {NoStop}%
\bibitem [{\citenamefont {Mott}\ and\ \citenamefont
  {Twose}(1961)}]{MottTwose_1961}%
  \BibitemOpen
  \bibfield  {author} {\bibinfo {author} {\bibfnamefont {N.}~\bibnamefont
  {Mott}}\ and\ \bibinfo {author} {\bibfnamefont {W.}~\bibnamefont {Twose}},\
  }\bibfield  {title} {\bibinfo {title} {The theory of impurity conduction},\
  }\href {https://doi.org/10.1080/00018736100101271} {\bibfield  {journal}
  {\bibinfo  {journal} {Advances in Physics}\ }\textbf {\bibinfo {volume}
  {10}},\ \bibinfo {pages} {107} (\bibinfo {year} {1961})}\BibitemShut
  {NoStop}%
\bibitem [{\citenamefont {Borland}(1963)}]{Borland_1963}%
  \BibitemOpen
  \bibfield  {author} {\bibinfo {author} {\bibfnamefont {R.~E.}\ \bibnamefont
  {Borland}},\ }\bibfield  {title} {\bibinfo {title} {The nature of the
  electronic states in disordered one-dimensional systems},\ }\href
  {http://www.jstor.org/stable/2414420} {\bibfield  {journal} {\bibinfo
  {journal} {Proc. R. Soc. London A}\ }\textbf {\bibinfo {volume} {274}},\ \bibinfo {pages}
  {529} (\bibinfo {year} {1963})}\BibitemShut {NoStop}%
\bibitem [{\citenamefont {Fleishman}\ and\ \citenamefont
  {Licciardello}(1977)}]{FleishmanLicciardello_JPC_1977}%
  \BibitemOpen
  \bibfield  {author} {\bibinfo {author} {\bibfnamefont {L.}~\bibnamefont
  {Fleishman}}\ and\ \bibinfo {author} {\bibfnamefont {D.~C.}\ \bibnamefont
  {Licciardello}},\ }\bibfield  {title} {\bibinfo {title} {Fluctuations and
  localization in one dimension},\ }\href
  {https://doi.org/10.1088/0022-3719/10/6/003} {\bibfield  {journal} {\bibinfo
  {journal} {Journal of Physics C: Solid State Physics}\ }\textbf {\bibinfo
  {volume} {10}},\ \bibinfo {pages} {L125} (\bibinfo {year}
  {1977})}\BibitemShut {NoStop}%
\bibitem [{\citenamefont {Inui}\ \emph {et~al.}(1994)\citenamefont {Inui},
  \citenamefont {Trugman},\ and\ \citenamefont
  {Abrahams}}]{Abrahamsetal_PRB_1994}%
  \BibitemOpen
  \bibfield  {author} {\bibinfo {author} {\bibfnamefont {M.}~\bibnamefont
  {Inui}}, \bibinfo {author} {\bibfnamefont {S.~A.}\ \bibnamefont {Trugman}},\
  and\ \bibinfo {author} {\bibfnamefont {E.}~\bibnamefont {Abrahams}},\
  }\bibfield  {title} {\bibinfo {title} {Unusual properties of midband states
  in systems with off-diagonal disorder},\ }\href
  {https://doi.org/10.1103/PhysRevB.49.3190} {\bibfield  {journal} {\bibinfo
  {journal} {Phys. Rev. B}\ }\textbf {\bibinfo {volume} {49}},\ \bibinfo
  {pages} {3190} (\bibinfo {year} {1994})}\BibitemShut {NoStop}%
\bibitem [{\citenamefont {Eggarter}\ and\ \citenamefont
  {Riedinger}(1978)}]{EggarterRiedinger_PRB_1978}%
  \BibitemOpen
  \bibfield  {author} {\bibinfo {author} {\bibfnamefont {T.~P.}\ \bibnamefont
  {Eggarter}}\ and\ \bibinfo {author} {\bibfnamefont {R.}~\bibnamefont
  {Riedinger}},\ }\bibfield  {title} {\bibinfo {title} {Singular behavior of
  tight-binding chains with off-diagonal disorder},\ }\href
  {https://doi.org/10.1103/PhysRevB.18.569} {\bibfield  {journal} {\bibinfo
  {journal} {Phys. Rev. B}\ }\textbf {\bibinfo {volume} {18}},\ \bibinfo
  {pages} {569} (\bibinfo {year} {1978})}\BibitemShut {NoStop}%
\bibitem [{\citenamefont {Theodorou}\ and\ \citenamefont
  {Cohen}(1976)}]{TheodorouCohen_PRB_1976}%
  \BibitemOpen
  \bibfield  {author} {\bibinfo {author} {\bibfnamefont {G.}~\bibnamefont
  {Theodorou}}\ and\ \bibinfo {author} {\bibfnamefont {M.~H.}\ \bibnamefont
  {Cohen}},\ }\bibfield  {title} {\bibinfo {title} {Extended states in a
  one-demensional system with off-diagonal disorder},\ }\href
  {https://doi.org/10.1103/PhysRevB.13.4597} {\bibfield  {journal} {\bibinfo
  {journal} {Phys. Rev. B}\ }\textbf {\bibinfo {volume} {13}},\ \bibinfo
  {pages} {4597} (\bibinfo {year} {1976})}\BibitemShut {NoStop}%
\bibitem [{\citenamefont {Anderson}\ \emph {et~al.}(1980)\citenamefont
  {Anderson}, \citenamefont {Thouless}, \citenamefont {Abrahams},\ and\
  \citenamefont {Fisher}}]{Anderson1980}%
  \BibitemOpen
  \bibfield  {author} {\bibinfo {author} {\bibfnamefont {P.~W.}\ \bibnamefont
  {Anderson}}, \bibinfo {author} {\bibfnamefont {D.~J.}\ \bibnamefont
  {Thouless}}, \bibinfo {author} {\bibfnamefont {E.}~\bibnamefont {Abrahams}},\
  and\ \bibinfo {author} {\bibfnamefont {D.~S.}\ \bibnamefont {Fisher}},\
  }\bibfield  {title} {\bibinfo {title} {New method for a scaling theory of
  localization},\ }\href {https://doi.org/10.1103/physrevb.22.3519} {\bibfield
  {journal} {\bibinfo  {journal} {Phys. Rev. B}\ }\textbf {\bibinfo {volume}
  {22}},\ \bibinfo {pages} {3519} (\bibinfo {year} {1980})}\BibitemShut
  {NoStop}%
\bibitem [{\citenamefont {Abrikosov}(1981)}]{Abrikosov1981}%
  \BibitemOpen
  \bibfield  {author} {\bibinfo {author} {\bibfnamefont {A.~A.}\ \bibnamefont
  {Abrikosov}},\ }\bibfield  {title} {\bibinfo {title} {The paradox with the
  static conductivity of a one-dimensional metal},\ }\href
  {https://doi.org/10.1016/0038-1098(81)91203-5} {\bibfield  {journal}
  {\bibinfo  {journal} {Solid State Commun.}\ }\textbf {\bibinfo {volume}
  {37}},\ \bibinfo {pages} {997} (\bibinfo {year} {1981})}\BibitemShut
  {NoStop}%
\bibitem [{\citenamefont {Markos}(2006)}]{markos_2006}%
  \BibitemOpen
  \bibfield  {author} {\bibinfo {author} {\bibfnamefont {P.}~\bibnamefont
  {Markos}},\ }\bibfield  {title} {\bibinfo {title} {Numerical analysis of the
  anderson localization},\ }\href@noop {} {\bibfield  {journal} {\bibinfo
  {journal} {Acta Physica Slovaca}\ }\textbf {\bibinfo {volume} {56}},\
  \bibinfo {pages} {p. 561–685} (\bibinfo {year} {2006})}\BibitemShut
  {NoStop}%
\bibitem [{\citenamefont {Soukoulis}\ and\ \citenamefont
  {Economou}(1981)}]{SoukoulisEconomou_PRB_1981}%
  \BibitemOpen
  \bibfield  {author} {\bibinfo {author} {\bibfnamefont {C.~M.}\ \bibnamefont
  {Soukoulis}}\ and\ \bibinfo {author} {\bibfnamefont {E.~N.}\ \bibnamefont
  {Economou}},\ }\bibfield  {title} {\bibinfo {title} {Off-diagonal disorder in
  one-dimensional systems},\ }\href {https://doi.org/10.1103/PhysRevB.24.5698}
  {\bibfield  {journal} {\bibinfo  {journal} {Phys. Rev. B}\ }\textbf {\bibinfo
  {volume} {24}},\ \bibinfo {pages} {5698} (\bibinfo {year}
  {1981})}\BibitemShut {NoStop}%
\bibitem [{\citenamefont {Mudry}\ \emph {et~al.}(2000)\citenamefont {Mudry},
  \citenamefont {Brouwer},\ and\ \citenamefont
  {Furusaki}}]{MudryFurusaki_PRB_2000}%
  \BibitemOpen
  \bibfield  {author} {\bibinfo {author} {\bibfnamefont {C.}~\bibnamefont
  {Mudry}}, \bibinfo {author} {\bibfnamefont {P.~W.}\ \bibnamefont {Brouwer}},\
  and\ \bibinfo {author} {\bibfnamefont {A.}~\bibnamefont {Furusaki}},\
  }\bibfield  {title} {\bibinfo {title} {Crossover from the chiral to the
  standard universality classes in the conductance of a quantum wire with
  random hopping only},\ }\href {https://doi.org/10.1103/PhysRevB.62.8249}
  {\bibfield  {journal} {\bibinfo  {journal} {Phys. Rev. B}\ }\textbf {\bibinfo
  {volume} {62}},\ \bibinfo {pages} {8249} (\bibinfo {year}
  {2000})}\BibitemShut {NoStop}%
\bibitem [{\citenamefont {Javan~Mard}\ \emph {et~al.}(2014)\citenamefont
  {Javan~Mard}, \citenamefont {Hoyos}, \citenamefont {Miranda},\ and\
  \citenamefont {Dobrosavljevi\ifmmode~\acute{c}\else
  \'{c}\fi{}}}]{Mard_PRB_2014}%
  \BibitemOpen
  \bibfield  {author} {\bibinfo {author} {\bibfnamefont {H.}~\bibnamefont
  {Javan~Mard}}, \bibinfo {author} {\bibfnamefont {J.~A.}\ \bibnamefont
  {Hoyos}}, \bibinfo {author} {\bibfnamefont {E.}~\bibnamefont {Miranda}},\
  and\ \bibinfo {author} {\bibfnamefont {V.}~\bibnamefont
  {Dobrosavljevi\ifmmode~\acute{c}\else \'{c}\fi{}}},\ }\bibfield  {title}
  {\bibinfo {title} {Strong-disorder renormalization-group study of the
  one-dimensional tight-binding model},\ }\href
  {https://doi.org/10.1103/PhysRevB.90.125141} {\bibfield  {journal} {\bibinfo
  {journal} {Phys. Rev. B}\ }\textbf {\bibinfo {volume} {90}},\ \bibinfo
  {pages} {125141} (\bibinfo {year} {2014})}\BibitemShut {NoStop}%
\bibitem [{\citenamefont {{Laflorencie, N.}}\ and\ \citenamefont {{Rieger,
  H.}}(2004)}]{LaflorencieRieger_EJPB_2004}%
  \BibitemOpen
  \bibfield  {author} {\bibinfo {author} {\bibnamefont {{Laflorencie, N.}}}\
  and\ \bibinfo {author} {\bibnamefont {{Rieger, H.}}},\ }\bibfield  {title}
  {\bibinfo {title} {Scaling of the spin stiffness in random spin-1/2 chains -
  crossover from pure-metallic behaviour to random singlet-localized regime},\
  }\href {https://doi.org/10.1140/epjb/e2004-00258-x} {\bibfield  {journal}
  {\bibinfo  {journal} {Eur. Phys. J. B}\ }\textbf {\bibinfo {volume} {40}},\
  \bibinfo {pages} {201} (\bibinfo {year} {2004})}\BibitemShut {NoStop}%
\bibitem [{\citenamefont {Quito}\ \emph {et~al.}(2015)\citenamefont {Quito},
  \citenamefont {Hoyos},\ and\ \citenamefont
  {Miranda}}]{Quito_PhysRevLett.115.167201}%
  \BibitemOpen
  \bibfield  {author} {\bibinfo {author} {\bibfnamefont {V.~L.}\ \bibnamefont
  {Quito}}, \bibinfo {author} {\bibfnamefont {J.~A.}\ \bibnamefont {Hoyos}},\
  and\ \bibinfo {author} {\bibfnamefont {E.}~\bibnamefont {Miranda}},\
  }\bibfield  {title} {\bibinfo {title} {Emergent su(3) symmetry in random
  spin-1 chains},\ }\href {https://doi.org/10.1103/PhysRevLett.115.167201}
  {\bibfield  {journal} {\bibinfo  {journal} {Phys. Rev. Lett.}\ }\textbf
  {\bibinfo {volume} {115}},\ \bibinfo {pages} {167201} (\bibinfo {year}
  {2015})}\BibitemShut {NoStop}%
\bibitem [{\citenamefont {Quito}\ \emph {et~al.}(2016)\citenamefont {Quito},
  \citenamefont {Hoyos},\ and\ \citenamefont {Miranda}}]{quitoprb2016}%
  \BibitemOpen
  \bibfield  {author} {\bibinfo {author} {\bibfnamefont {V.~L.}\ \bibnamefont
  {Quito}}, \bibinfo {author} {\bibfnamefont {J.~A.}\ \bibnamefont {Hoyos}},\
  and\ \bibinfo {author} {\bibfnamefont {E.}~\bibnamefont {Miranda}},\
  }\bibfield  {title} {\bibinfo {title} {Random su(2)-symmetric spin-$s$
  chains},\ }\href {https://doi.org/10.1103/PhysRevB.94.064405} {\bibfield
  {journal} {\bibinfo  {journal} {Phys. Rev. B}\ }\textbf {\bibinfo {volume}
  {94}},\ \bibinfo {pages} {064405} (\bibinfo {year} {2016})}\BibitemShut
  {NoStop}%
\bibitem [{\citenamefont {Lieb}\ \emph {et~al.}(1961)\citenamefont {Lieb},
  \citenamefont {Schultz},\ and\ \citenamefont {Mattis}}]{Lieb1961}%
  \BibitemOpen
  \bibfield  {author} {\bibinfo {author} {\bibfnamefont {E.}~\bibnamefont
  {Lieb}}, \bibinfo {author} {\bibfnamefont {T.}~\bibnamefont {Schultz}},\ and\
  \bibinfo {author} {\bibfnamefont {D.}~\bibnamefont {Mattis}},\ }\bibfield
  {title} {\bibinfo {title} {{Two soluble models of an antiferromagnetic
  chain}},\ }\href {https://doi.org/10.1016/0003-4916(61)90115-4} {\bibfield
  {journal} {\bibinfo  {journal} {Ann. Phys.}\ }\textbf {\bibinfo {volume}
  {16}},\ \bibinfo {pages} {407} (\bibinfo {year} {1961})}\BibitemShut
  {NoStop}%
\bibitem [{\citenamefont {Yang}\ and\ \citenamefont
  {Bhatt}(1998)}]{PhysRevLett.80.4562}%
  \BibitemOpen
  \bibfield  {author} {\bibinfo {author} {\bibfnamefont {K.}~\bibnamefont
  {Yang}}\ and\ \bibinfo {author} {\bibfnamefont {R.~N.}\ \bibnamefont
  {Bhatt}},\ }\bibfield  {title} {\bibinfo {title} {Generation of large moments
  in a spin-1 chain with random antiferromagnetic couplings},\ }\href
  {https://doi.org/10.1103/PhysRevLett.80.4562} {\bibfield  {journal} {\bibinfo
   {journal} {Phys. Rev. Lett.}\ }\textbf {\bibinfo {volume} {80}},\ \bibinfo
  {pages} {4562} (\bibinfo {year} {1998})}\BibitemShut {NoStop}%
\bibitem [{\citenamefont {Lin}\ \emph {et~al.}(1998)\citenamefont {Lin},
  \citenamefont {Balents},\ and\ \citenamefont {Fisher}}]{linetal98}%
  \BibitemOpen
  \bibfield  {author} {\bibinfo {author} {\bibfnamefont {H.-H.}\ \bibnamefont
  {Lin}}, \bibinfo {author} {\bibfnamefont {L.}~\bibnamefont {Balents}},\ and\
  \bibinfo {author} {\bibfnamefont {M.~P.~A.}\ \bibnamefont {Fisher}},\
  }\bibfield  {title} {\bibinfo {title} {Exact so(8) symmetry in the
  weakly-interacting two-leg ladder},\ }\href
  {https://doi.org/10.1103/PhysRevB.58.1794} {\bibfield  {journal} {\bibinfo
  {journal} {Phys. Rev. B}\ }\textbf {\bibinfo {volume} {58}},\ \bibinfo
  {pages} {1794} (\bibinfo {year} {1998})}\BibitemShut {NoStop}%
\bibitem [{\citenamefont {Quito}\ \emph {et~al.}(2019)\citenamefont {Quito},
  \citenamefont {Lopes}, \citenamefont {Hoyos},\ and\ \citenamefont
  {Miranda}}]{QuitoLopes2019PRB}%
  \BibitemOpen
  \bibfield  {author} {\bibinfo {author} {\bibfnamefont {V.~L.}\ \bibnamefont
  {Quito}}, \bibinfo {author} {\bibfnamefont {P.~L.~S.}\ \bibnamefont {Lopes}},
  \bibinfo {author} {\bibfnamefont {J.~A.}\ \bibnamefont {Hoyos}},\ and\
  \bibinfo {author} {\bibfnamefont {E.}~\bibnamefont {Miranda}},\ }\bibfield
  {title} {\bibinfo {title} {Highly symmetric random one-dimensional spin
  models},\ }\href {https://doi.org/10.1103/PhysRevB.100.224407} {\bibfield
  {journal} {\bibinfo  {journal} {Phys. Rev. B}\ }\textbf {\bibinfo {volume}
  {100}},\ \bibinfo {pages} {224407} (\bibinfo {year} {2019})}\BibitemShut
  {NoStop}%
\bibitem [{\citenamefont {Quito}\ \emph {et~al.}(2020)\citenamefont {Quito},
  \citenamefont {Lopes}, \citenamefont {Hoyos},\ and\ \citenamefont
  {Miranda}}]{QuitoLopes2020EJPB}%
  \BibitemOpen
  \bibfield  {author} {\bibinfo {author} {\bibfnamefont {V.~L.}\ \bibnamefont
  {Quito}}, \bibinfo {author} {\bibfnamefont {P.~L.~S.}\ \bibnamefont {Lopes}},
  \bibinfo {author} {\bibfnamefont {J.~A.}\ \bibnamefont {Hoyos}},\ and\
  \bibinfo {author} {\bibfnamefont {E.}~\bibnamefont {Miranda}},\ }\bibfield
  {title} {\bibinfo {title} {Emergent su(n) symmetry in disordered so(n) spin
  chains},\ }\href {https://doi.org/10.1140/epjb/e2019-100576-6} {\bibfield
  {journal} {\bibinfo  {journal} {The European Physical Journal B}\ }\textbf
  {\bibinfo {volume} {93}},\ \bibinfo {pages} {17} (\bibinfo {year}
  {2020})}\BibitemShut {NoStop}%
\bibitem [{\citenamefont {Cohen-Tannoudji}\ \emph {et~al.}(1977)\citenamefont
  {Cohen-Tannoudji}, \citenamefont {Diu},\ and\ \citenamefont
  {Lalo{\"e}}}]{cohen1977quantum}%
  \BibitemOpen
  \bibfield  {author} {\bibinfo {author} {\bibfnamefont {C.}~\bibnamefont
  {Cohen-Tannoudji}}, \bibinfo {author} {\bibfnamefont {B.}~\bibnamefont
  {Diu}},\ and\ \bibinfo {author} {\bibfnamefont {F.}~\bibnamefont
  {Lalo{\"e}}},\ }\href@noop {} {\emph {\bibinfo {title} {Quantum
  Mechanics}}},\ \bibinfo {number} {v. 2}\ (\bibinfo  {publisher} {Wiley, New York},\ \bibinfo {year} {1977})\BibitemShut {NoStop}%
\bibitem [{\citenamefont {Edmonds}(1996)}]{Edmondsbook}%
  \BibitemOpen
  \bibfield  {author} {\bibinfo {author} {\bibfnamefont {A.~R.}\ \bibnamefont
  {Edmonds}},\ }\href@noop {} {\emph {\bibinfo {title} {Angular Momentum in
  Quantum Mechanics}}}\ (\bibinfo  {publisher} {Princeton University Press},\
  \bibinfo {address} {Princeton, NJ},\ \bibinfo {year} {1996})\BibitemShut
  {NoStop}%
\bibitem [{\citenamefont {Stein}\ and\ \citenamefont {Krey}(1979)}]{Stein1979}%
  \BibitemOpen
  \bibfield  {author} {\bibinfo {author} {\bibfnamefont {J.}~\bibnamefont
  {Stein}}\ and\ \bibinfo {author} {\bibfnamefont {U.}~\bibnamefont {Krey}},\
  }\bibfield  {title} {\bibinfo {title} {Numerical studies on the anderson
  localization problem},\ }\href {https://doi.org/10.1007/BF01325624}
  {\bibfield  {journal} {\bibinfo  {journal} {Zeitschrift f{\"u}r Physik B
  Condensed Matter}\ }\textbf {\bibinfo {volume} {34}},\ \bibinfo {pages} {287}
  (\bibinfo {year} {1979})}\BibitemShut {NoStop}%
\bibitem [{\citenamefont {Fisher}\ and\ \citenamefont
  {Lee}(1981)}]{Fisher1981}%
  \BibitemOpen
  \bibfield  {author} {\bibinfo {author} {\bibfnamefont {D.~S.}\ \bibnamefont
  {Fisher}}\ and\ \bibinfo {author} {\bibfnamefont {P.~A.}\ \bibnamefont
  {Lee}},\ }\bibfield  {title} {\bibinfo {title} {Relation between conductivity
  and transmission matrix},\ }\href {https://doi.org/10.1103/physrevb.23.6851}
  {\bibfield  {journal} {\bibinfo  {journal} {Phys. Rev. B}\ }\textbf {\bibinfo
  {volume} {23}},\ \bibinfo {pages} {6851} (\bibinfo {year}
  {1981})}\BibitemShut {NoStop}%
\bibitem [{\citenamefont {Thouless}\ and\ \citenamefont
  {Kirkpatrick}(1981)}]{Thouless_IOP_1981}%
  \BibitemOpen
  \bibfield  {author} {\bibinfo {author} {\bibfnamefont {D.~J.}\ \bibnamefont
  {Thouless}}\ and\ \bibinfo {author} {\bibfnamefont {S.}~\bibnamefont
  {Kirkpatrick}},\ }\bibfield  {title} {\bibinfo {title} {Conductivity of the
  disordered linear chain},\ }\href
  {https://doi.org/10.1088/0022-3719/14/3/007} {\bibfield  {journal} {\bibinfo
  {journal} {Journal of Physics C: Solid State Physics}\ }\textbf {\bibinfo
  {volume} {14}},\ \bibinfo {pages} {235} (\bibinfo {year} {1981})}\BibitemShut
  {NoStop}%
\bibitem [{\citenamefont {Del~Maestro}\ \emph {et~al.}(2010)\citenamefont
  {Del~Maestro}, \citenamefont {Rosenow}, \citenamefont {Hoyos},\ and\
  \citenamefont {Vojta}}]{DelMaestro_PRL_2010}%
  \BibitemOpen
  \bibfield  {author} {\bibinfo {author} {\bibfnamefont {A.}~\bibnamefont
  {Del~Maestro}}, \bibinfo {author} {\bibfnamefont {B.}~\bibnamefont
  {Rosenow}}, \bibinfo {author} {\bibfnamefont {J.~A.}\ \bibnamefont {Hoyos}},\
  and\ \bibinfo {author} {\bibfnamefont {T.}~\bibnamefont {Vojta}},\ }\bibfield
   {title} {\bibinfo {title} {Dynamical conductivity at the dirty
  superconductor-metal quantum phase transition},\ }\href
  {https://doi.org/10.1103/PhysRevLett.105.145702} {\bibfield  {journal}
  {\bibinfo  {journal} {Phys. Rev. Lett.}\ }\textbf {\bibinfo {volume} {105}},\
  \bibinfo {pages} {145702} (\bibinfo {year} {2010})}\BibitemShut {NoStop}%
\bibitem [{\citenamefont {Damle}\ and\ \citenamefont {Huse}(2002)}]{Damle2002}%
  \BibitemOpen
  \bibfield  {author} {\bibinfo {author} {\bibfnamefont {K.}~\bibnamefont
  {Damle}}\ and\ \bibinfo {author} {\bibfnamefont {D.~A.}\ \bibnamefont
  {Huse}},\ }\bibfield  {title} {\bibinfo {title} {Permutation-symmetric
  multicritical points in random antiferromagnetic spin chains},\ }\href
  {https://doi.org/10.1103/PhysRevLett.89.277203} {\bibfield  {journal}
  {\bibinfo  {journal} {Phys. Rev. Lett.}\ }\textbf {\bibinfo {volume} {89}},\
  \bibinfo {pages} {277203} (\bibinfo {year} {2002})}\BibitemShut {NoStop}%
\bibitem [{\citenamefont {Hoyos}\ and\ \citenamefont
  {Miranda}(2004)}]{HoyosMiranda_PRB_2004}%
  \BibitemOpen
  \bibfield  {author} {\bibinfo {author} {\bibfnamefont {J.~A.}\ \bibnamefont
  {Hoyos}}\ and\ \bibinfo {author} {\bibfnamefont {E.}~\bibnamefont
  {Miranda}},\ }\bibfield  {title} {\bibinfo {title} {Random antiferromagnetic
  $\mathrm{SU}(n)$ spin chains},\ }\href
  {https://doi.org/10.1103/PhysRevB.70.180401} {\bibfield  {journal} {\bibinfo
  {journal} {Phys. Rev. B}\ }\textbf {\bibinfo {volume} {70}},\ \bibinfo
  {pages} {180401} (\bibinfo {year} {2004})}\BibitemShut {NoStop}%
\bibitem [{\citenamefont {Hoyos}\ \emph
  {et~al.}(2007{\natexlab{a}})\citenamefont {Hoyos}, \citenamefont {Vieira},
  \citenamefont {Laflorencie},\ and\ \citenamefont {Miranda}}]{Hoyos2007PRB}%
  \BibitemOpen
  \bibfield  {author} {\bibinfo {author} {\bibfnamefont {J.~A.}\ \bibnamefont
  {Hoyos}}, \bibinfo {author} {\bibfnamefont {A.~P.}\ \bibnamefont {Vieira}},
  \bibinfo {author} {\bibfnamefont {N.}~\bibnamefont {Laflorencie}},\ and\
  \bibinfo {author} {\bibfnamefont {E.}~\bibnamefont {Miranda}},\ }\bibfield
  {title} {\bibinfo {title} {Correlation amplitude and entanglement entropy in
  random spin chains},\ }\href {https://doi.org/10.1103/PhysRevB.76.174425}
  {\bibfield  {journal} {\bibinfo  {journal} {Phys. Rev. B}\ }\textbf {\bibinfo
  {volume} {76}},\ \bibinfo {pages} {174425} (\bibinfo {year}
  {2007}{\natexlab{a}})}\BibitemShut {NoStop}%
\bibitem [{\citenamefont {Hoyos}\ \emph
  {et~al.}(2007{\natexlab{b}})\citenamefont {Hoyos}, \citenamefont {Kotabage},\
  and\ \citenamefont {Vojta}}]{Hoyos_Vojta_PRL_2007}%
  \BibitemOpen
  \bibfield  {author} {\bibinfo {author} {\bibfnamefont {J.~A.}\ \bibnamefont
  {Hoyos}}, \bibinfo {author} {\bibfnamefont {C.}~\bibnamefont {Kotabage}},\
  and\ \bibinfo {author} {\bibfnamefont {T.}~\bibnamefont {Vojta}},\ }\bibfield
   {title} {\bibinfo {title} {Effects of dissipation on a quantum critical
  point with disorder},\ }\href {https://doi.org/10.1103/PhysRevLett.99.230601}
  {\bibfield  {journal} {\bibinfo  {journal} {Phys. Rev. Lett.}\ }\textbf
  {\bibinfo {volume} {99}},\ \bibinfo {pages} {230601} (\bibinfo {year}
  {2007}{\natexlab{b}})}\BibitemShut {NoStop}%
\bibitem [{\citenamefont {Del~Maestro}\ \emph {et~al.}(2008)\citenamefont
  {Del~Maestro}, \citenamefont {Rosenow}, \citenamefont {M\"uller},\ and\
  \citenamefont {Sachdev}}]{Sachdev_2008_PRL}%
  \BibitemOpen
  \bibfield  {author} {\bibinfo {author} {\bibfnamefont {A.}~\bibnamefont
  {Del~Maestro}}, \bibinfo {author} {\bibfnamefont {B.}~\bibnamefont
  {Rosenow}}, \bibinfo {author} {\bibfnamefont {M.}~\bibnamefont {M\"uller}},\
  and\ \bibinfo {author} {\bibfnamefont {S.}~\bibnamefont {Sachdev}},\
  }\bibfield  {title} {\bibinfo {title} {Infinite randomness fixed point of the
  superconductor-metal quantum phase transition},\ }\href
  {https://doi.org/10.1103/PhysRevLett.101.035701} {\bibfield  {journal}
  {\bibinfo  {journal} {Phys. Rev. Lett.}\ }\textbf {\bibinfo {volume} {101}},\
  \bibinfo {pages} {035701} (\bibinfo {year} {2008})}\BibitemShut {NoStop}%
\bibitem [{\citenamefont {Lindgren}(1974)}]{lindgren1974BW}%
  \BibitemOpen
  \bibfield  {author} {\bibinfo {author} {\bibfnamefont {I.}~\bibnamefont
  {Lindgren}},\ }\bibfield  {title} {\bibinfo {title} {The rayleigh-schrodinger
  perturbation and the linked-diagram theorem for a multi-configurational model
  space},\ }\href {https://doi.org/10.1088/0022-3700/7/18/010} {\bibfield
  {journal} {\bibinfo  {journal} {Journal of Physics B: Atomic and Molecular
  Physics}\ }\textbf {\bibinfo {volume} {7}},\ \bibinfo {pages} {2441}
  (\bibinfo {year} {1974})}\BibitemShut {NoStop}%
\bibitem [{\citenamefont {{E. Westerberg}}\ \emph {et~al.}(1997)\citenamefont
  {{E. Westerberg}}, \citenamefont {{A. Furusaki}}, \citenamefont {{M.
  Sigrist}},\ and\ \citenamefont {{P. A. Lee}}}]{Westerberg1997}%
  \BibitemOpen
  \bibfield  {author} {\bibinfo {author} {\bibnamefont {{E. Westerberg}}},
  \bibinfo {author} {\bibnamefont {{A. Furusaki}}}, \bibinfo {author}
  {\bibnamefont {{M. Sigrist}}},\ and\ \bibinfo {author} {\bibnamefont {{P. A.
  Lee}}},\ }\bibfield  {title} {\bibinfo {title} {Low-energy fixed points of
  random quantum spin chains},\ }\href
  {https://doi.org/10.1103/PhysRevB.55.12578} {\bibfield  {journal} {\bibinfo
  {journal} {Phys. Rev. B}\ }\textbf {\bibinfo {volume} {55}},\ \bibinfo
  {pages} {12578} (\bibinfo {year} {1997})}\BibitemShut {NoStop}%
\end{thebibliography}

%

\end{document}